\documentclass{aa}
\pdfoutput=1 
\usepackage{amsmath,amstext}
\usepackage[T1]{fontenc}
\usepackage{txfonts}
\usepackage{graphicx,subcaption}
\usepackage{color}
\usepackage{multirow}
\usepackage{rotating}
\usepackage{pdflscape}
\usepackage{float}
\usepackage{afterpage}
\usepackage{appendix}
\usepackage{placeins}
\usepackage{capt-of}
\usepackage{adjustbox}

\usepackage{natbib,twoopt}
\usepackage[breaklinks=true]{hyperref} 
\bibpunct{(}{)}{;}{a}{}{,}             
\makeatletter
  \newcommandtwoopt{\citeads}[3][][]{\href{http://adsabs.harvard.edu/abs/#3}%
    {\def\hyper@linkstart##1##2{}%
     \let\hyper@linkend\@empty\citealp[#1][#2]{#3}}}
  \newcommandtwoopt{\citepads}[3][][]{\href{http://adsabs.harvard.edu/abs/#3}%
    {\def\hyper@linkstart##1##2{}%
     \let\hyper@linkend\@empty\citep[#1][#2]{#3}}}
  \newcommandtwoopt{\citetads}[3][][]{\href{http://adsabs.harvard.edu/abs/#3}%
    {\def\hyper@linkstart##1##2{}%
     \let\hyper@linkend\@empty\citet[#1][#2]{#3}}}
  \newcommandtwoopt{\citeyearads}[3][][]%
    {\href{http://adsabs.harvard.edu/abs/#3}
    {\def\hyper@linkstart##1##2{}%
     \let\hyper@linkend\@empty\citeyear[#1][#2]{#3}}}
\makeatother

\newcommand{\xmm}{{\textit{XMM-Newton}}\/}
\newcommand{\swift}{{\textit{Swift}}\/}
\newcommand{\cha}{{\textit{Chandra}}\/}
\newcommand{\hst}{{\textit{Hubble Space Telescope}}\/}

\newcommand\T{\rule{0pt}{2.6ex}}       
\newcommand\B{\rule[-1.2ex]{0pt}{0pt}} 

\begin{document}

\title{Repeating tidal disruptions in GSN~069: Long-term evolution and constraints on quasi-periodic eruptions' models}

   \author{G. Miniutti\inst{1}
          \and
          M. Giustini\inst{1}
          \and
          R. Arcodia\inst{2}
          \and
          R. D. Saxton\inst{3}
          \and
          A. M. Read\inst{4}
          \and
          S. Bianchi\inst{5}
          \and
          K. D. Alexander\inst{6,7}
}

   \institute{Centro de Astrobiolog\'ia (CAB), CSIC-INTA, Camino Bajo del Castillo s/n, Campus ESAC, E-28692, Villanueva de la Ca\~nada, Madrid, Spain 
\and Max-Planck-Institut f\"{u}r
     extraterrestrische Physik (MPE), Giessenbachstrasse 1, 85748
     Garching bei M\"{u}nchen, Germany 
\and Telespazio-Vega UK for
     ESA, Operations Department, European Space Astronomy Centre
     (ESAC), Villanueva de la Ca\~nada, E-28692 Madrid, Spain 
\and Department of Physics \& Astronomy, University of Leicester, Leicester, LE1 7RH, UK
\and Dipartimento di Matematica e Fisica,
     Universit\`a degli Studi Roma Tre, via della Vasca Navale 84,
     00146 Roma, Italy 
\and Steward Observatory, University of Arizona, 933 North Cherry Avenue, Tucson, AZ 85721-0065, USA     
\and Center for Interdisciplinary Exploration and Research in Astrophysics (CIERA) and Department of Physics and Astronomy, Northwestern University, Evanston, IL 60208, USA
}

   \date{Received  / Accepted  }

\abstract
{GSN~069 is the first galactic nucleus where quasi-periodic eruptions
  (QPEs) have been identified in December 2018. These are
  high-amplitude, soft X-ray bursts recurring every $\sim 9$~hr,
  lasting $\sim 1$~hr, and during which the X-ray count rate increases
  by up to two orders of magnitude with respect to an otherwise stable
  quiescent level. The X-ray spectral properties and the long-term
  evolution of GSN~069 in the first few years since its first X-ray
  detection in 2010 are consistent with a long-lived tidal disruption
  event (TDE).}
{We aim to derive the properties of QPEs and of the long-term X-ray
  evolution in GSN~069 over the past $12$~yr.}
{We analyse timing and spectral X-ray data from 11 \xmm, one \cha, and
  34 \swift\ observations of GSN~069 on timescales ranging from
  minutes to years. }
{QPEs in GSN~069 are a transient phenomenon with a lifetime of
  $\gtrsim 1.05$~yr. The QPE intensity and recurrence time oscillate
  and allow for alternating strong-weak QPEs and long-short recurrence
  times to be defined. In observations with QPEs, the quiescent level
  exhibits a quasi-periodic oscillation with a period equal to the
  average separation between consecutive QPEs. The QPE spectral
  evolution is consistent with thermal emission from a very compact
  region that heats up quickly and subsequently cools down via X-ray
  emission while expanding by a factor of $\sim 3$ in radius. The
  long-term evolution of the quiescent level is characterised by two
  repeating TDEs $\sim 9$~yr apart. We detect a precursor X-ray flare
  prior to the second TDE that may be associated with the
  circularisation phase during disc formation. A similar precursor
  flare is tentatively detected just before the first TDE. }
{We provide a comprehensive summary of observational results that can
  be used to inform further theoretical and numerical studies on the
  origin of QPEs in GSN~069 and we discuss our results in terms of
  currently proposed QPE models. Future X-ray observations of GSN~069
  promise that the QPE origin and the relation between QPEs and
  repeating TDEs in this galactic nucleus will be constrained, with
  consequences for the other sources where QPEs have been
  identified. }


\titlerunning{Repeating TDEs in GSN~069: long-term evolution and constraints on QPE models}
\authorrunning{Miniutti et al.}

\maketitle

\section{Introduction}
\label{sec:intro}

GSN~069 was first detected in July 2010 during an \xmm\ slew
\citep{2011arXiv1106.3507S} at a flux level more than a factor 240
above previous upper limits from {\it{ROSAT}} observations performed
16 years earlier. Subsequent observations with the {\it Neil Gehrels}
\swift\ observatory (hereafter \swift) show a relatively constant
X-ray flux for the first $\sim 1$~yr \citep{2013MNRAS.433.1764M}, and
further observations with \swift\ and \xmm\ reveal a smooth flux decay
during the following $\sim 7$-$8$~yr. This X-ray long-term evolution
is best interpreted as the result of a tidal disruption event (TDE)
whose long-lived nature and UV spectral properties suggest the
disruption of an evolved star
\citep{2018ApJ...857L..16S,2021ApJ...920L..25S}. During \xmm\ and
\cha\ observations between December 2018 and February 2019, the X-ray
light curve of GSN~069 exhibits high-amplitude, short-lived X-ray
flares recurring every $\sim 9$~hr \citep{2019Natur.573..381M}. These
quasi-periodic eruptions (QPEs) produce an increase in the X-ray count
rate by up to two orders of magnitude in the hardest energy
bands. They are characterised by thermal-like X-ray spectra with a
typical temperature evolving from $\simeq 50$~eV up to $\sim
100-120$~eV, superimposed to an otherwise stable blackbody-like
quiescent level, most likely due to the emission from the accretion
disc resulting from the initial TDE.

Following the discovery of QPEs in GSN~069, this peculiar phenomenon
has been identified in a select number of additional sources: RX~J1301.9+2747
\citep{2020A&A...636L...2G}, eRO-QPE1 and eRO-QPE2
\citep{2021Natur.592..704A}, and most likely XMMSL1~J024916.6-041244
\citep{2021ApJ...921L..40C}. So far, QPEs have been consistent with being
associated with the nuclei of galaxies harboring relatively small mass
black holes ($10^5-{\rm{few}}\times10^6~{\rm{M}}_\odot$) as derived
from a variety of different methods
\citep{2019Natur.573..381M,2021Natur.592..704A,2022A&A...659L...2W}. None
of the sources show the broad optical emission lines that are
associated with active nuclei; however, the presence of a narrow line
region (most evident in GSN~069), combined with the lack of intrinsic
absorption capable of obscuring the broad line region, suggests some
level of past (or present, but weak) nuclear activity
\citep{2013MNRAS.433.1764M,2022A&A...659L...2W}.

As discussed by \citet{2019Natur.573..381M} the X-ray and optical
spectra of the Seyfert~2 galaxy 2XMM~J123103.2+110648
\citep{2012ApJ...752..154T} closely match those of GSN~069. Its X-ray
variability is also peculiar with a $\sim 3.8$~hr quasi-periodic
oscillation detected in different \xmm\ exposures
\citep{2012ApJ...752..154T,2013ApJ...776L..10L}, and with overall
long- and short-timescale variability properties that suggest a
possible connection with both TDEs and QPEs
\citep{2017MNRAS.468..783L,2022MNRAS.tmp.3139W}. Recurrent X-ray
flares sharing some common characteristics with QPEs have also been
recently reported in the ultra-luminous X-ray source
XMMU~J122939.7+075333 located in the globular cluster RZ~2109 in the
Virgo galaxy NGC~4472 \citep{2022A&A...661A..68T}. Other sources that
may perhaps be associated with QPEs or with the
associated phenomenology are ESO~249-39 HLX-1
\citep{2009Natur.460...73F,2014ApJ...793..105G}, ASASSN-14ko
\citep{2021ApJ...910..125P,2022arXiv220613494L}, and
eRASSt~J045650.3-203750 \citep{2022arXiv220812452L}. 

\begin{table}
        \centering
        \caption{Summary of the 11 pointed \xmm\ observations used in
          this work.}
        \label{tab:obs}
        \begin{tabular}{lccc} 
          \hline
\T  & ObsID & Date (start) & Exposure  \B \\
\hline
\T XMM1  & 0657820101 & 2010-12-02 & 13 \\
\T XMM2  & 0740960101 & 2014-12-05 & 92 \\
\T XMM3  & 0823680101 & 2018-12-24 & 50 \\
\T XMM4  & 0831790701 & 2019-01-16 & 134 \\
\T XMM5  & 0851180401 & 2019-05-31 & 132  \\
\T XMM6  & 0864330101 & 2020-01-10 & 131 \B  \\
\T XMM7  & 0864330201 & 2020-05-28 & 125 \B  \\
\T XMM8  & 0863330301 & 2020-06-03 & 126 \B  \\
\T XMM9  & 0863330401 & 2020-06-13 & 118 \B  \\
\T XMM10 & 0884970101 & 2021-06-30 & 48 \B  \\
\T XMM11 & 0884970101 & 2021-12-03 & 45 \B  \\
          \hline
        \end{tabular}
\tablefoot{The usable exposure (in ks) for all \xmm\ observations
          refers to the EPIC-pn camera. GSN~069 was first detected in
          the X-rays by an \xmm\ slew on 14 July 2010, about five months
          before XMM1.}
\end{table}

\begin{figure*}
\centering
\includegraphics[width=18cm]{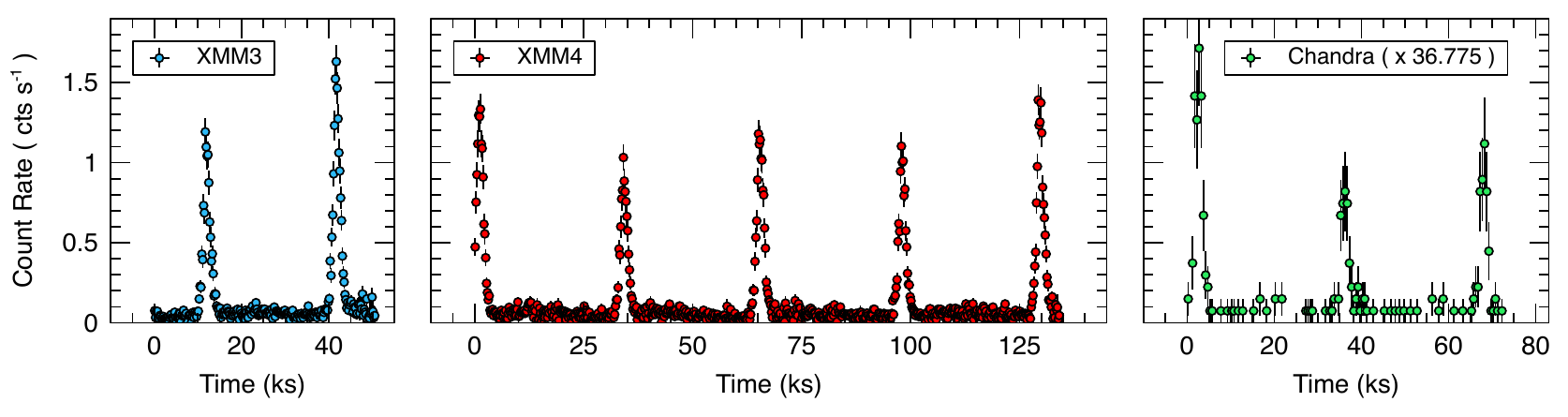}
\includegraphics[width=18cm]{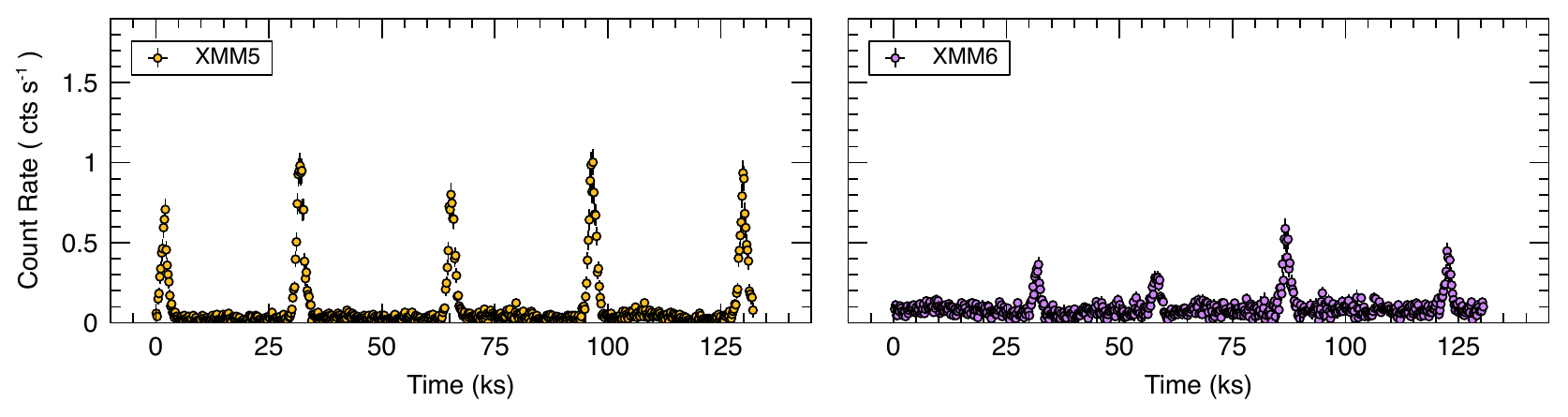}
\caption{\xmm\ (EPIC-pn) and \cha\ (ACIS-S) background-subtracted
  light curves from all observations with QPEs in a common 0.4-1~keV
  band. The \cha\ light curve has been re-scaled to the \xmm\ pn
  effective area to better display the QPE amplitude evolution and to
  ease comparison between data from different
  detectors. We use time bins of 200~s and 500~s
  duration for \xmm\ and \cha\ data respectively. All panels have a
  common y-axis range. }
\label{fig:lcs}
\end{figure*}

Several theoretical models have been considered to explain the QPE
phenomenon. Proposed scenarios include (i) modified disc instability
models
\citep{2020A&A...641A.167S,2021ApJ...909...82R,2022ApJ...928L..18P,2022arXiv221100704K};
(ii) gravitational lensing in a nearly equal-mass supermassive black
hole (SMBH) binary \citep{2021MNRAS.503.1703I}; (iii) mass transfer
from one or more orbiting bodies in a variety of different
configurations as discussed by
\citet{2020MNRAS.493L.120K,2022MNRAS.515.4344K},
\citet{2022ApJ...930..122C}, \citet{2022ApJ...926..101M},
\citet{2022A&A...661A..55Z}, \citet{2022ApJ...933..225W},
\citet{2022arXiv220902786K}, and \citet{2022arXiv221008023L}; or (iv)
collisions between an orbiting secondary body and the accretion disc
that is formed following the initial TDE
\citep{2021ApJ...917...43S,2021ApJ...921L..32X}. Most proposed models
focus on QPEs ignoring the long-term evolution of the quiescent level
emission which is particularly well monitored in GSN~069.

In this work, we present results obtained from a series of X-ray
observations of GSN~069 (11 by \xmm, one by \cha, and 34 by \swift)
and we discuss the short- and long-timescale properties of both QPEs
and continuum (quiescent) emission over the past $12$~yr. Our results
represent a significant step in deriving observational constraints
that can inform further theoretical and numerical work on the QPE
formation process and long-term evolution in GSN~069, as well as in
the other QPE-emitting galactic nuclei identified so far.

\section{Observations and data analysis}
\label{sec:obs}

GSN~069 was observed in the X-rays on several occasions with the \xmm,
\swift, and \cha\ X-ray observatories. Most of our work is based on
the 11 pointed \xmm\ observations that are reported in
Table~\ref{tab:obs}. Data also comprise one 73~ks \cha\ observation
performed on 14 February 2019; we only used \cha\ data for X-ray
variability analysis (light curve). Due to the severe contamination
affecting the low-energy response of the ACIS detector, we considered
\cha\ data above $0.4$~keV only, and we ignored them in our spectral
analysis as the X-ray emission in GSN~069 is super-soft and cannot be
reliably constrained by \cha. \xmm\ and \cha\ source and background
products were extracted from circular regions on the same detector
chip using the latest versions of the SAS (\xmm) and CIAO (\cha)
dedicated software. X-ray light curves were background subtracted, as
well as corrected for various effects (bad pixels, quantum efficiency,
vignetting, dead time) using the SAS {\texttt{epiclccorr}} and CIAO
{\texttt{dmextract}} tasks. For \xmm, we used data from the EPIC-pn
camera only for simplicity. We also analysed 34 observations of GSN
069 made with the \swift\ X-ray telescope, XRT
\citep{2005SSRv..120..165B} in photon counting mode. The \swift-XRT
observations were analysed following the procedure outlined in
\citet{2009MNRAS.397.1177E}, which uses fully calibrated data and
corrects for effects such as pile-up and the bad columns on the CCD,
to obtain count-rates on an observation-by-observation
basis. Occasionally, when the source counts in two neighbouring
observations were insufficient to obtain detection of the source, the
observations were combined. Finally, although only a minuscule effect,
the photon arrival times from all observations were
barycentre-corrected in the DE405-ICRS reference system.

\section{The transient nature and properties of QPEs in GSN~069}
\label{sec:QPEprop}

Background-subtracted light curves from all observations with QPEs are
shown in Fig.~\ref{fig:lcs} in a common 0.4-1~keV band. The
\cha\ light curve has been re-scaled to match the \xmm\ EPIC-pn
effective area to better show the QPE intensity evolution regardless
of the detector in use\footnote{We derived the correction factor by
  assuming the best-fitting spectral model for QPEs
  \citep{2019Natur.573..381M} from \xmm\ data, retrieving the model
  expected count rates from the \xmm\ EPIC-pn and \cha\ ACIS-S
  detectors in the common 0.4-1~keV band. The resulting correction
  factor by which we multiply the \cha\ light curve is $36.775$.}.
QPEs are consistently detected from December 2018 (XMM3) to January
2020 (XMM6), although they appear much weaker and more irregularly
spaced in the latter observation. No QPEs are detected during XMM2, a
$\sim 92$~ks \xmm\ observation in December 2014, that is  QPEs first
appeared in GSN~069 sometime between December 2014 and December
2018. No clear QPEs are detected in a $\sim 125$~ks observation
performed in May 2020 (XMM7) and in none of the subsequent
observations. We hence conclude that QPEs in GSN~069 are a transient
phenomenon with an observed life-time $\gtrsim 1.05$~yr. Weak QPEs
might have been present at other epochs, but undetected because of the
reduced contrast against a higher X-ray flux quiescent level, which
may increase the actual QPE life-time.

\begin{figure}
\centering
\includegraphics[width=8.8cm]{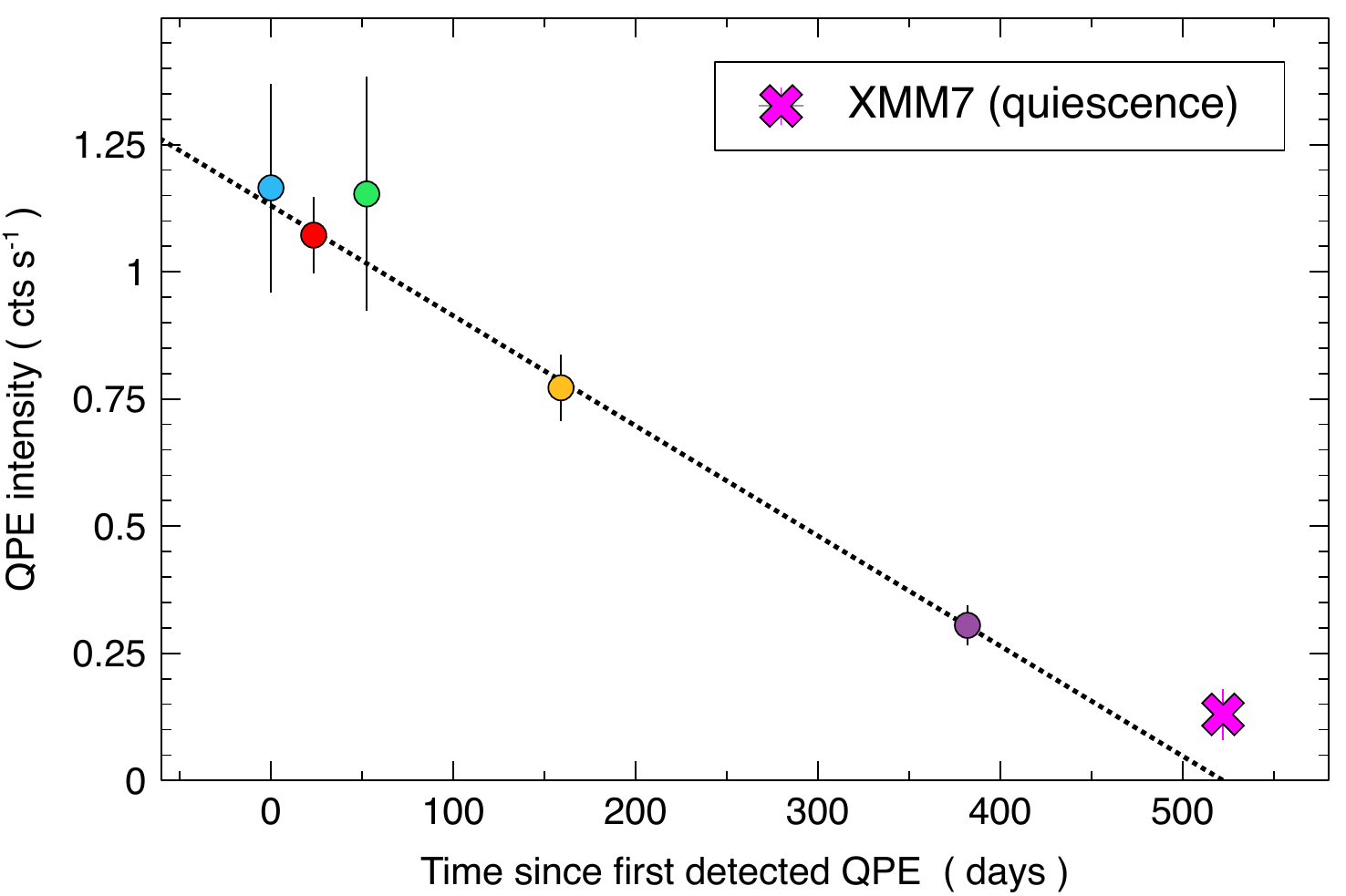}
\caption{Time evolution of the QPE intensity since first
  detection. The QPE intensity is measured as the average of the
  best-fitting Gaussian normalisation in each observation (i.e. we
  ignore here the difference between strong and weak QPEs). The colour
  scheme is the same as in Fig.~\ref{fig:lcs}. The dotted line
  represents the best-fitting linear relation (corresponding to a
  decay of about $0.22$~cts~s$^{-1}$ per 100~d). We also show (cross)
  the continuum level during the first observation with no detected
  QPEs (XMM7).  }
\label{fig:PeakEvol}
\end{figure}

We fitted each 0.4-1~keV light curve with the simplest possible model
comprising an observation-dependent constant C, representing the
average quiescent level during each exposure, and a series of Gaussian
functions with normalisation N and width $\sigma$ describing QPEs. We
define the recurrence time T$_{\mathrm{rec}}$ as the time interval
between the peak of two consecutive QPEs. The best-fitting results for
all observations are reported in Table~\ref{tab:qpefit}. We point out
that results are only valid in the 0.4-1~keV band, all quantities
being energy-dependent as discussed by \citet{2019Natur.573..381M}. In
Fig.~\ref{fig:PeakEvol}, we show the evolution of the QPE intensity,
as measured from the best-fitting normalisation of the Gaussian
functions (see Table~\ref{tab:qpefit}), averaged over each
observation. The QPE intensity decays monotonically by
$0.22$~cts~s$^{-1}$ per 100~d in the 0.4-1~keV band. We also show the
continuum level of the first observation with no QPEs (XMM7). If the
decaying trend continued after XMM6, our ability to detect QPEs was
seriously compromised already $\simeq 20$~d before the XMM7
observation.

\begin{figure*}
\centering
\includegraphics[width=18cm]{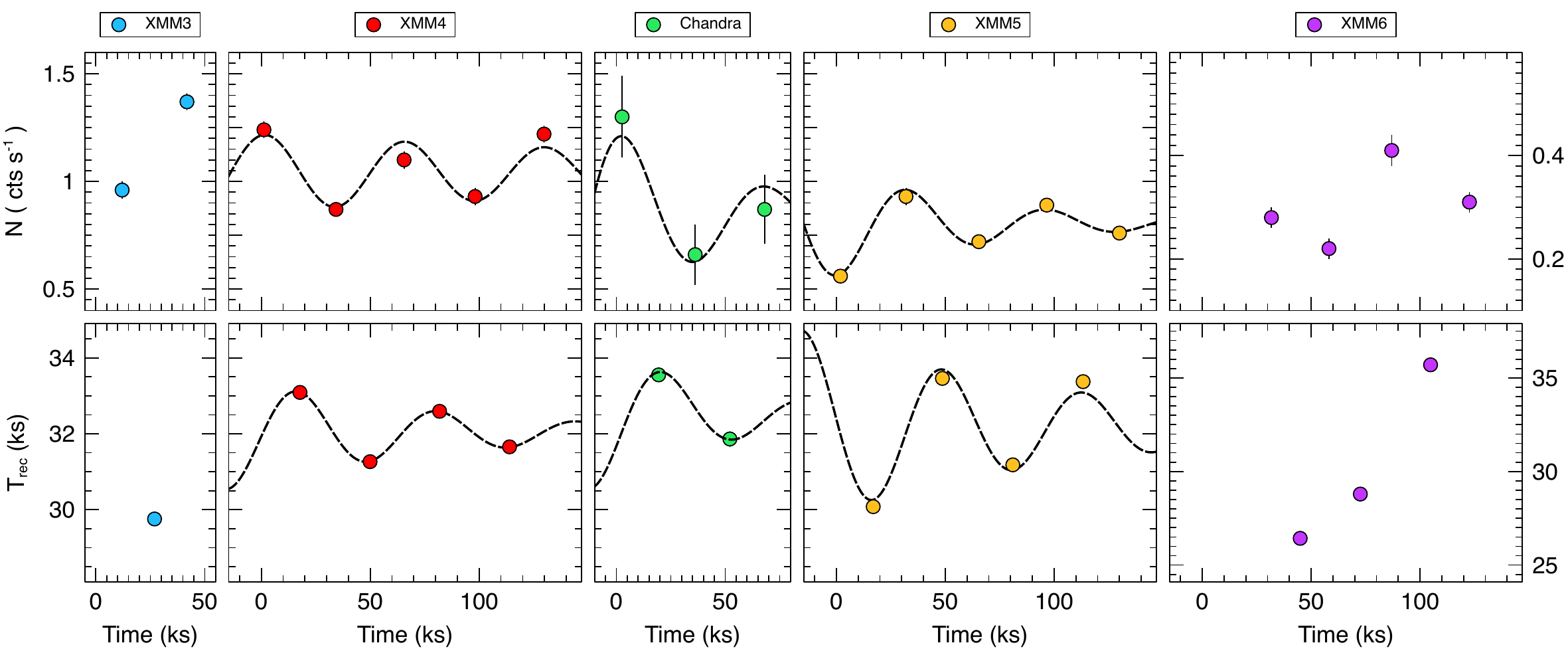}
\caption{Alternating QPE intensities and recurrence times. The QPE
  intensity (N) for all observations with QPEs is shown in the upper
  panels. The recurrence time between consecutive QPEs is shown in the
  lower panels (each data point is placed at a time corresponding to
  half the separation between consecutive QPEs). The y-axis is always
  in the same range, except for the irregular XMM6 observation. Dashed
  lines in the XMM4, Chandra, and XMM5 panels are sine functions with
  period equal to twice the averaged, observation-dependent recurrence
  time, and with an exponentially decaying amplitude that most likely
  indicates a common long-term modulation. The lines are intended to
  guide the eye rather than to provide fits to the data.}
\label{fig:ATrec}
\end{figure*}

As is visually clear from Fig.~\ref{fig:lcs} and quantitatively shown in
Table~\ref{tab:qpefit}, the QPE intensity oscillates in long enough
observations allowing for alternating strong and weak QPE
types to be defined. The same is true for the recurrence time between QPEs, with
strong (weak) QPEs being systematically followed by longer (shorter)
recurrence times to the next QPE (with the exception of the more
irregular XMM6 observation). The QPE intensity and recurrence time
evolution during all observations comprising QPEs is shown in
Fig.~\ref{fig:ATrec}. The dashed lines are sine functions with period
fixed at the observation-dependent average separation between QPEs of
the same type (strong or weak), that is at about twice the average
recurrence time between consecutive QPEs. As the difference between
strong and weak QPE intensities and between long and short T$_{\rm{rec}}$
decreases in all observations, we also impose an exponentially
decaying amplitude of the sine functions. This signals that the
intensity and recurrence time are likely modulated on a longer
timescale and in a similar way. In fact, all exponential folding times
are consistent with each other (although with very large
uncertainties) and are of the order of $1$-$2$~d. The origin of such a
longer-term modulation is being analysed, and results will be
presented in a forthcoming publication (Miniutti et al. in
preparation). The dashed lines in Fig.~\ref{fig:ATrec} are not to be
considered as actual fits, and they are only included to guide the
eye.

The behaviour is different during XMM6: while intensities still
oscillate, they do so in a much less regular manner to the extent that
it is difficult to unambiguously define the QPE type (especially for
the first and last QPE). Most importantly, the recurrence time during
XMM6 does not oscillate as in all other cases, but increases
monotonically. As is reported in Table~\ref{tab:qpefit}, the typical QPE
duration, defined here arbitrarily as twice the best-fitting Gaussian
FWHM, is $\simeq$~3780~s ($\simeq$~1.05~hr), and there is no clear
difference between strong and weak QPEs duration, nor evident
long-term evolution.

\begin{figure}
\centering
\includegraphics[width=8.8cm]{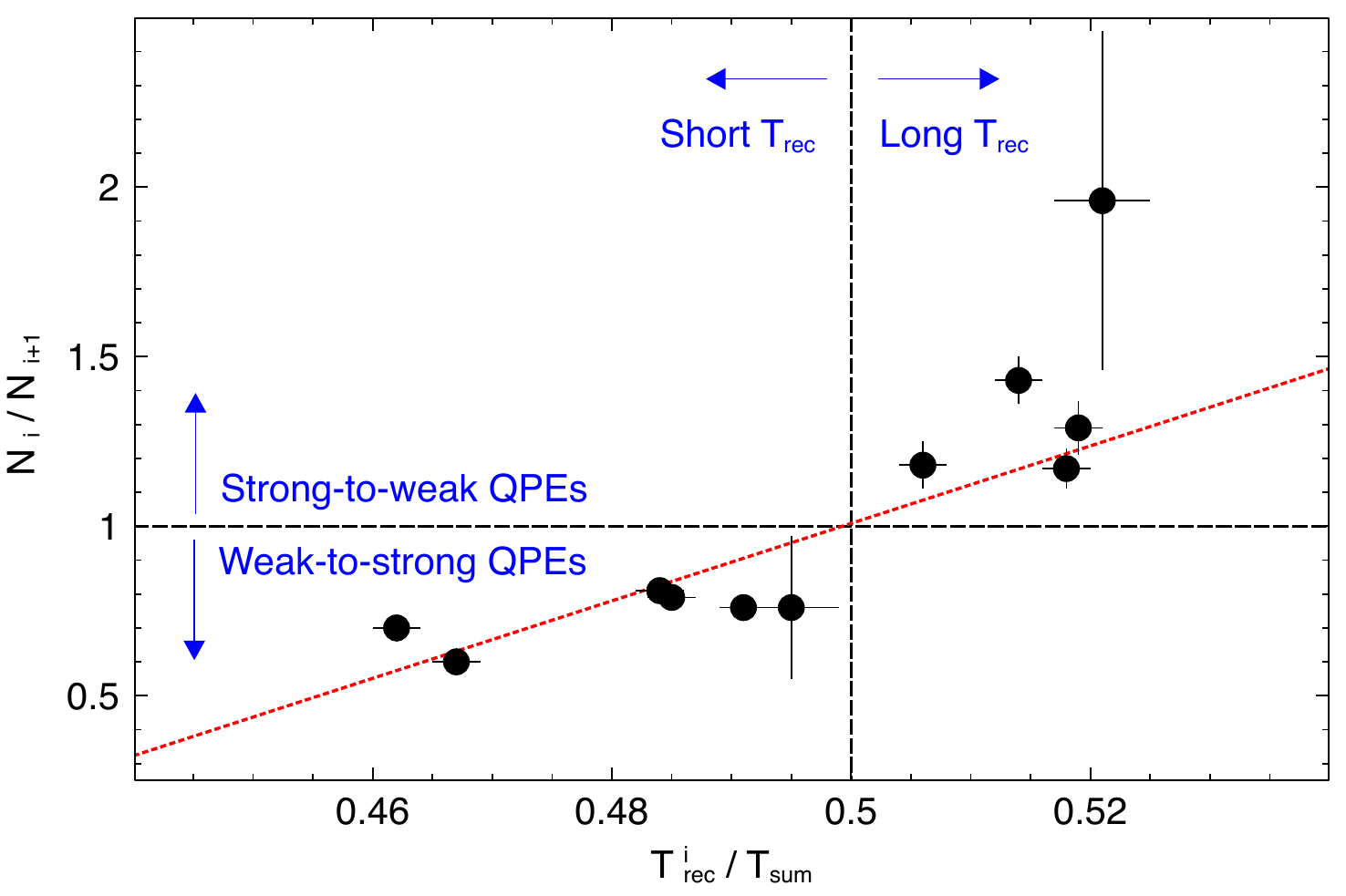}
\caption{Consecutive QPEs intensity ratio and recurrence time
  correlation. The ratio between the intensity of consecutive QPEs is
  shown as a function of the recurrence time between them (normalised
  to the sum of consecutive long and short intervals $T_{\rm{sum}}$). The
  horizontal line ($N_{i}~/N_{i+1} = 1$) separates events where
  strong QPEs precede weak ones (upper half of the Figure), from those
  where weak QPEs precede strong ones (lower half). The
  $T_{\rm{rec}}^{i}~/ T_{\rm{sum}} = 0.5$ vertical line separates
  short from long recurrence times. The (red) dotted line show the
  best-fitting linear relation between the two quantities. }
\label{fig:AversusT}
\end{figure}

In Fig.~\ref{fig:AversusT}, we plot the ratio between the intensity of
consecutive QPEs as a function of the recurrence time between
them. The recurrence time is shown as fraction of the average
separation between QPEs of the same strong or weak type (or,
equivalently, of the sum of consecutive long and short recurrence
times), $T_{\rm{sum}} \simeq64$~ks.  The two quantities are well
correlated, although significant scatter is present. The upper-left
and lower-right quadrants in Fig.~\ref{fig:AversusT} are not
populated. This means that (at least for the events that we have
observed) long recurrence times always follow strong QPEs. Remarkably,
the ratio between the intensity of consecutive QPEs is consistent with being the same
($N_{i}~/N_{i+1} = 1$) when long and short recurrence times have
equal duration ($T_{rec}^{i}~/ T_{\rm{sum}} = 0.5$), as is shown by the
intersection of the vertical and horizontal lines with the
best-fitting linear relation. Fig.~\ref{fig:AversusT} suggests an
oscillatory behaviour of consecutive QPE intensity and separation
around a mean in which intensities and separations of consecutive QPEs
are equal. This configuration is never observed exactly, and is almost
reached for the 3rd-4th QPEs during XMM4 when $N_{i}~/N_{i+1} =
1.18\pm0.07$ and $T_{rec}^{i}~/ T_{\rm{sum}} = 0.506\pm 0.002$. The
QPE X-ray spectral evolution is presented in
Sect.~\ref{sec:QPEspectral} together with estimates of the total
radiated energy from QPEs in GSN~069.

\begin{figure*}
\centering
\includegraphics[width=18cm]{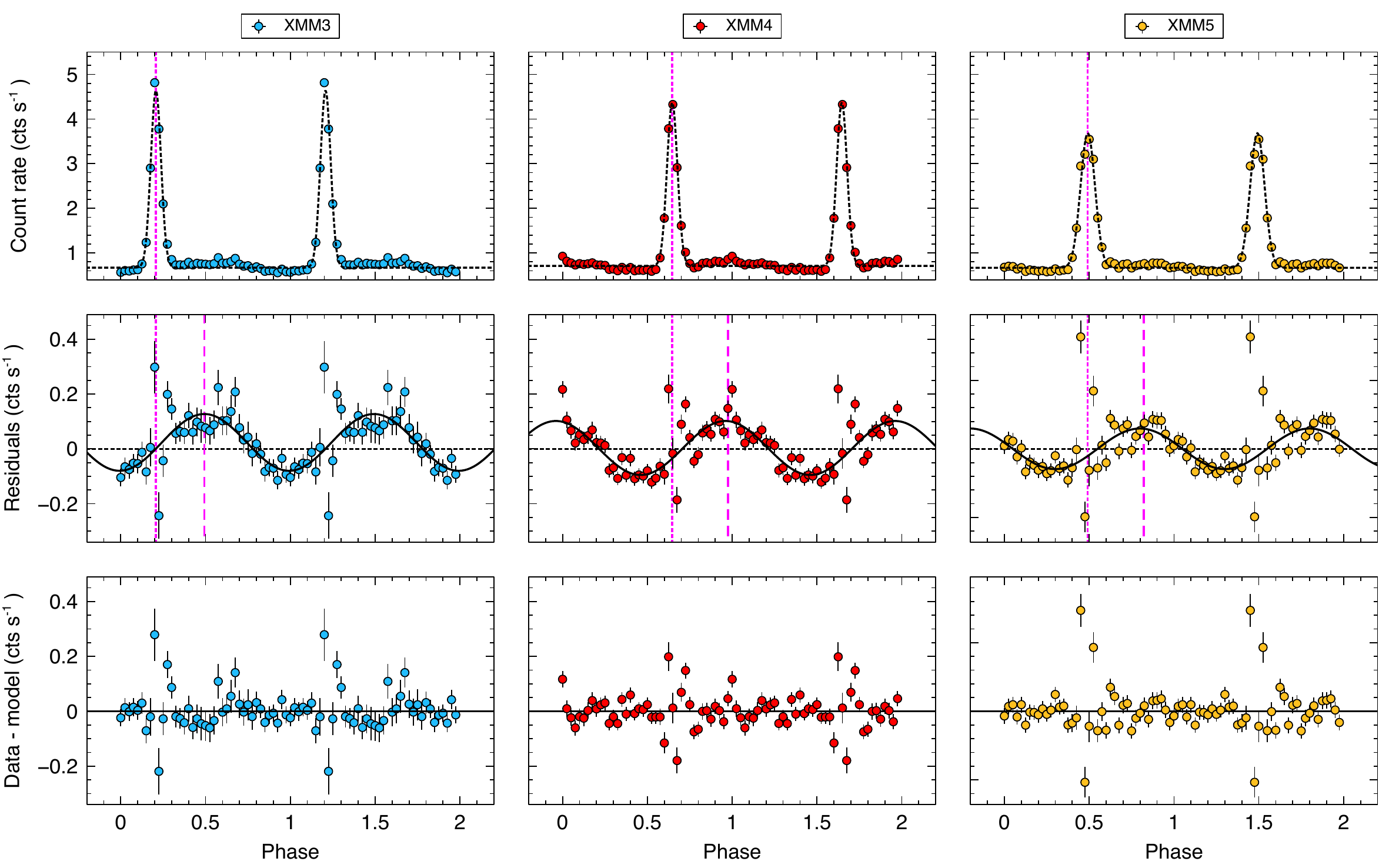}
\caption{Quasi-periodic oscillation of the quiescent level. In the
  upper panels, we show the 0.2-1~keV light curves from the XMM3,
  XMM4, and XMM5 folded at the average observation-dependent
  recurrence time, and the corresponding best-fitting baseline model
  (constant plus Gaussian functions, dotted). We show the 0-2 phase
  interval for visual clarity. In the middle panels, we show the
  resulting residual light curves together with a sine function with
  period fixed to 1 (in phase space) to guide the eye. The vertical
  lines represent the phase of the primary QPE (dotted) and of the
  peak of the sine function (dashed). In the lower panel, we show the
  residual for a fit of the original folded light curve that comprises
  a sine function with period fixed to 1 (i.e. to the folding
  timescale) in addition to the baseline model.  }
\label{fig:fold}
\end{figure*}

\section{Quasi-periodic oscillation of the quiescent level (or secondary QPEs)}
\label{sec:QPO}

The statistical quality of the fits to the individual light curves
with the simple constant plus Gaussian model is fair, but not
excellent (see Table~\ref{tab:qpefit}). This is not due to inaccurate
modelling of the QPE profiles, but rather to residual quiescent level
variability. Visual inspection of the best-fitting residuals reveals
that excess emission is systematically present $\sim$~10~ks after most
QPEs. As discussed in Sect.~\ref{app:QPE-QPO} and Fig.~\ref{fig:qpos},
and excluding the irregular XMM6 observation, the excess is quite
clear after nine of the ten QPEs observed with \xmm\ for which sufficient
data following a QPE exist (the \cha\ data do not have sufficient
quality to reveal variability of the quiescent level, as this is
barely detected). In fact, the quiescent level appears to vary
systematically with a characteristic timescale similar to the average
observation-dependent recurrence time between consecutive QPEs, except
during the XMM6 observation where excess emission is tentatively seen
following QPEs, but with distinct properties (e.g. duration) with
respect to all others (see Fig.~\ref{fig:qpos}).

In order to search for systematic trends of the quiescent level
variability with better signal-to-noise, we folded the light curves from
the XMM3, XMM4, and XMM5 observations at the average observation-dependent
recurrence time. We ignored the XMM6 observation whose increasing
recurrence times prevented us from defining a clear folding timescale
(see Fig.~\ref{fig:ATrec}). As \cha\ data were ignored, there was no
need to restrict the energy band above 0.4~keV, and we used the full
0.2-1~keV band in this analysis.

The 0.2-1~keV folded light curves are shown in the upper panels of
Fig.~\ref{fig:fold} together with their best-fitting baseline
(constant plus Gaussian functions) model. The middle panels show the resulting
residual light curves where a sinusoidal trend is evident, strongly
suggesting the presence of a quiescent level quasi-periodic
oscillation (QPO) with period equal to the average
observation-dependent recurrence time. Re-fitting the original folded
light curves with the addition of a sine function (with period fixed
at the folding timescale) improves very significantly the fits in all
cases, and best-fitting residuals are shown in the lower panels. The
best-fitting statistics for the model comprising the QPO provides and
improvement of $\Delta\chi^2= -134$ (XMM3), $\Delta\chi^2 = -302$
(XMM4), and $\Delta\chi^2 = - 133$ (XMM5) for $2$ degrees of
freedom. The remaining residuals (see lower panel of
Fig.~\ref{fig:fold}) can be attributed to the fact that QPE profiles
are not exactly Gaussian, especially once folded at the average
recurrence time since the periodicity is not perfect and the folding
process introduces small distortions. We conclude that a QPO is
highly significant in the folded light curves. Although deriving
periodicity from arbitrarily folded light curves is not always robust,
the extremely similar shape of the residual light curves in all
observations makes it highly unlikely that the derived sinusoidal
trend is spurious. The vertical lines in Fig.~\ref{fig:fold} mark the
phase of the QPE (dotted) and of the peak of the QPO (dashed).

\begin{figure}
\centering
\includegraphics[width=8.8cm]{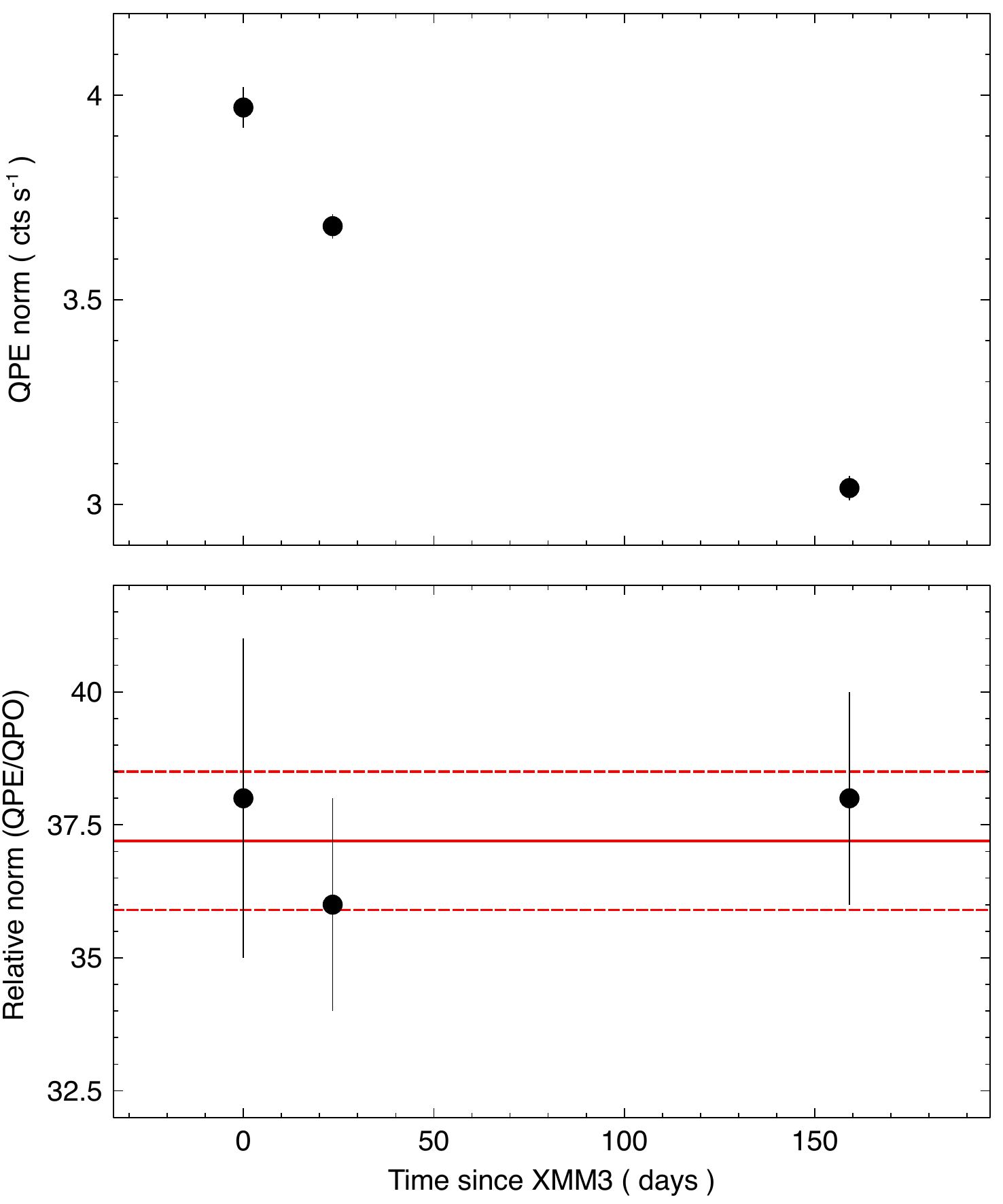}
\caption{Evolution of the QPE intensity (upper panel) and of the ratio
  between the QPE and QPO intensities (lower panel). The latter show
  that the QPE and QPO decay together preserving a ratio of $37.2\pm
  1.3$ (horizontal lines).}
\label{fig:qpo_prop}
\end{figure}

\begin{figure}
\centering
\includegraphics[width=8.8cm]{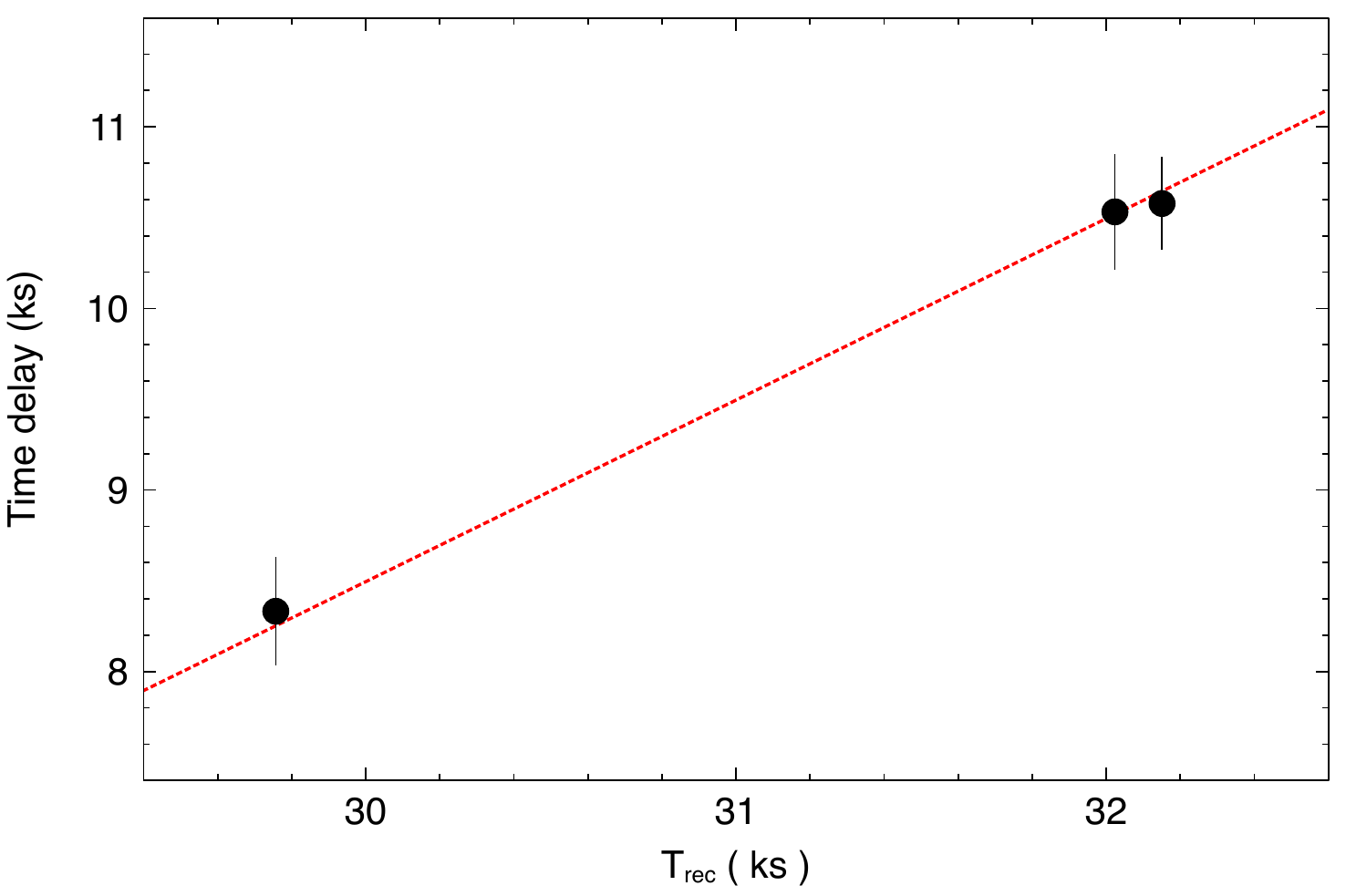}
\caption{QPE-QPO time delay. We show the time delay of the QPO peak
  with respect to the preceding QPE as a function of the average
  observation-dependent recurrence time. The dotted line is a linear
  function with slope fixed to 1, the best-fitting one being $0.95\pm
  0.15$.  }
\label{fig:delaycorr}
\end{figure}

In Fig.~\ref{fig:qpo_prop}, we show the time-evolution of the 
QPE intensity (here the average between strong and weak ones, as we
folded on the recurrence time between consecutive QPEs) as well as of
the ratio between the QPE and QPO best-fitting normalisation. The
intensity of both QPEs and QPOs decays with time, but
their ratio is consistent with being constant (and equal to $\sim
37.2$). The peak of the QPO emission lags the preceding QPE by $8.3\pm
0.3$~ks, $10.6\pm 0.3$~ks, and $10.5\pm 0.3$~ks in XMM3, XMM4, and
XMM5 respectively. Although the number of data points is limited, it is
tempting to associate shorter time delays to observations with shorter
recurrence times. In fact, the delay and observation-dependent
recurrence (folding) time appear to obey a 1:1 correlation, as shown
in Fig.~\ref{fig:delaycorr}.

The detection of a quiescent level QPO with characteristic timescale
of $\sim 9$~hr, equal to the QPE recurrence
time, suggests that the QPO production mechanism is linked to every single
QPE regardless of its intrinsic type (strong or weak), and
likely indicates that every QPE induces either an oscillation
of the quiescent (disc) emission, or the production of a weaker and
broader (longer-lasting) secondary QPE. A common production mechanism
(or at least a link between the QPE and the QPO) is also suggested by
their common time evolution which preserves their intensity ratio as
well as by the 1:1 correlation between the QPO time delay (with respect to the
preceding QPE) and the observation-dependent recurrence time.

A QPO of the quiescent level was also reported for another
QPE-source, RX~J1301.9+2747 \citep{2020A&A...636L...2G} by
\citet{2020A&A...644L...9S}. In RX~J1301.9+2747, the QPO is detected
in two observations $\sim 18,5$~yr apart (both of which presenting
X-ray QPEs), with a stable timescale of $\sim 1\,500$~s. The QPO
timescale in GSN~069 ($\sim 30$-$32$~ks) and RX~J1301.9+2747 ($\sim
1.5$~ks) are remarkably different despite similar QPE
recurrence times in the two sources (within a factor of $2$) which may
indicate a different production mechanism. In eRO-QPE1
no QPO is detected, although QPEs there are significantly more complex
than in GSN~069 and it may be difficult to separate QPOs from
multi-component QPEs \citep{2022A&A...662A..49A}. In eRO-QPE2, 
the quiescent level does not have high enough signal-to-noise
to enable us to investigate the presence of any QPO in detail (Arcodia
et al. in preparation).

\section{Long-term evolution of the quiescent level emission}
\label{sec:longterm}

Following its first X-ray detection during an \xmm\ slew (14 July
2010), GSN~069 has been consistently detected in the soft X-rays. Its
X-ray flux evolution during the first few yr shows a smooth decay by a
factor of a few, consistent with a long-lived TDE, as first noted by
\citet{2018ApJ...857L..16S}.  UV spectra from \hst\ observations
obtained in 2014 and 2018 show anomalous N-rich abundance
\citep{2021ApJ...920L..25S}, also suggesting an association with TDEs
\citep{2016MNRAS.458..127K}. 

\subsection{\xmm\ spectral analysis}
\label{sec:spectral}

We extracted EPIC-pn spectra from the available 11
\xmm\ observations excluding time intervals comprising QPEs, as our
goal here is to present the properties and evolution of the quiescent
emission, most likely due to the accretion flow formed by the initial
TDE. We assumed a model comprising thermal disc emission using the
{\tt{DISKBB}} model in {\tt{XSPEC}} \citep{1984PASJ...36..741M} and a
power law component, both at the redshift of GSN~069 ($z=0.0181$). We
also included neutral absorption with $N_{\rm H} = 2.3\times
10^{20}$~cm$^{-2}$ fixed at the Galactic value
\citep{2016A&A...594A.116H}.

Most spectra exhibit an absorption structure around $0.7$~keV, first
noted by \citet{2013MNRAS.433.1764M} in the XMM1 spectrum. Indeed, the
feature is clearer in the XMM1 highest-flux observation, but appears
possibly present in all others as well. In order to first establish
whether the quiescent level in observations with QPEs is 
different from the others in terms of absorption, we combined X-ray
spectra from different observations to increase the signal-to-noise
and we produced: (i) a merged spectrum of the quiescent level from all
observations comprising QPEs (XMM3 to XMM6); (ii) a merged
spectrum from all other observations except XMM1 (where the feature is
already clear). The two merged spectra include observations with
similar flux levels, so that none of the two spectra is dominated by
one particular observation, but represents a sensible average in both
cases. Both merged spectra show clear signs of an absorption feature
around $0.7$~keV, consistent with being the same.

\begin{figure}
\centering
\includegraphics[width=8.8cm]{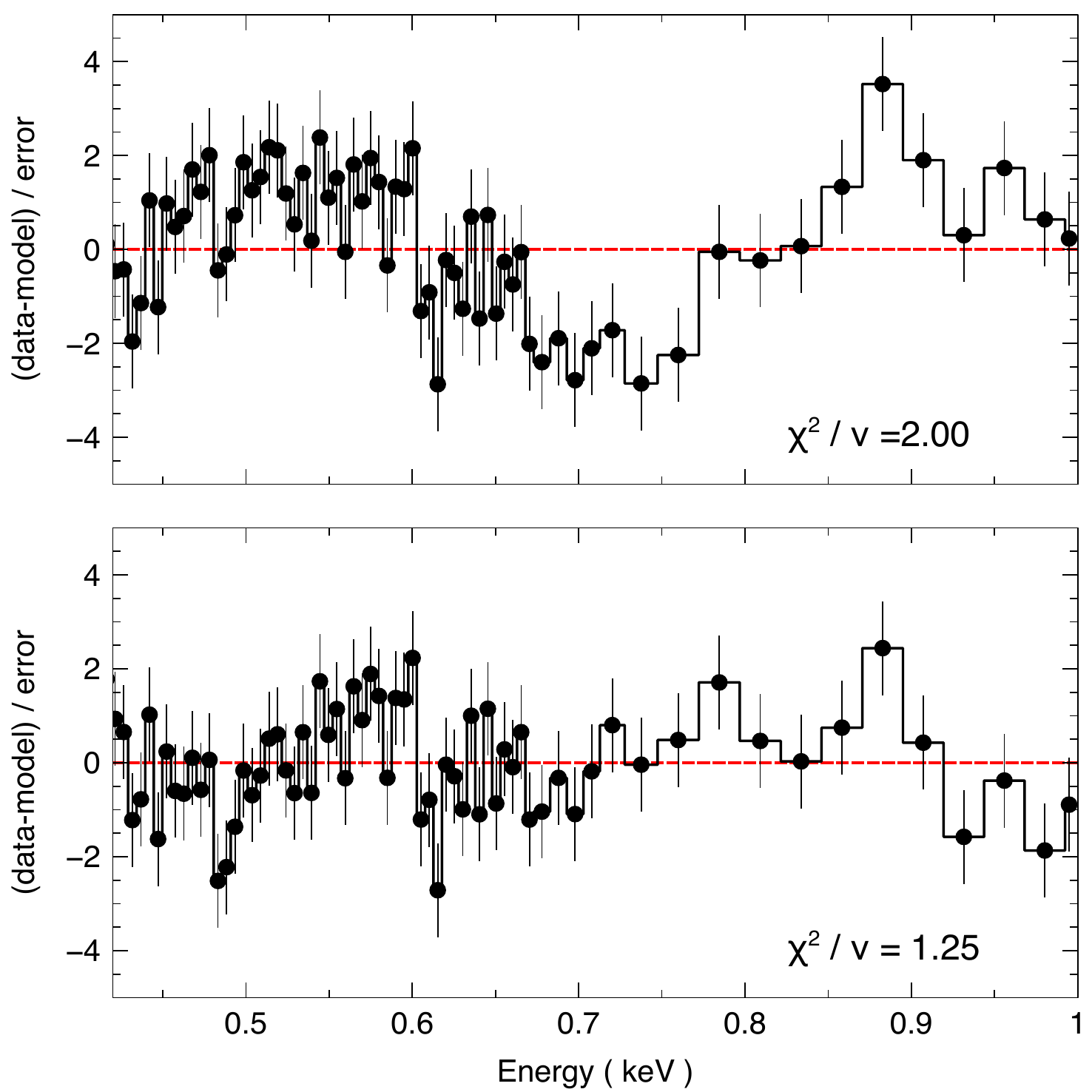}
\caption{Residuals (in terms of
  $\sigma$) when the merged XMM2 to XMM11 spectrum is fitted without
  (upper panel) and with (lower panel) a warm absorber component. We
  only show the most interesting energy range close to the most
  prominent absorption feature at $\simeq 0.7$~keV for better visual
  clarity. Data have been slightly re-binned for visualisation
  purposes only.}
\label{fig:merged}
\end{figure}

As the two merged spectra (from observations with and without QPEs)
are consistent with each other in terms of absorption, we then produced
a single merged spectrum by combining the XMM2 to XMM11 spectra, and we
compared it to the XMM1 spectrum (the one with the clearest absorption
feature). In order to account for the extra-absorption (most likely
from ionised gas), we included a warm absorber model using a
custom-built {\tt{XSTAR}} grid \citep{2001ApJS..133..221K} that
assumes a black body with $kT=50$~eV as irradiating spectral energy
distribution (SED), as appropriate for GSN~069
\citep{2019Natur.573..381M}. We assumed a turbulent velocity of
$100$~km~s$^{-1}$. Residuals for the merged XMM2 to XMM11 spectrum
with and without the warm absorber model are presented in
Fig.~\ref{fig:merged}. The model with no warm absorber
(upper panel) does require a higher neutral column density ($\simeq
5.3\times 10^{20}$~cm$^{-2}$) than the Galactic one ($\simeq 2.3\times
10^{20}$~cm$^{-2}$), while the model with warm absorption does
not. The warm absorber best-fitting parameters for the merged spectrum
are N$_{\rm H} = (6.5\pm 0.1)\times 10^{21}$~cm$^{-2}$ and $\log\xi =
0.49\pm 0.01$, consistent with those derived from XMM1, N$_{\rm
  H}^{\rm (XMM1)} = (7.0\pm 2.0)\times 10^{21}$~cm$^{-2}$ and
$\log\xi^{\rm (XMM1)} = 0.52\pm 0.03$, although the latter have larger
uncertainties due to the much lower signal-to-noise. We conclude that
a warm absorber is always present in GSN~069 and that its properties
(column density and ionisation) are consistent with being the same at
all epochs.

\begin{figure}
\centering 
\includegraphics[width=8.8cm]{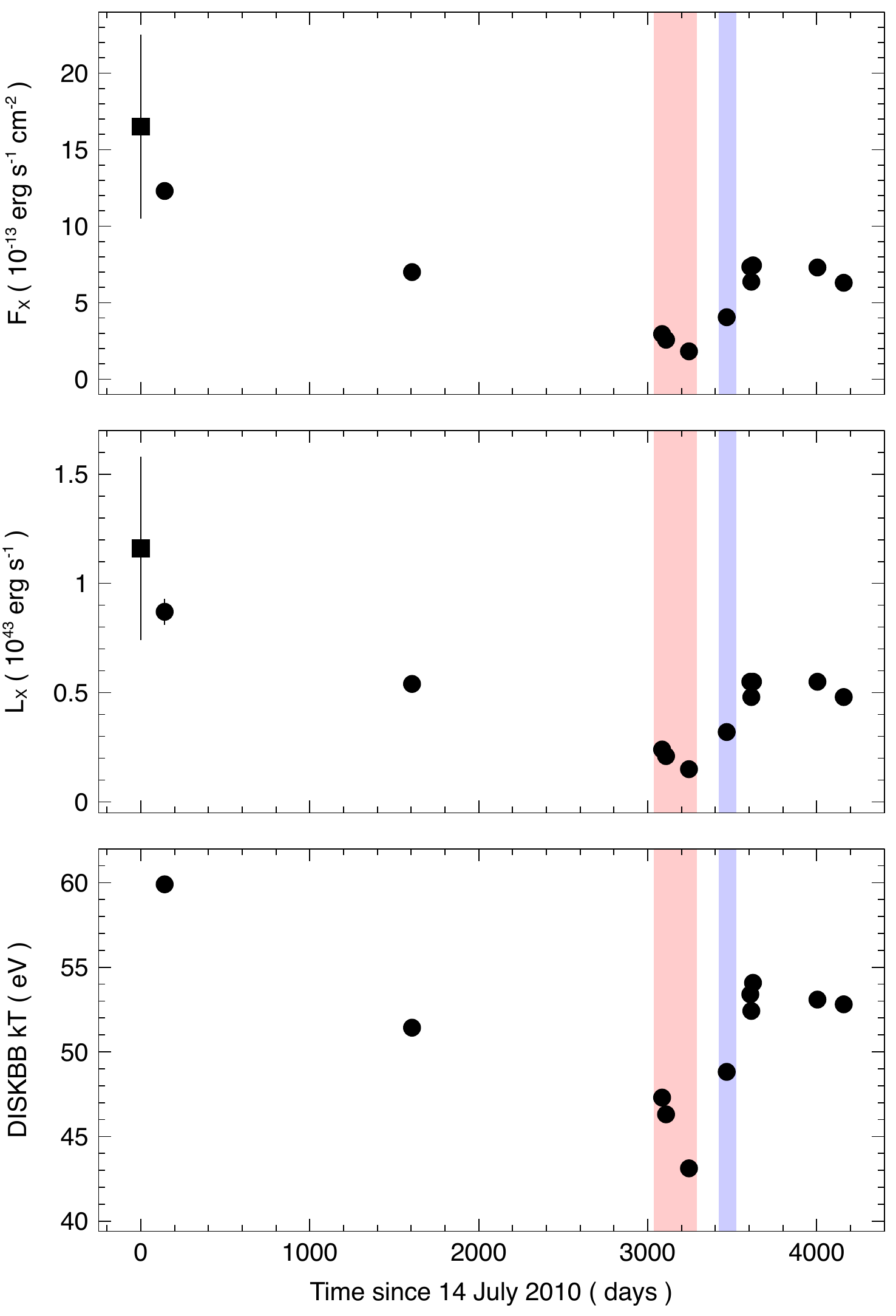}
\caption{X-ray flux, luminosity, and temperature evolution. From upper
  to lower panel, we show the observed 0.3-1~keV X-ray flux, the
  0.3-1~keV intrinsic X-ray luminosity, and the (rest-frame) disc
  temperature evolution, as derived from our best-fitting {\tt DISKBB}
  models to the 11 \xmm\ pointed observations. The first data point in
  the upper and middle panels refers to the slew observation whose
  observed count rate has been converted into X-ray flux and
  luminosity by using the best-fitting model from the nearest
  \xmm\ pointed observation (XMM1). In all panels, the shaded areas
  highlights the period during which regular (red, XMM3 to XMM5) and
  irregular (blue, XMM6) QPEs are detected.}
\label{fig:xmmevolution}
\end{figure}

We then fitted jointly the 11 individual spectra (XMM1 to XMM11)
including the warm absorber in the spectral model. After a few initial
tests, and as suggested by the previous analysis, we found that the
quality of the fits did not worsen significantly by forcing the warm
absorber parameters to be the same at all epochs. The final fit is
acceptable with $\chi^2 = 922$ for 814 degrees of freedom. The
(common) warm absorber column density is N$_{\rm{H}}=(~6.6\pm
0.1~)\times 10^{21}$~cm$^{-2}$ with $\log\xi = 0.49\pm 0.01$. Allowing
for an outflow/inflow velocity of the absorber does not improve the
fit, as expected because of the moderate resolution of the EPIC-pn
detector. The hard power law component is always detected, but it is
so weak that its contribution below 1~keV is negligible
\citep{2019Natur.573..381M}. Excluding the power law component from
the spectral model slightly worsen the fitting statistics but does not
change significantly any of the relevant parameters.

The resulting observed 0.3-1~keV X-ray flux, intrinsic X-ray
luminosity, and (rest-frame) disc temperature evolution are shown in
Fig.~\ref{fig:xmmevolution}. As is clear from the upper two panels, the
X-ray emission of GSN~069 exhibits an initial decay during the first
$\sim 9$~yr, followed by a highly significant X-ray
re-brightening that appears to start just after XMM5, that is the last
observation where regular QPEs are detected. Here
we focus first on the overall spectral evolution, deferring the
discussion of this interesting behaviour to
Sect.~\ref{sec:FLevolution}. Besides the \xmm\ slew data point that is
affected by very large uncertainties, the highest 0.3-1~keV X-ray
luminosity, $L_X \simeq 8.7\times 10^{42}$~erg~s$^{-1}$, is observed
during XMM1 ($\sim 140$~d after the initial slew detection), while the
lowest is reached during XMM5 ($\sim 3\,240$~d after the slew
detection) and is $L_X \simeq 1.5\times 10^{42}$~erg~s$^{-1}$, namely
a factor of $\simeq 5.8$ lower. We point out that the bolometric luminosity
 variation is likely to be significantly smaller because part
of the variability amplitude in the $L_X$ (and $F_X$) light curve is due to the
shift of the overall SED to lower energies as the temperature decays
(see lower panel in Fig.~\ref{fig:xmmevolution}). 

\begin{figure}
\centering 
\includegraphics[width=8.8cm]{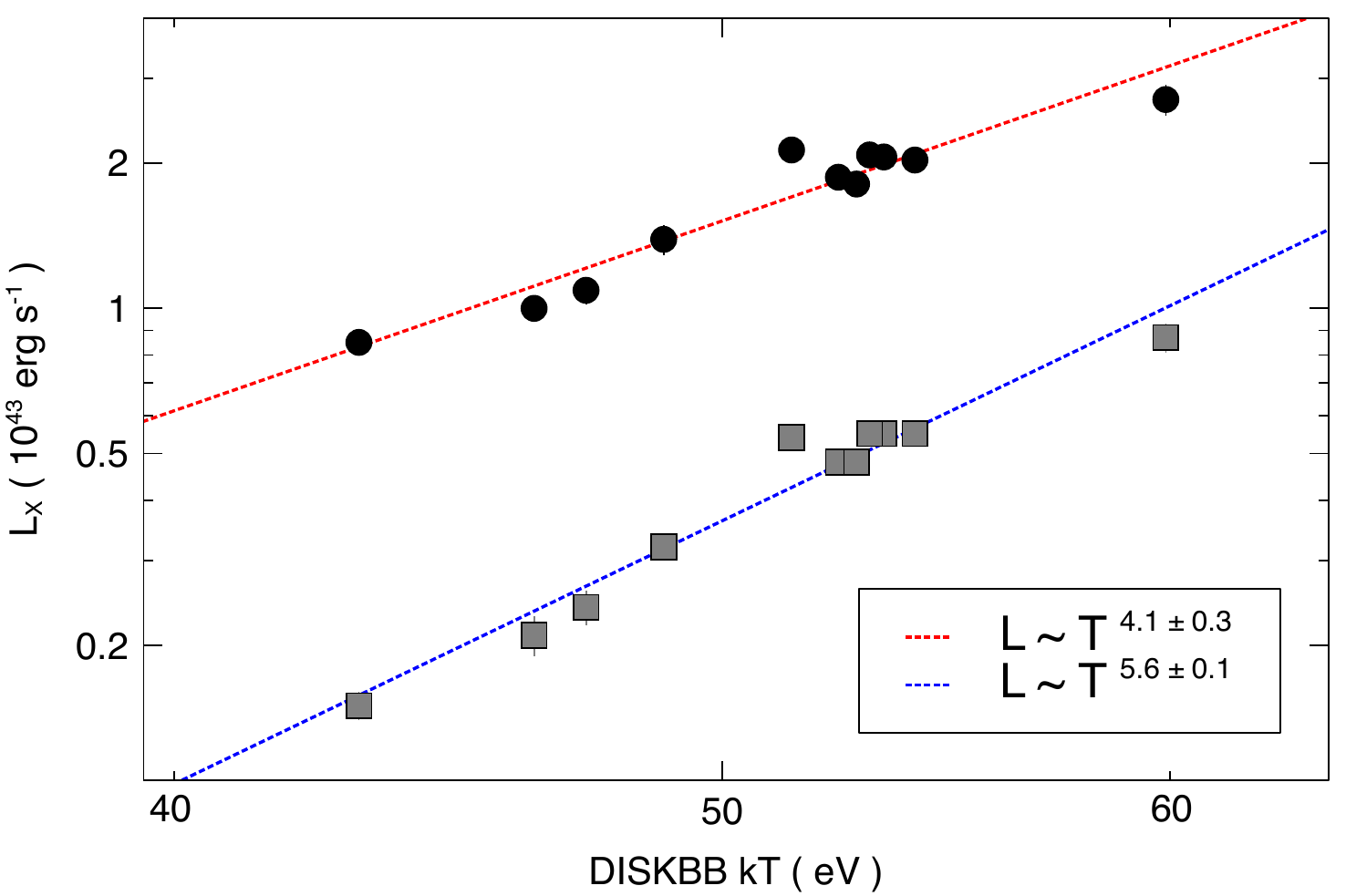}
\caption{Intrinsic $0.2-2$~keV (black circles) and $0.3-1$~keV (grey
  squares) X-ray luminosity as a function of rest-frame
  temperature for all 11 \xmm\ observations. The dotted lines are
  best-fitting relations assuming $L_X\propto T^q$.}
\label{fig:kTLumin}
\end{figure}

The relation between intrinsic disc X-ray luminosity and rest-frame
temperature is shown in Fig.~\ref{fig:kTLumin}. We show the relation
for the observed $0.3-1$~keV $L_X$ (grey squares) as well as that for
the $0.2-2$~keV band (black circles), obtained by extrapolating the
best-fitting model. The dotted lines are best-fitting relations of the
form $L_X \propto T^q$. The $0.3-1$~keV $L_X-T$ relation is steeper
($q\simeq 5.6$) than the $0.2-2$~keV one ($q\simeq 4.1$). The
different slope is expected because the narrower $0.3-1$~keV band is
more severely affected by the shift of the SED to lower energies as
the temperature decays. What is surprising, however, is that the
$0.2-2$~keV luminosity is consistent with the $L\propto T^4$ relation
that, for constant-area blackbody emission, is naturally expected to
be satisfied by the bolometric, rather than narrow-band,
luminosity. As, for a standard disc, the X-ray bolometric correction
($K_{\rm{0.2-2~keV}} = L_{\rm bol}/L_{\rm{0.2-2~keV}}$) is a function
of temperature\footnote{The range of $K_{\rm{0.2-2~keV}}$ reported
  here for the considered range of temperatures has been computed
  using the {\tt DISKBB} model.} and varies in the range
$\simeq[4.9-11.6]$ for the observed range of $kT \simeq [43-60]$~eV,
the $L \propto T^q$ relation cannot be satisfied by $L_{\rm bol}$ with
$q=4$, and the slope is necessarily shallower. Forcing the $L_{\rm
  bol}\propto T^4$ relation instead\footnote{This can be done by
  forcing the normalisation of the {\tt DISKBB} model to be always the
  same.}, produces a lower quality fit with $\chi^2=1002$ for 824
degrees of freedom (to be compared with $\chi^2=922$ for 814 degrees
of freedom for the case in which the {\tt DISKBB} normalisation is
free to vary).

The $L_{\rm{0.2-2~keV}}\propto T^4$ relation suggests that $L_{\rm 0.2-2~keV}$ is
a good proxy of the bolometric luminosity, that is that the bolometric
correction is small and approximately constant in the observed range
of temperatures. This cannot be achieved with a standard disc, but it
can approximately be obtained with a compact accretion flow with small
outer radius ($R_{\rm out}$) significantly suppressing any disc
optical/UV emission, that is a disc emitting most of its energy in the
soft X-rays. We point out that, while a small $R_{\rm out}$ is
implausible in active galactic nuclei (AGNs), the accretion flow in
TDEs is at least initially expected to be compact, as the stellar
debris circularise at radii comparable (within a factor of a few) to the star's pericenter
distance that can be as small as few-tens gravitational radii ($R_g =
G~M_{\rm BH}/c^2$).

\subsubsection{UV emission}
\label{sec:UV}

\begin{figure}
\centering 
\includegraphics[width=8.8cm]{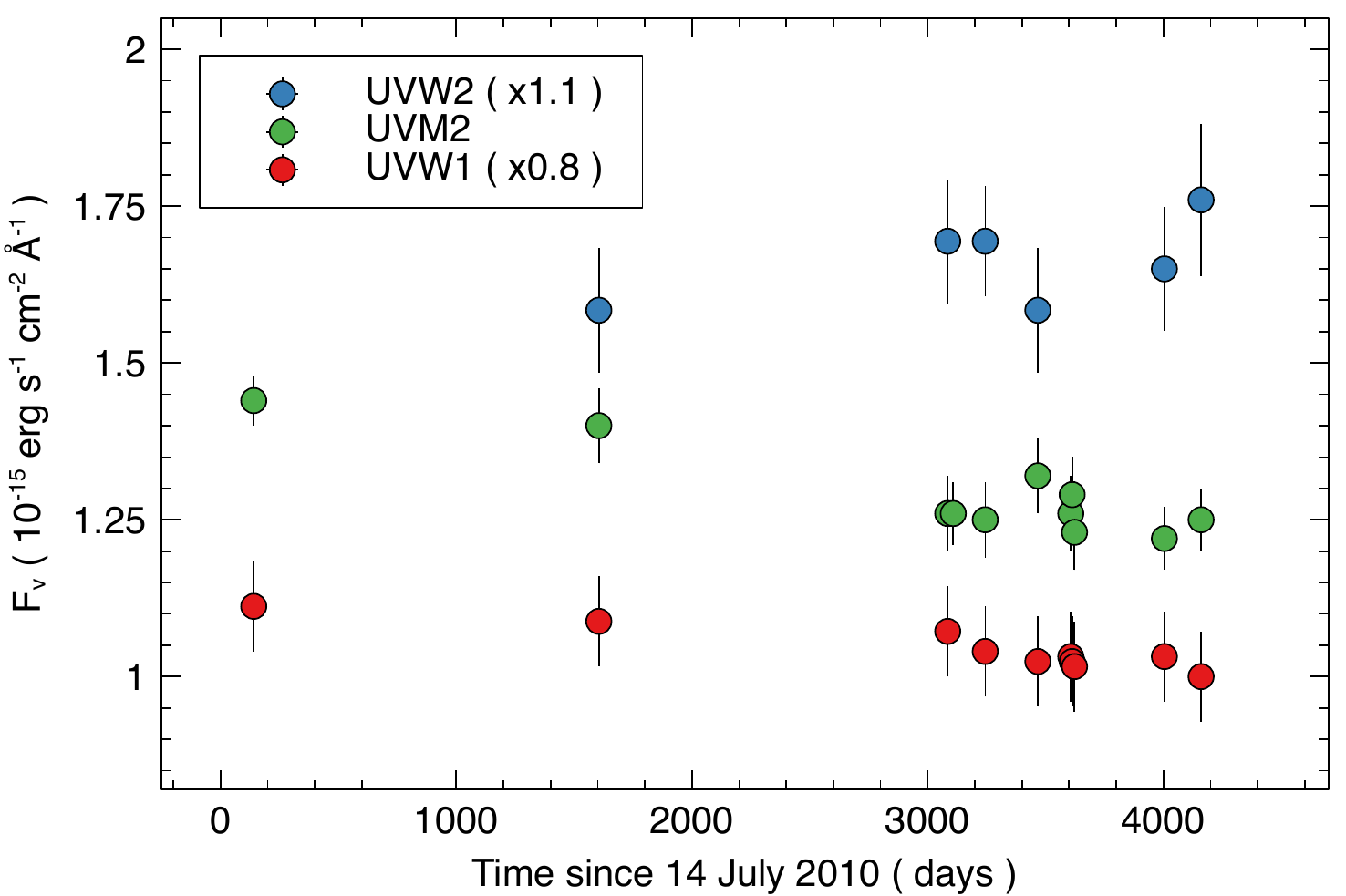}          
\caption{UV light curves from OM data obtained during the
  \xmm\ observations. The UVW2 ($\lambda_{\rm eff} \simeq
  2\,120$~\AA), and UVW1 ($\simeq 2\,910$~\AA) fluxes are re-scaled by
  factors of $1.1$ and $0.8$ respectively for visual clarity, while
  the UVM2 ($\simeq 2\,310$~\AA) remains as is. }
\label{fig:UVlc}
\end{figure}

In principle, the disc contribution in the UV can be derived from data
obtained with the Optical Monitor (OM) on board \xmm. Most UV fluxes
from the OM, obtained simultaneously with the X-ray data, do not show
any variability despite the large X-ray variation, which suggests that
the UV emission in GSN~069 is quite severely contaminated by stellar
light or not directly associated with disc emission, but rather to
another process (for example, UV reprocessing). The UV flux light
curves for the three available OM filters are shown in
Fig.~\ref{fig:UVlc}. The UVM2 filter (at $2\,310$~\AA) is the only one
where some variability is observed with higher fluxes during the first
two observations (XMM1 and XMM2) than during all others (XMM3 to
XMM11). Fitting a constant to the light curves results in $\chi^2_\nu
\simeq 0.2$~($0.4$) for UVW1 (UVW2), and $\chi^2_\nu \simeq 2.2$ for
UVM2 (due to the first two data points only, XMM1 and XMM2). We must
however point out that the XMM7 to XMM11 observations (last five data
points in Fig.~\ref{fig:UVlc}) have very similar X-ray flux to XMM2,
but the UVM2 flux does not recover, casting some doubts on the
reliability of the UVM2 fluxes (and therefore putative variability) in
XMM1 and XMM2. Moreover, UV fluxes from the bluer UVW2 filter
($2\,120$~\AA) and the redder UVW1 one ($2\,910$~\AA) do not show any
significant variability, which is surprising if the UV are dominated
by disc emission.

Another estimate of the UV flux can be obtained by making use of two
\hst\ spectroscopic observations that have been performed
quasi-simultaneously with XMM2 and XMM3. The 2014 (XMM2) flux is
slightly higher than that measured in 2018 (XMM3) by $\lesssim 7$\%,
while the X-ray flux decayed by $\sim 60$\%. Milky Way's
absorption appears to be sufficient to account for the slight
reddening in the \hst\ spectra, and the de-reddened UV flux in the
relatively features-free range around $1\,400$~\AA\ is $\simeq
1.4\times 10^{-15}$~erg~s$^{-1}$~cm$^{-2}$~\AA$^{-1}$ which, after
subtracting appropriate stellar continuum, reduces to $\simeq 1\times
10^{-15}$~erg~s$^{-1}$~cm$^{-2}$~\AA$^{-1}$
\citep{2021ApJ...920L..25S}. Our best-fitting {\tt DISKBB} model
over-predicts the intrinsic $1\,400$~\AA\ flux by a factor of $4-5$
during XMM2 and XMM3, which may be taken as an indication of a compact
accretion flow with suppressed UV emission (as suggested by
the $L_{\rm{0.2-2~keV}}\propto T^4$ relation), or of the overall inadequacy of the
model to reproduce the broadband SED.
  
A conservative analysis should thus look into solutions in which the
UV disc emission is constrained between two extremal cases, namely: (i)
the UV flux is entirely due to disc emission and the disc model only
needs not to over-predict it; (ii) the UV flux is entirely due to a
different phenomenon (e.g. reprocessing) and the disc model is
allowed to have negligible contribution in the UV. These extremal
cases will set the systematic uncertainty on the disc bolometric
luminosity derived below. In our analysis, we took as UV flux
that measured by \hst\ after correcting for stellar contamination and
Milky Way reddening, that is $\simeq 1\times
10^{-15}$~erg~s$^{-1}$~cm$^{-2}$~\AA$^{-1}$ at $1\,400$~\AA, and
compared it with the low-energy extrapolation of unabsorbed X-ray
accretion disc models at the same epoch (XMM2 and XMM3).

\subsubsection{Bolometric luminosity evolution}
\label{sec:Lbol}

We considered an accretion disc model that is constructed with the
goal of reproducing the broadband SED of AGNs, namely the {\tt
  OPTXAGNF} model in {\tt XSPEC} \citep{2012MNRAS.420.1848D}. As we
are only interested in accretion disc emission, we switched off all
Comptonisation zones that are originally introduced to account for the
high-energy power law and soft excess X-ray components in AGNs. The
treatment of the disc emission is more physical than in the {\tt
  DISKBB} model used so far. In particular, each annulus on the disc
thermalises to the local blackbody temperature, but a
temperature-dependent colour-correction factor is introduced to
account for electron scattering within the disc. Such correction
factor basically shifts the emission towards higher energies and
produces higher X-ray-to-UV flux ratios with respect to models where
the correction is ignored, which may potentially account for the
depressed UV flux discussed above. The model naturally
enforces the $L_{\rm bol}\propto T^4$ relation. When all
Comptonisation components are switched off, the model only depends on
black hole mass $M_{\rm BH}$ and spin $a$, Eddington ratio, and outer
disc radius $R_{\rm out}$. The black hole spin sets the inner disc
radius, assumed to be coincident with the innermost stable circular
orbit at radius $R_{\rm isco}$. In our analysis, we assumed $a=0$, because the
corresponding $R_{\rm isco} = 6~R_g$ is intermediate between those
associated with accretion around a maximally spinning black hole for a
prograde or retrograde accretion flow ($R_{\rm isco} \simeq 1.24~R_g$
and $R_{\rm isco}\simeq 9~R_g$ respectively). We initially fixed $R_{\rm
  out} = 10^3~R_g$.

\begin{figure}
\centering 
\includegraphics[width=8.8cm]{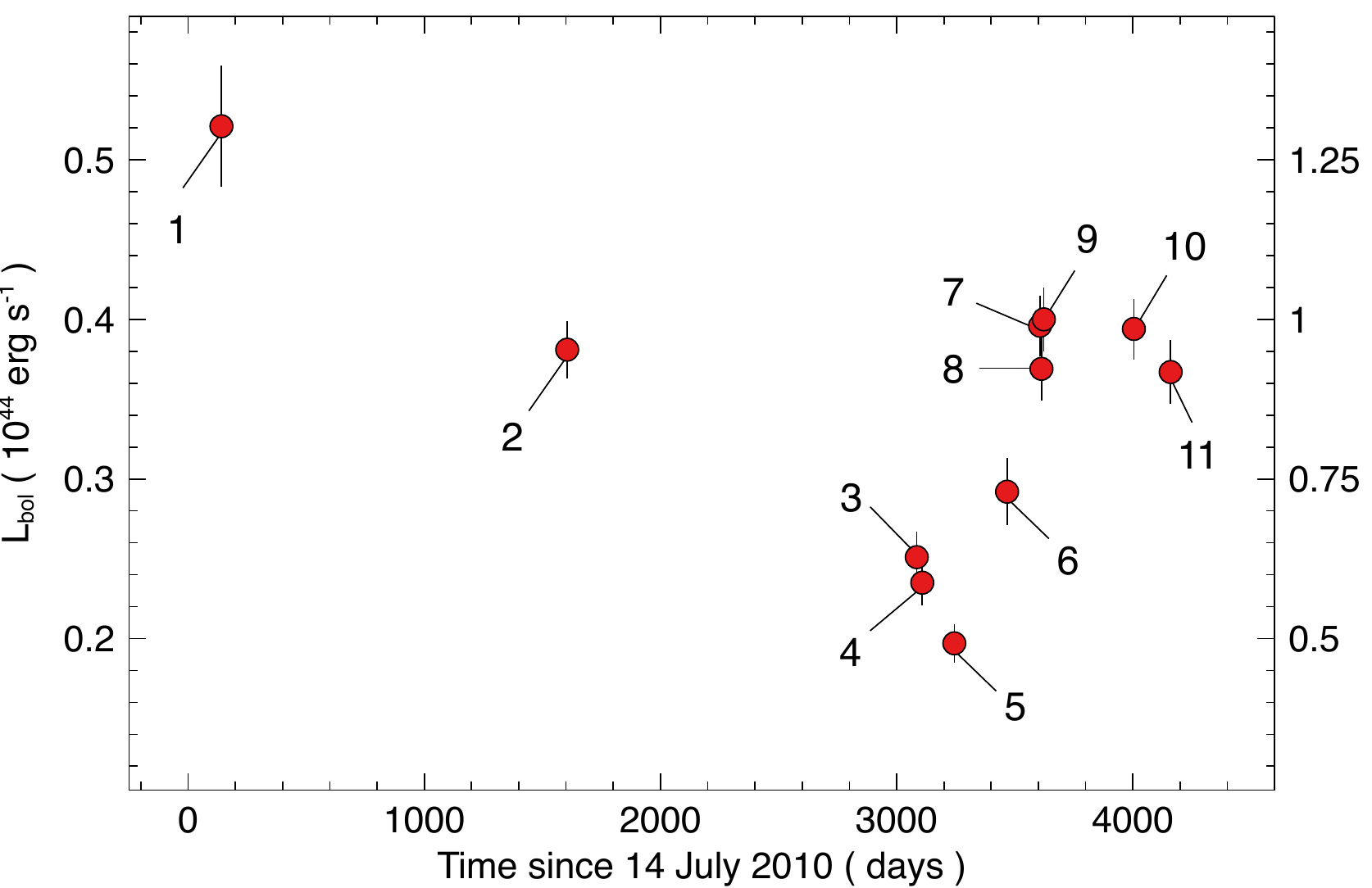}           
\caption{Bolometric luminosity evolution as inferred from {\tt
    OPTXAGNF} best-fitting models to the 11 \xmm\ observations. The
  left y-axis corresponds to the compact disc ($R_{\rm out}\simeq
  15~R_g$), and the right y-axis to the more extended one ($R_{\rm
    out}\simeq 140~R_g$). The conversion between the two is a factor
  of $\simeq 2.5$. For each data point, we also show the corresponding
  \xmm\ observation number (XMM1 to XMM11) as reference.}
\label{fig:xmmLbolRout}
\end{figure}

We fitted  all \xmm\ spectra jointly, keeping the warm absorber and power
law components as before (with warm absorber parameters forced to be
the same in all spectra, and power law photon index fixed at $\Gamma =
1.9$). The fit is relatively good with $\chi^2 =999$ for 823 degrees
of freedom and has similar quality to that obtained with {\tt DISKBB}
forcing the normalisation to always be the same (i.e. enforcing the $L_{\rm
  bol}\propto T^4$ relation), which resulted in $\chi^2 = 1002$ for
824 degrees of freedom. The unabsorbed best-fitting model, extrapolated to the UV, still over-predicts the $1\,400$~\AA\ flux by a large factor. We then
decreased the outer disc radius $R_{\rm out}$ until a good match with
the measured \hst\ flux was found, and we derive $R_{\rm out} \simeq
140~R_g$, that is the maximum allowed
disc outer radius that is needed not to over-predict the UV flux. As for the
minimum one, corresponding to negligible UV contribution, it cannot be
constrained by using the UV flux level. Hence, we re-fitted the X-ray
data by allowing $R_{\rm out}$ to vary, forcing it to be the same for
all observations (i.e. at all epochs). The statistical quality of the
fit improves significantly, by $\Delta\chi^2 = -30$ for 1 degree of freedom, and we
obtain $R_{\rm out} = 15\pm 5~R_g$, that is the X-ray data alone tend to
prefer a compact accretion disc with small outer radius. 

The $L_{\rm bol}$
evolution from the 11 \xmm\ pointed observations is shown in
Fig.~\ref{fig:xmmLbolRout} where we also indicate the
\xmm\ observation number corresponding to each data point for
reference. We use two different y-axis scales: the left axis is
$L_{\rm bol}$ as measured by the compact disc model ($R_{\rm
  out}\simeq 15~R_g$), while the right axis  refers to the more
extended one ($R_{\rm out}\simeq 140~R_g$). The bolometric luminosity
ranges from $L_{\rm bol} = 0.5-1.3\times 10^{44}$~erg~s$^{-1}$ (XMM1)
to $0.2-0.5\times 10^{44}$~erg~s${-1}$ (XMM5) where the lower
luminosity is associated with the compact disc. 

As for the remaining best-fitting parameters, the warm absorber and
power law are unchanged with respect to previous fits using the {\tt
  DISKBB} model, while we obtain a black hole mass of $M_{\rm BH} =
2-4\times 10^6~M_\sun$ depending on the adopted model, the lower mass
being associated with the larger $R_{\rm out}$.  The derived
$M_{\rm BH}$ is significantly higher than that estimated by
\citet{2019Natur.573..381M} using {\tt DISKBB} fits ($\simeq 4\times
10^5~M_\sun$), and this is due to the different disc model used here
that includes the colour-correction factor, shifting the SED to higher
energies and therefore requiring higher black hole mass. Indeed, if we
remove the colour-correction, adopting the similar but not
colour-corrected {\tt AGNSED} model \citep{2018MNRAS.480.1247K}, we
recover a best-fitting black hole mass of $(4\pm 1)\times
10^5~M_\sun$. Hence, while our analysis allows us to estimate the disc
bolometric luminosity of GSN~069, at least within the systematic
uncertainty associated with the different possible $R_{\rm out}$, the
black hole mass (and hence Eddington ratio evolution) remains highly
model-dependent and thus largely unconstrained.

As a final note on black hole mass, the stellar velocity dispersion in
GSN~069 is $\sigma_\star = (63\pm 4)$~km~s$^{-1}$ which can be used to
infer a black hole mass of $M_{\rm BH} = 3-30\times 10^5~M_\sun$
\citep{2022A&A...659L...2W}, consistent with the range of black hole
masses that we infer from the X-ray spectral analysis using models
with and without colour-correction. When needed, hereafter we consider
the full range of plausible black hole masses as obtained from our
spectral analysis using models with ({\tt OPTXAGNF} model) and without
({\tt AGNSED} model) colour-correction, namely $M_{\rm BH}
=0.3-4\times 10^6~M_\sun$. The estimated range assumes a non-rotating
Schwarzschild black hole. Significant black hole spin associated with
either prograde or retrograde accretion induces an even broader range
of allowed black hole mass. For reference, assuming (quite
arbitrarily) that the historical peak luminosity of GSN~069
corresponds to its Eddington luminosity, the black hole mass range
shrinks to $0.4-1\times 10^6~M_\sun$ depending on the assumed outer
disc radius.

As a caveat, we note that the X-ray luminosity in
Fig.~\ref{fig:xmmevolution} and Fig.~\ref{fig:kTLumin} are different
from those reported in \citet{2019Natur.573..381M} for the common
observations. This is due to the different absorption model adopted
here. The warm absorber induces a significantly higher X-ray
luminosity than the simpler neutral absorber used by
\citet{2019Natur.573..381M} as it is associated with a higher column
density and only moderate ionisation. While more physical, our
specific absorption model uses an {\tt{XSTAR}} grid obtained by
assuming a standard disc SED (with significant optical/UV emission) as
well as Solar abundances, which may not be accurate if the accretion
flow if confined to small radii or if the absorbing matter is
associated with the unbound debris from the TDE which may well have
different abundances, depending on the properties of the disrupted
star. While the relationship between luminosity and temperature shown
in Fig.~\ref{fig:kTLumin} is preserved for whatever specific
absorption model (provided that its parameters do not evolve with
time, as the data suggest), the absolute value of the reported X-ray
luminosities, as well as the estimate on radiated energy and accreted
mass given below, should then be taken with caution.

\begin{figure}
\centering
\includegraphics[width=8.8cm]{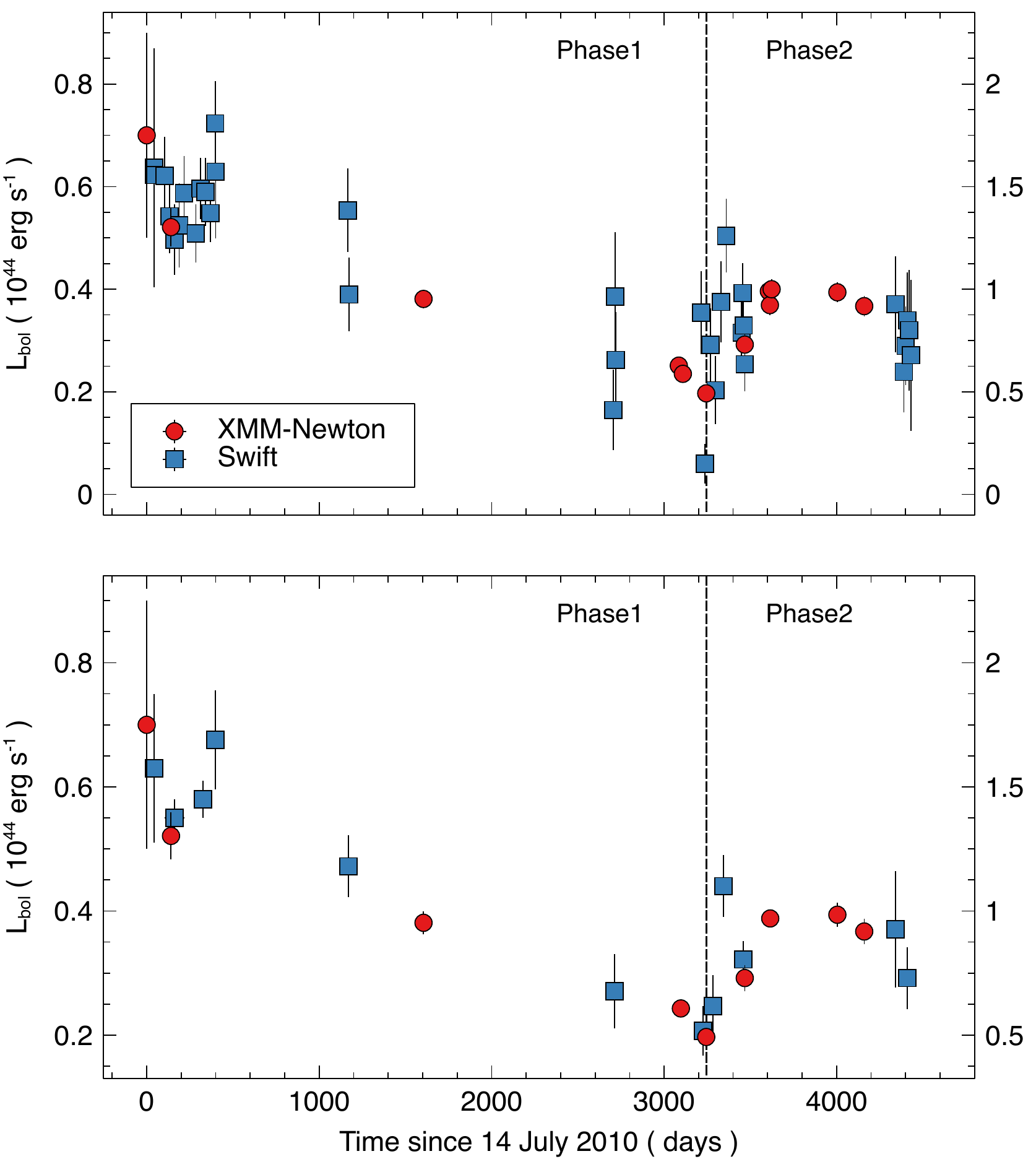}
\caption{Long-term evolution of the (quiescent level) bolometric
  luminosity. The y-axis scales have been defined in the caption of
  Fig.~\ref{fig:xmmLbolRout}. In the upper panel we show the original
  light curve from all \xmm\ and \swift\ observations. The lower panel
  shows a re-binned version of the same light curve (see text for
  details). In both panels, a vertical dashed line separates two
  different phases of the evolution: Phase1 comprises the initial
  X-ray peak and subsequent decay during the first $\simeq 9$~yr,
  while Phase2 is associated with the late-time X-ray re-brightening.}
\label{fig:Lbolevolution}
\end{figure}

\section{GSN~069 as a repeating TDE candidate}
\label{sec:FLevolution}

In order to study the long-term evolution of GSN~069, we considered,
besides the 11 \xmm\ pointed observations discussed above, an
additional set of data provided by long-term X-ray monitoring
observations of GSN~069 with the \swift\ observatory. As
\swift\ observations (as well as the initial \xmm\ slew detection) are
shallow, only few-tens of X-ray counts per pointing are typically
collected, so that spectral information is limited or absent. We then
used best-fitting models from appropriate (i.e. close in time) deep
\xmm\ exposures to convert count rates into
luminosities. Fig.~\ref{fig:Lbolevolution} shows the $L_{\rm bol}$
evolution over the whole $\sim 12$~yr period probed by our X-ray
campaign. The upper panel shows the original light curve from all
\xmm\ and \swift\ observations. The left (right) y-axis refers to
$L_{\rm bol}$ as measured with the compact (extended) disc
model. Hereafter, when discussing the $L_{\rm bol}$ evolution, we
refer to measurements obtained with the compact disc. However, in our
estimates of physical quantities (e.g. radiated energy and accreted
mass), we always consider the full range of possible $L_{\rm bol}$. As
shown in the upper panel of Fig.~\ref{fig:Lbolevolution}, significant
short-term variability is present on timescales of the order of
few-tens of days. In order to study the long-term evolution reducing
this source of noise as well as the uncertainties of the \swift\ data
points, we slightly re-binned the original $L_{\rm bol}$ light curve
by combining observations that were performed close in time (we
combined \swift\ and \xmm\ observations separately, while the
\xmm\ slew detection remained as is). The slightly re-binned $L_{\rm
  bol}$ light curve is shown in the lower panel of
Fig.~\ref{fig:Lbolevolution}.

The $L_{\rm bol}$ light curve in Fig.~\ref{fig:Lbolevolution} (lower
panel) is characterised by an initial rise peaking at $L_{\rm
  bol}\simeq 6-7\times 10^{43}$~erg~s$^{-1}$, and subsequent decay
lasting $\simeq 9$~yr. A significant X-ray re-brightening is seen
after that, peaking at $\simeq 3\,360$~d with $L_{\rm bol} \simeq
4-5\times 10^{43}$~erg~s$^{-1}$. The re-brightening peak is followed
by a fast luminosity drop by a factor of $\simeq 1.5$ in less than
$\sim 100$~d. The final evolution is characterised by a further rise
and subsequent decay. We define Phase1 as the initial $9$~yr-long
decay, and Phase2 as the subsequent X-ray re-brightening.

\subsection{Phase1: The initial decay}
\label{sec:DecayFits}

\begin{figure}
\centering 
\includegraphics[width=8.8cm]{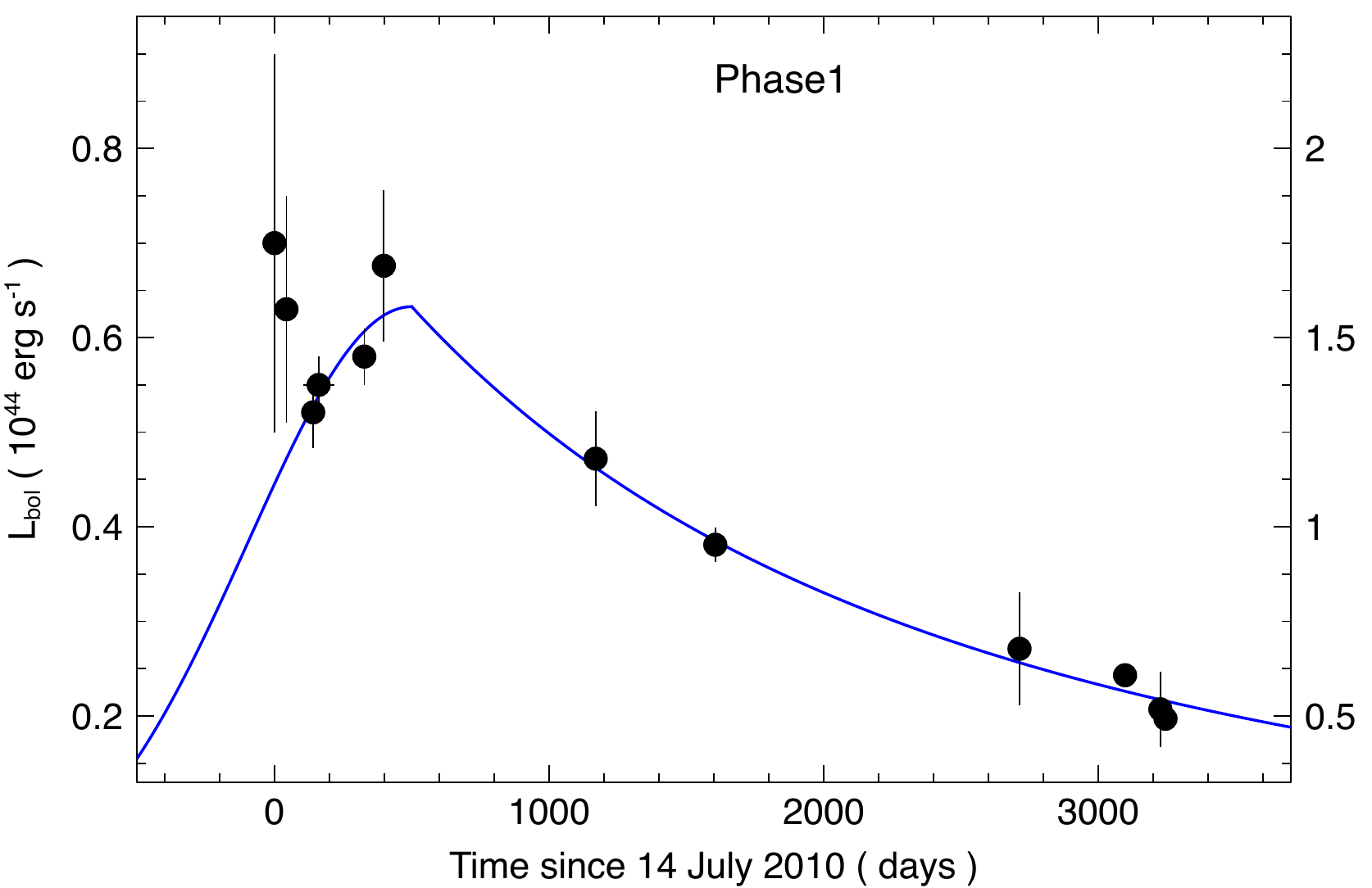}           
\caption{$L_{\rm bol}$ light curve during Phase1. The y-axis scales
  have been defined in the caption of Fig.~\ref{fig:xmmLbolRout}. We
  show a simple model comprising a Gaussian rise and a power law decay
  with $t^{-9/4}$. }
\label{fig:PlatDecay}
\end{figure}

The  \swift\ and \xmm\ $L_{\rm bol}$ light curve
during Phase1 is shown in Fig.~\ref{fig:PlatDecay}.  The 
evolution is highly suggestive of the typical rise and decay light
curve of TDEs. We adopted a simple parametrisation of the $L_{\rm
  bol}=L_{\rm bol} (t)$ light curve in which the rise is described by
a Gaussian function, while the decay has a power law form 
\citep{2021ApJ...908....4V}:
\begin{align}
L_{\rm bol} (t) &= L_{\rm peak}
    \times \begin{cases} e^{-(t-t_{\rm peak})^2/2\sigma^2} & t\leq t_{\rm peak} \\
     [(t-t_{\rm peak}+t_0)/t_0]^{n} & t>t_{\rm peak}\\
    \end{cases}
\label{eq:LXmodel}
\end{align}
where $t_{\rm peak}$ is the time when the peak luminosity $L_{\rm
  peak}$ is reached, $\sigma$ is the Gaussian rise width, $t_0$ is the
normalisation of the power law decay, and $n$ is its power law
index. 

The decay power law index cannot be constrained well by the data. In
particular, both canonical values of $n =-5/3$ and $n=-9/4$,
describing the decay for full or partial TDEs respectively based on
fallback rate calculations
\citep{1988Natur.333..523R,1989IAUS..136..543P,2019ApJ...883L..17C,2020ApJ...899...36M,2021ApJ...922..168N},
provide a good representation of the decay and cannot be distinguished
on a statistical basis. We point out, however, that even an exponential decay with
e-folding timescale $\tau \simeq 7.6$~yr can describe the data well
with similar statistical quality. For simplicity, and for reasons that
will become clear in the subsequent Sects., we assume $n = -9/4$, as
expected from the fallback rate in the case of a partial TDE
(pTDE). As is clear from Fig.~\ref{fig:PlatDecay}, our description is not
unique. In particular, the rise and decay phases may be connected by a
plateau that we ignore here for simplicity. As for the rise phase, the
Gaussian width is $\sigma \simeq 500-700$~d, peaking at $t_{\rm peak}
\simeq 500-800$~d with $L_{\rm bol} \simeq 6-7\times 10^{43}
$~erg~s$^{-1}$~cm$^{-2}$. The best-fitting model is shown as a (blue)
solid line in Fig.~\ref{fig:PlatDecay}. We note that the first two
data points (from the \xmm\ slew and first two \swift\ observations
combined) lie above the model. We shall discuss this small
discrepancy further in Sect.~\ref{sec:evolutionall}.

By integrating $L_{\rm bol} (t)$, the total radiated
energy during Phase1 is
$E^{\rm (obs)} = (2.3\pm 1.0) \times 10^{52}$~erg, where the
uncertainty (here and hereafter) reflects the systematic uncertainty on
$L_{\rm bol}$ associated to the compact or extended disc. By assuming a
radiative efficiency $\eta = 0.1$, this corresponds to an accreted mass
of $M_{\rm accr}^{\rm (obs)} = 0.13\pm 0.06~M_\sun$. Extending to $t
= \infty$, one has $E^{\rm (\infty)} = (4.1\pm 1.8) \times
10^{52}$~erg and $M_{\rm accr}^{(\infty )} = 0.23\pm
0.10~M_\sun$. As bound and unbound debris in TDEs likely represent
$\sim 50$\% of the disrupted mass, the TDE in Phase1 likely disrupted of the
order of $0.46\pm 0.20~M_\sun$.

\begin{figure}
\centering
\includegraphics[width=8.8cm]{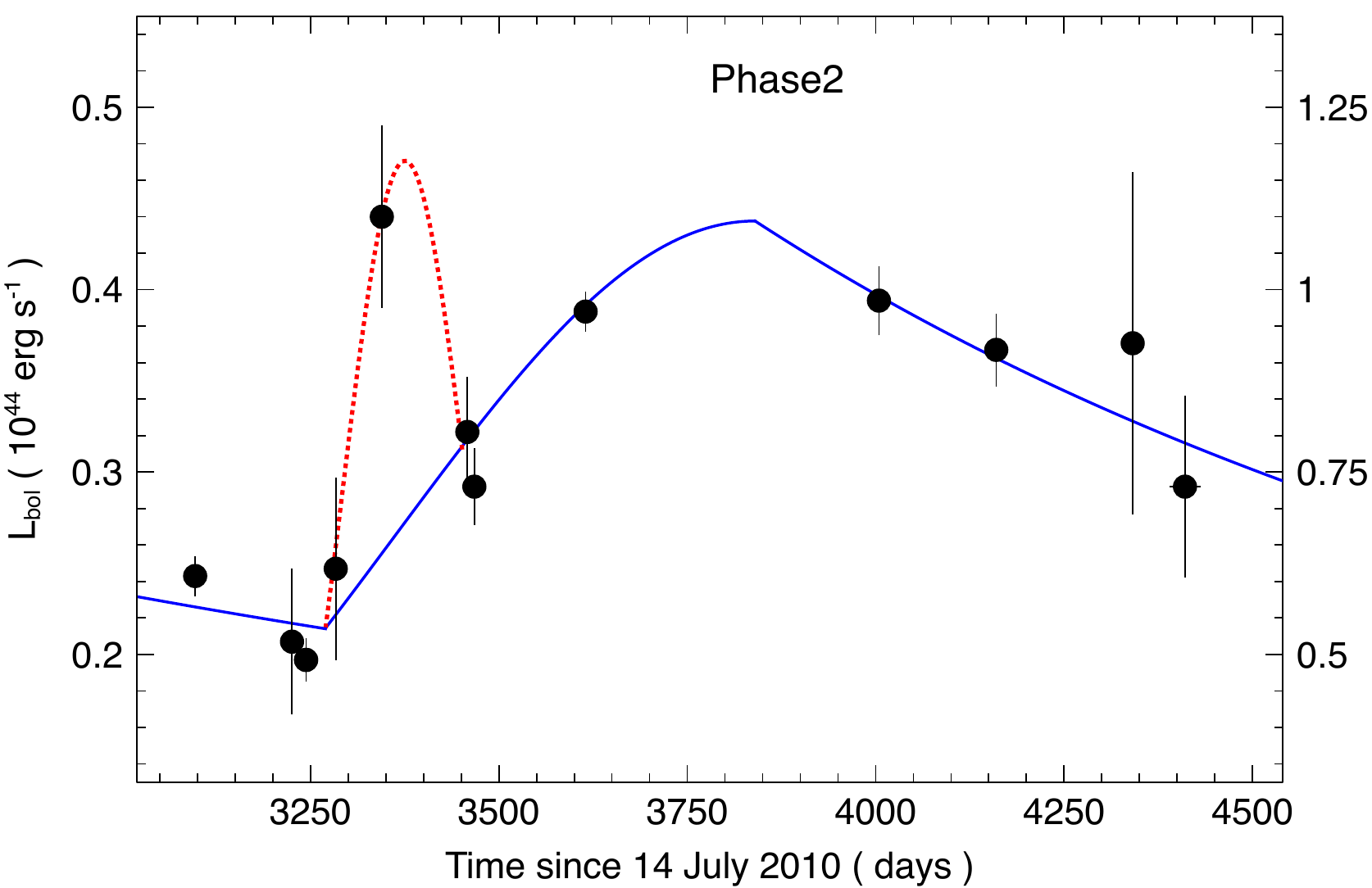}
\caption{$L_{\rm bol}$ light curve during Phase2. The y-axis
  scales have been defined in the caption of
  Fig.~\ref{fig:xmmLbolRout}. The first three data points are from the
  final decay of Phase1 and we also show the best-fitting decay model
  until Phase2 begins. A simple model comprising a Gaussian rise and a
  power law decay with $t^{-9/4}$ for Phase2 is shown as a solid blue
  line, together with a Gaussian X-ray flare precursor (red dotted
  line).}
\label{fig:rebrightening}
\end{figure}

\subsection{Phase2: The X-ray re-brightening}
\label{sec:RebrighteningFits}

The $L_{\rm bol}$ light curve during Phase2 is shown in
Fig.~\ref{fig:rebrightening}. The first three data points in
Fig.~\ref{fig:rebrightening} are the last three of the Phase1 decay in
Fig.~\ref{fig:PlatDecay}, and we also re-plot the best-fitting decay
law ($n=-9/4$) until Phase2 begins. Besides an apparent high-amplitude
X-ray flare at the beginning of Phase2 (red dotted line in
Fig.~\ref{fig:rebrightening}), the evolution is very
similar to that during Phase1. As a first step, we treated Phase2 as
independent from Phase1. This is likely inaccurate, as part of the
observed luminosity is still associated with residual accretion of debris from
Phase1. We adopted a simple parametrisation of Phase2 by describing the
initial X-ray flare with a Gaussian function, and the subsequent
evolution with the same model that successfully describes Phase1,
namely a Gaussian rise and subsequent power law decay with index
$n$. We first discuss the TDE-like evolution (Gaussian rise and
power law decay), ignoring for a moment the X-ray flare precursor.

As was the case for Phase1, power law indices of $n=-5/3$ and $n=-9/4$
are both acceptable for the decay, and produce the same statistical
quality with reduced $\chi^2 \simeq 1$. We adopt $n=-9/4$ as during
Phase1. An exponential is also an excellent fir to the decay of Phase2
and results in an e-folding timescale of $\simeq 4.8$~yr, slightly
shorter than that characterising Phase1 ($\simeq 7.6$~yr). The
Gaussian rise has width $\sigma \simeq 400$~d, slightly shorter than
the rise of Phase1. The peak luminosity is $\simeq 4-5\times
10^{43}$~erg~s$^{-1}$ (the luminosity for the extended disc being
$\simeq 2.5$ times higher) and is reached at $t_{\rm peak} \simeq
3\,750-3\,900$~d. By integrating over the observed time-interval, and
accounting for the systematic uncertainty in $L_{\rm bol}$, the total
radiated energy is $E^{\rm (obs)} = (4.5\pm 0.9) \times 10^{51}$~erg,
corresponding to $M_{\rm accr}^{\rm (obs)} = 0.025\pm
0.005~M_\sun$. Extending to $t = \infty$, $E^{\rm (\infty)} = (2.3\pm
1.0) \times 10^{52}$~erg with $M_{\rm accr}^{\rm (\infty)} = 0.13\pm
0.06~M_\sun$.

However, part of the emitted radiation in Phase2 is
likely associated with the accretion of residual mass disrupted in
Phase1 and not yet accreted. In order to account for it (at least
qualitatively), we re-fitted Phase2 by including, as baseline level, the
extrapolation of the best-fitting Phase1 model. This significantly
reduces the luminosity associated solely to the Phase2 event, as well
as the associated timescales. The decay e-folding time shortens from
$\sim 4.8$~yr to $\simeq 3.4$~yr, and the Gaussian rise width from
$\simeq 400$~d to $\simeq 270$~d. The total radiated energy lowers to
$E^{\rm (obs)} = (2.1\pm 0.4) \times 10^{51}$~erg, and $E^{\rm
  (\infty)} = (6.5\pm 1.3) \times 10^{51}$~erg, corresponding to
$M_{\rm accr}^{\rm (obs)} = 0.0125\pm 0.002~M_\sun$ and $M_{\rm
  accr}^{\rm (\infty)} = 0.036\pm 0.007~M_\sun$. The disrupted mass
during Phase2 is therefore highly uncertain and strongly depends on
whether we consider or not that part of the X-ray luminosity comes
from the accretion of debris from the previous Phase1. We conclude
that sensible lower and upper limits are set by the two extremal
situations we have considered, so that we conservatively estimate a
disrupted mass in the range of $0.06-0.38~M_\sun$ during Phase2
(assuming $M_{\rm disr} \simeq 2\, M_{\rm accr}$). We point out,
however, that the disrupted mass is likely closer to the lower rather
than upper limit because the luminosity from the accretion of the
remaining debris from Phase1 cannot be completely ignored.

\begin{figure}
\centering
\includegraphics[width=8.8cm]{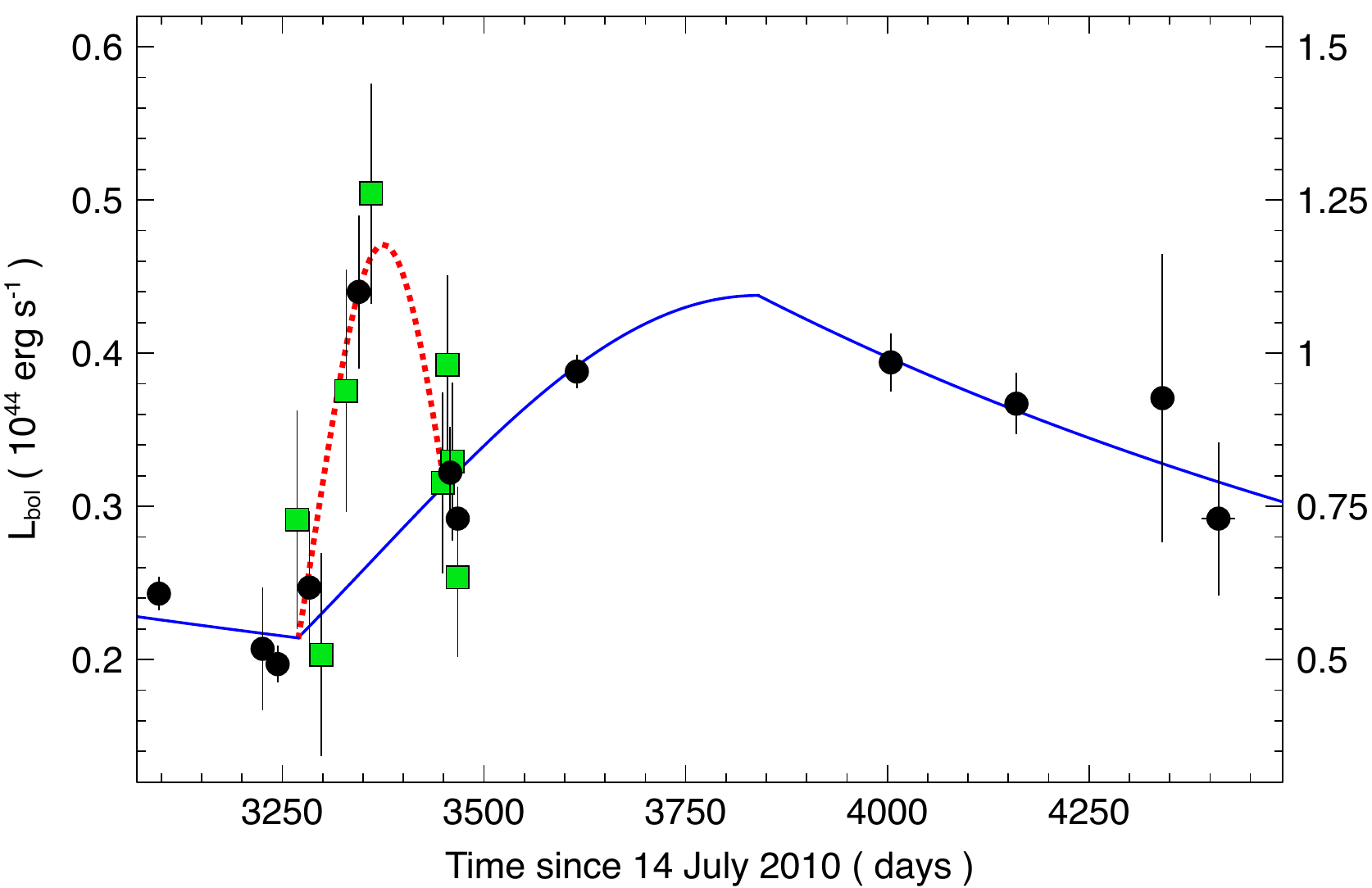}
\caption{Re-binned light curve during Phase2 (black
  circles), together with its best-fitting model (see text for
  details). The green squares are the original \swift\ data points
  (before re-binning) around the precursor flare.}
\label{fig:precursor}
\end{figure}

As is shown in Fig.~\ref{fig:rebrightening}, an X-ray precursor flare is
seen before the peak of Phase2. When described with a Gaussian
function, the precursor flare peaks at $t\simeq 3\,380$~d with peak
luminosity in the range of of $3-5\times
10^{43}$~erg~s$^{-1}$~cm$^{-2}$ (depending on whether we include the
Phase1 model extrapolation in the fit or not). The Gaussian width is
$50-90$~d, and the radiated energy is $E = 0.3-1.8 \times 10^{51}$~erg
once all uncertainties are included. Under the (strong) assumption
that the precursor flare is associated with a radiative efficiency $\eta
\simeq 0.1$, one has $M_{\rm accr} = 0.002-0.01~M_\sun$. In principle,
as the precursor flare is only defined by one single data point in the
re-binned light curve of Fig.~\ref{fig:rebrightening}, one should
consider its detection as merely tentative. However, in
Fig.~\ref{fig:precursor} we re-plot the same Phase2 re-binned light
curve shown in Fig.~\ref{fig:rebrightening} but with the original
\swift\ data points around the precursor flare superimposed. The
precursor flare appears to be confirmed by at least another couple of
data points, and we therefore consider it as a real feature
of the Phase2 light curve.

\section{QPE spectral evolution}
\label{sec:QPEspectral}

Having studied in detail the spectral properties of the quiescent
emission of GSN~069, we are now in a position to describe the QPE
spectral evolution.  We considered as case study QPEs during the XMM4
observation. The same conclusions discussed here were reached when the
XMM3 and XMM5 observations were studied, the only difference being that
the QPE peak is slightly more (less) luminous in XMM3 (XMM5) with
respect to XMM4 (see e.g. Fig.~\ref{fig:PeakEvol}). As for the weak,
irregular QPEs during the XMM6 observation, the low signal-to-noise
prevented us from reaching a sufficiently high spectral quality to
perform a detailed spectral analysis, and we can only report that the
typical peak luminosity if a factor of $\sim 2$ lower than during
XMM4.

\begin{figure}
\centering
\includegraphics[width=8.8cm]{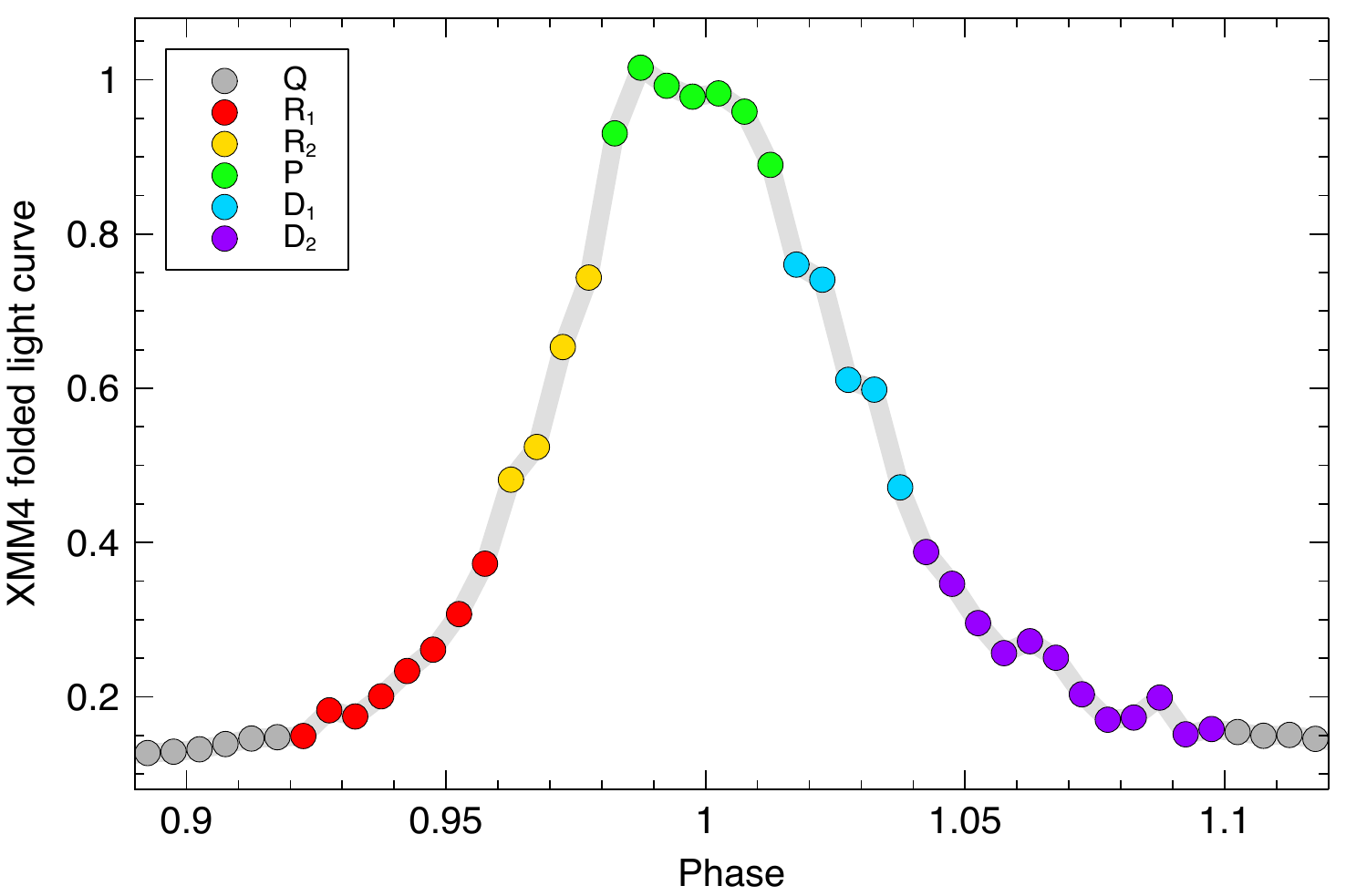}
\caption{XMM4 folded light curve (at the average recurrence time
  between QPEs). The light curve is normalised to the peak amplitude
  and shifted so that the peak is at phase 1. We show with different
  colours the phase intervals used to accumulate the spectra during
  the rise (R1 and R2), peak (P), and decay (D1 and D2). The quiescent
  spectrum (Q) is accumulated throughout the exposure (excluding
  time-intervals when QPE are present) and only a small fraction of
  the corresponding phase-interval is shown here. A phase interval
  $\Delta p = 0.05$ corresponds to a time interval $\Delta t \simeq
  1.6$~ks.}
\label{fig:QPEphase}
\end{figure}

To increase the signal-to-noise during the XMM4 observation, we
accumulated spectra during the rise, peak, and decay of QPEs without
distinguishing between strong and weak ones\footnote{The difference
  between strong and weak QPEs is marginal,
  with strong QPEs being slightly more luminous and hotter (at peak)
  than weak ones.}. We defined five different spectra during the QPE
evolution, namely: the quiescent spectrum (Q), two spectra defined
during the QPE rise (hereafter R1 and R2), the peak spectrum (P), and
two spectra during the QPE decay (D1 and D2). The phase-resolved
spectra were accumulated during the intervals shown in
Fig.~\ref{fig:QPEphase}, as in \citet{2019Natur.573..381M}. We then
subtracted the Q spectrum from all others to study the evolution of the
intrinsic QPE spectrum. The resulting difference spectra were fitted
jointly, and are all well described by a blackbody model (at the
redshift of GSN~069) with only Galactic absorption. The blackbody
model we used is not unique, and the spectral evolution could also be
discussed in terms of other (thermal-like) spectral models. However,
we used the simplest possible spectral model as it allows us to
highlight the main properties of the QPE spectral evolution.

\begin{figure}
\centering
\includegraphics[width=8.8cm]{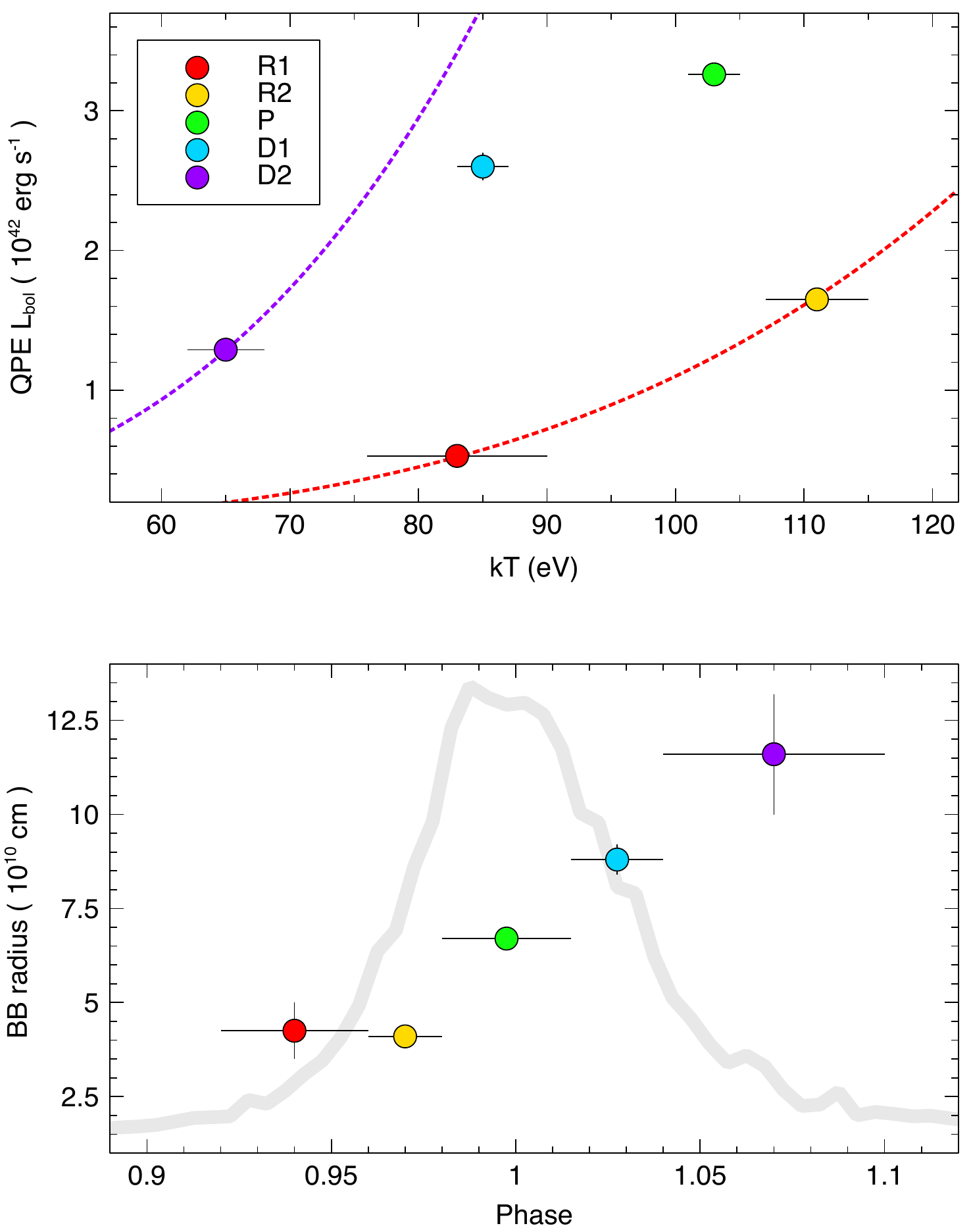}
\caption{QPE spectral evolution during XMM4. In the upper
  panel we show $L_{\rm bol}$ as a function of blackbody rest-frame
  temperature. The colour-code is that defined in
  Fig.~\ref{fig:QPEphase}. The evolution starts at R1 (lowest $L_{\rm
    bol}$ at $\simeq 83$~eV, red data point) and proceeds
  counter-clockwise. The two dotted lines are $L_{\rm bol} \propto
  T^4$ relations normalised to the extremal of the QPE evolution (R1
  and D2). The lower panel shows the estimated radius of the blackbody
  emitting region as a function of QPE phase. The re-scaled QPE
  profile is also shown for reference (light grey).}
\label{fig:QPEBB}
\end{figure}

A joint fit to the five QPE difference spectra results in $\chi^2 =
342$ for 305 degrees of freedom.  No signatures for a warm absorber
(nor for any excess absorption with respect to the Galactic column
density) is seen during the QPE evolution. As already discussed in
\citet{2019Natur.573..381M} QPEs exhibit a rather smooth temperature
and luminosity evolution peaking at $L_{\rm bol} \simeq 3.3\times
10^{42}$~erg~s$^{-1}$. The maximum temperature ($kT = 111\pm 4$~eV) is
reached slightly earlier than the peak ($kT= 103\pm 2$~eV). In the
upper panel of Fig.~\ref{fig:QPEBB} we show the resulting relation
between the QPE bolometric luminosity and the derived blackbody
(rest-frame) temperature. We also show, as dotted lines, two $L_{\rm
  bol} \propto T^4$ relations normalised to the extremal data points
in the evolution, R1 and D2. While the initial rise (R1 to R2) lies on
the $L_{\rm bol} \propto T^4$ relation, the overall evolution is not
consistent with constant-area blackbody emission but shows instead
hysteresis. The hysteresis behaviour in the $L_{\rm bol}-T$ plane is
fully consistent with that observed in the other QPE source eRO-QPE~1
\citep{2022A&A...662A..49A} despite its very different timing
properties, as well as in RX~J1301.9+2747 (Giustini et al. in
preparation), suggesting it is a common property of QPEs.

Under the assumption of blackbody emission, the area $A$ of the
emitting region can be estimated for each data point (e.g. at each
phase during the QPE evolution) as $A=L_{\rm bol} \, \sigma^{-1}\,
T^{-4}$, where $\sigma$ is the Stefan-Boltzmann constant. In the lower
panel of Fig.~\ref{fig:QPEBB}, we show the estimated blackbody radius
(assuming $A=2\pi R^2$) as a function of QPE phase. The derived
blackbody radius is extremely small at all phases, as noted by
\citet{2020MNRAS.493L.120K} and \citet{2022arXiv220902786K}. The range
of observed radii ($4-12\times 10^{10}$~cm) corresponds to $\simeq
0.04-0.13~R_g$ ($M_{\rm BH}\simeq 4\times 10^6~M_\sun$) and $\simeq
1-3~R_g$ ($M_{\rm BH}\simeq 3\times 10^5~M_\sun$) for the extremal
$M_{\rm BH}$ values we derive from the spectral analysis in
Sect.~\ref{sec:spectral}. We point out, however, that the derived
blackbody size could be under-estimated by as much as $\sim$ one order
of magnitude as we use a pure blackbody spectral model ignoring
scattering, that is ignoring any X-ray colour-correction.

As was noted, the QPE spectra do not require any absorption in excess
of the Galactic one, that is the warm absorber that is clearly present
during the quiescent period disappears during QPEs. This could in
principle be attributed to a sudden increase in ionisation during
QPEs, or to the lower signal-to-noise of the phase-resolved QPE
spectra with respect to the quiescent level one. In the former case,
the bolometric luminosity associated with QPEs is (even at peak) only
a tiny fraction ($\leq 15$\%) of that of the quiescent level. However,
the QPE luminosity fraction is drastically higher above $\sim
0.7$~keV, which is the relevant band in terms of ionising luminosity
as far as absorption features around that energy are concerned (see
Fig.~\ref{fig:merged}), with $L_{\rm QPE} \simeq 200 \times L_{\rm
  quiescence}$ at QPE peak. Such a sudden increase in ionising
luminosity could then be responsible for the disappearance of the warm
absorber during QPEs.

The total energy radiated during one QPE is of the order of $E_{\rm
  QPE} \simeq 6.7\times 10^{45}$~erg. We do not have enough
observations with QPEs to constrain the peak luminosity evolution on
timescales longer than $\sim 1$~yr. Hence, we assume that $E_{\rm QPE}$ is representative
of the average QPE radiated energy at all times. Although QPEs cannot
be interpreted as accretion-related without ambiguity (see discussion
in Sect.~\ref{sec:QPEemission}), we estimate, for future reference, the
corresponding accreted mass assuming an efficiency $\eta =0.1$ which
results into $M_{\rm accr}^{\rm (QPE)}\simeq 3.7\times 10^{-8}~M_\sun$
per QPE or, considering $\sim 971$ QPEs per year, $M_{\rm accr}^{\rm
  (QPE)}\simeq 3.6\times 10^{-5}~M_\sun$~yr$^{-1}$. If
QPEs were always present but undetected against higher quiescent
level emission, the total mass accreted from QPEs in $\sim 12$~yr
would be $\leq 4.3\times 10^{-4}~M_\sun$, a very
conservative upper limit.

\section{Discussion}
\label{sec:discussion}

Before discussing some of the physical consequences of our
findings, we provide a summary of all results reported
here and in previous studies of GSN~069 to inform the discussion. The
body of results presented here can be used in future studies to constrain
the different theoretical models that have been (and likely will be)
proposed for QPEs.

\subsection{Summary of the main results}
\label{sec:summary}

\begin{itemize}

\item QPEs in GSN~069 are abrupt flares occurring approximately every
  $\sim 9$~hr and lasting $\sim 1$~hr during which the X-ray
  count rate increases by up to two orders of magnitude with respect
  to the quiescent level, depending on the considered energy band
  \citep{2019Natur.573..381M}. They were first discovered in December
  2018 and last detected in January 2020. QPEs are then a transient
  phenomenon in GSN~069 with an observed life-time of $\sim
  1.05$~yr. However, considering that QPEs may have been missed
  in observations with high (quiescent) flux level, as well accounting
  for periods that were not covered by long enough exposures, the
  actual QPE life-time could be significantly longer.

\item With the exception of the last observation where they are
  detected (XMM6), QPEs are characterised by alternating recurrence
  times and intensities. The contrast between the intensity of
  consecutive QPEs is correlated with the recurrence time between
  them, as shown in Fig.~\ref{fig:AversusT}. Strong QPEs are always
  followed by long recurrence times (and vice-versa). Remarkably, the
  correlation implies that consecutive QPEs have the same intensity
  when long and short recurrence times have equal duration. During the
  XMM6 observation, the alternating intensity is (roughly) preserved,
  but recurrence times increase monotonically suggesting that the
  system was experiencing some major change at the epoch of the XMM6
  observation (see Fig.~\ref{fig:ATrec}).

\item QPEs measured in harder energy bands are stronger (with respect
  to the quiescent level), peak earlier, and have shorter duration than
  when measured at softer energies \citep{2019Natur.573..381M}. When
  modelled with blackbody emission, the QPE spectral evolution shows
  hysteresis in the $L_{\rm bol}-kT$ plane (Fig.~\ref{fig:QPEBB}) and
  the maximum temperature, of the order of $\simeq 110$~eV, is reached
  slightly earlier than the luminosity peak. The blackbody emitting
  area is extremely small with radius evolving from $1$ to $3~R_g$ at
  most if $M_{\rm BH}\sim {\rm few} \times 10^5~M_\sun$ (although the
  size may be larger if scattering is important). The total energy
  radiated away is $\simeq 6.7\times 10^{45}$~erg per QPE during XMM4,
  and slightly higher (lower) during XMM3 (XMM5).

\item During observations with QPEs, the quiescent level emission
  varies with a characteristic timescale equal to the average
  observation-dependent recurrence time, and folded light curves
  exhibit a characteristic quasi-periodic oscillation, or QPO (see
  Fig.~\ref{fig:fold}). The fact that the quiescent level QPO period
  is equal to the recurrence time between consecutive QPEs suggests
  that the QPO is associated with every single QPE, independently on
  its strong or weak type. The peak of the quiescent level QPO lags the
  preceding QPE by $8$-$10$~ks, and the time delay is well correlated
  with the observation-dependent average recurrence time. The QPE-QPO
  intensity ratio in the folded light curves is constant over the XMM3
  to XMM5 observations spanning $\simeq 160$~d (and is of the order of
  $\sim37$, see Fig~\ref{fig:qpo_prop}).

\item The quiescent level X-ray spectrum is consistent with accretion
  disc emission satisfying $L_X\propto T^4$ in the 0.2-2~keV band
  which suggests that $L_X$ is a good proxy of the bolometric
  luminosity. This most likely implies a compact accretion flow with
  suppressed optical/UV emission. Indeed, by combining the X-ray data
  with two \hst\ observations performed $\sim 4$ years apart and
  quasi-simultaneously with two \xmm\ exposures, we infer an outer
  disc radius $R_{\rm out} \leq 140~R_g$, with a statistical preference
  for a very compact disc with $R_{\rm out} \simeq 15~R_g$. If the
  disc is really only $\sim 15~R_g$ in size, the UV emission must be
  due to a process different from intrinsic disc emission such as,
  for example, reprocessing. The black hole mass can only be constrained in
  the range $M_{\rm BH} = 0.3-4\times 10^6~M_\sun$ (for a non-spinning
  Schwarzschild black hole) and mostly depends on the assumed
  colour-correction for the disc emission.

\item A warm absorber (mostly detected through a broad absorption
  feature around $0.7$~keV) is present in the quiescent spectra
  throughout the $12$~yr evolution, and it is consistent with being
  constant in both column density and ionisation. The warm absorber
  disappears during QPEs, possibly because of the $\sim 200$ times
  higher X-ray luminosity above $\sim 0.7$~keV at QPE peak, likely
  inducing a much higher ionisation and thus reduced absorption
  features equivalent width.

\item The $12$~yr-long evolution of the quiescent emission is
  consistent with repeating TDEs in GSN~069, with two events $\sim
  9$~yr apart peaking at $L_{\rm bol}\simeq 6-7\times
  10^{43}$~erg~s$^{-1}$ and $\simeq 4-5\times 10^{43}$~erg~s$^{-1}$
  respectively when assuming a compact disc with $R_{\rm out} \sim
  15~R_g$ (the luminosity for the more extended case being $\sim 2.5$
  times higher). Assuming that $\sim 50$\% of the disrupted mass is
  bound to the SMBH and subsequently accreted, we estimate a disrupted
  mass of $0.46\pm 0.20~M_\sun$ for the first TDE. The disrupted mass
  during the second event is more uncertain and it depends on how we
  consider the effect of the debris from the first event that still
  have to be accreted. We conservatively estimate a disrupted mass in
  the range of $0.06-0.38~M_\sun$ for the second TDE.

\item We detect a precursor X-ray flare before the second TDE,
  peaking at $L_{\rm bol} \simeq 3-5\times 10^{43}$~erg~s$^{-1}$. The
  precursor flare lasts $\simeq 150-200$~d in total, and precedes the
  peak of the second TDE by $400-500$~d. A possible precursor
  flare is present also before the first TDE, but its detection is
  only tentative. If interpreted as an accretion flare, the radiated energy
  of the precursor flare in TDE~2 corresponds to an
  accreted mass of $\simeq 0.006\pm 0.004~M_\sun$.

\end{itemize}

\subsection{Constraints on the QPE emission mechanism}
\label{sec:QPEemission}

As is discussed by \citet{2022A&A...662A..49A}, models based on pure
gravitational lensing \citep{2021MNRAS.503.1703I} are disfavoured due
to the QPE energy-dependence (lensing being achromatic) and the
difficulties in explaining simultaneously both the duration and
amplitude of QPEs. Classical disc instability models are difficult to
reconcile with the fast QPE timescales even for a black hole mass at
the lower end of the allowed range (${\rm few}\times 10^5~M_\sun$)
unless the disc properties are significantly different than
standard. Moreover, the alternating pattern of QPE intensities and
recurrence times is not naturally predicted, and may need to be
introduced by an ad-hoc feedback mechanism. Instability-based models
generally produce light curves with slow rise and fast decay, while
QPEs in GSN~069 appear symmetric and, if anything, with a slower
exponential decay as shown in Fig.~2 of
\citet{2019Natur.573..381M}. However, modified models that
significantly reduce the size of the unstable region and include
disc-corona interaction or the presence of magnetic fields are worth
exploring in the future
\citep{2020A&A...641A.167S,2022ApJ...928L..18P,2022arXiv221100704K}
and, in fact, the QPE spectral evolution is consistent with
expectations from instability models \citep{2022ApJ...928L..18P}.

Another type of instability has been suggested by
\citet{2021ApJ...909...82R} as a potential source of AGN variability,
including QPEs. The idea is that accretion flows that are misaligned
to the black hole spin can become warped due to Lense-Thirring
precession, to the point that strongly warped discs may be unstable
and break up into discrete rings. The rings then precess basically
independently from each other, and their misalignment can give rise to
shocks which may result in radiative losses and also in an enhancement
of mass accretion rate on timescales that can be shorter than the
viscous one. As shown by \citet{2021ApJ...909...82R}, if the warp is
unstable in the innermost regions, quasi periodic shocks as well as
mass accretion rate variations are expected, which may give rise to
QPEs (and, possibly, to QPOs). In order for the predicted timescales
of the instability to match the observed QPE recurrence time in
GSN~069, relatively large values of the black hole mass are favoured
($\sim 10^6~M_\sun$), which is consistent with our
findings (see Sect.~\ref{sec:spectral}). As the properties of the
instability depend on the disc structure (e.g. viscosity $\alpha$ and
scale-height $H/R$), the proposed mechanism might explain why QPEs are
not always present in GSN~069 by assuming, for instance, that the
accretion flow is stabilised against disc tearing at high mass
accretion rates.

Most other QPE models invoke the interaction between a stellar-like
body and the SMBH (or SMBH plus accretion flow) identifying the QPE
quasi-periodicity with the orbital period of the stellar
companion. The proposed models can be roughly separated into two
classes in which QPEs are either due to collisions between the stellar
companion and the accretion flow in a nearly circular orbit
\citep{2021ApJ...921L..32X}, or to episodic mass transfer events at
each pericenter passage in a mildly or highly eccentric orbit
\citep{2020MNRAS.493L.120K,2022ApJ...930..122C,2022MNRAS.515.4344K,2022ApJ...933..225W,2022A&A...661A..55Z,2022arXiv220902786K,2022arXiv221008023L}. Below
we discuss, with no ambition of completeness, some of the most
relevant aspects of the two classes of proposed models in comparison
with the rich phenomenology of GSN~069.

As shown in Fig.~\ref{fig:QPEBB}, the QPE bolometric luminosity does
not follow the $L\propto T^4$ relation except, possibly, during the
initial rise. Rather, hysteresis is clearly present in the $L_{\rm
  bol}-T$ plane, a likely general QPE property
\citep{2022A&A...662A..49A}. Moreover, assuming blackbody emission,
the resulting emitting area is very small, although we must point out
that the inferred sizes may be underestimated by a factor of a few as
the spectral model (a simple blackbody one) does not include
scattering. At face value, this suggests that QPEs originate from a
compact region that heats up rapidly, inflates, and cools down via
X-ray emission while expanding. Qualitatively, QPEs appear then to be
more consistent with shocks than with accretion onto the SMBH.

One class of QPE models where the X-ray emission is powered by shocks
is that of collisions between an orbiting stellar body and the
accretion flow occurring twice per orbit. \citet{2021ApJ...921L..32X}
show that such a model can explain the irregularities of the QPE
arrival times and the alternating pattern of their
recurrence times via the combination of small eccentricity and
relativistic precession. Their orbital model successfully reproduces
the QPE arrival times for an orbit with semi-major axis $a\simeq
350-400$~R$_g$ and small eccentricity ($e\simeq 0.06$) crossing the
accretion disc around a $3\times 10^5$~M$_\sun$ SMBH twice per
orbit. The derived SMBH mass is consistent with the very low end of
the allowed mass range in
Sect.~\ref{sec:spectral}. \citet{2021ApJ...921L..32X} propose that the
orbiter is the remnant of the pTDE of a red giant star, in line with
our analysis (see Sect.~\ref{sec:evolutionall}).

Further work is needed to account for the time delay between TDE~1 and
the appearance of QPEs $4.5-8.5$~yr later. Moreover, that shocks
produced by impacts at such large radii can produce relatively high
soft X-rays luminosity with negligible associated UV emission
\citep{2019Natur.573..381M} needs to be studied in
detail. \citet{2021ApJ...921L..32X} do not discuss the alternating
QPE intensity pattern, nor its relation with the alternating 
recurrence times. If, as would appear natural, strong (weak) QPEs are
due to impact towards (away from) the observer, precession inevitably
means that, depending on precession phase, strong QPEs would be
followed by either long or short recurrence times as shown in the
right panel of Fig.~4 of \citet{2021ApJ...921L..32X}. This behaviour
is not consistent with QPE data, as strong QPEs are always followed by
long recurrence times, at least for the events that we have observed
(see Fig.~\ref{fig:AversusT}).

A different class of models is based on the idea that QPEs are instead
due to episodic mass transfer from a donor star at each pericenter
passage through Roche lobe overflow, which can be further
differentiated in models assuming QPE emission from shocks or from
accretion onto the SMBH. Circularisation shocks are proposed as a
potential QPE emission mechanism by \citet{2022arXiv220902786K} and
\citet{2022arXiv221008023L}. While \citet{2022arXiv220902786K} ignore
the presence of the accretion flow that gives rise to the quiescent
level emission and invoke stream-stream interactions as a source of
QPEs, \citet{2022arXiv221008023L} consider the stream-flow interaction
where they assume that the flow is due to previous mass transfer
events. Stream self-interactions, as proposed by
\citet{2022arXiv220902786K}, may also play an important role in this
context, but it seems difficult that the incoming streams would always
miss the existing accretion flow and interact first between
themselves. As shown by \citet{2022arXiv220902786K}, shocks involving
velocities of the order of the orbiter's velocity at pericenter
produce emission with a typical temperature that is consistent with
that observed in QPEs. For both scenarios (stream-flow and
stream-stream interaction) the energy budget also appears to be
consistent with observations.  Both proposed models invoke episodic
mass transfer of a Sun-like star with $M_*\simeq 0.5-1~M_\sun$ in a
mildly eccentric orbit. As discussed in Sect.~\ref{sec:evolutionall},
however, the properties of TDE~1 suggest that the remnant is unlikely
to be a Sun-like star, and that a compact core, potentially surrounded
by a remaining envelope, is more likely. Hence, if a (low-mass)
Sun-like star is responsible for QPEs, it is likely unrelated to TDE~1
and TDE~2 and the system may be a triple. 

QPEs due to episodic mass
transfer from a compact object (e.g. a white dwarf or a compact
stellar core) in a highly eccentric orbit, possibly captured through
the Hills mechanism \citep{2022ApJ...929L..20C}, are proposed instead
by \citet{2020MNRAS.493L.120K,2022MNRAS.515.4344K},
\citet{2022ApJ...930..122C}, and \citet{2022ApJ...933..225W}, while
\citet{2022A&A...661A..55Z} consider mass transfer from the He
envelope of a H-deficient post-AGB star. In all of these cases cases, the QPE X-ray
emission is assumed to be powered directly by accretion onto the SMBH
of the tidally stripped mass, leading to self-consistent solutions for
the stellar structure and orbit.

The hysteresis in Fig.~\ref{fig:QPEBB} is not necessarily inconsistent
with an accretion origin for QPEs considering their highly dynamical
nature. If a mass accretion rate perturbation propagates inwards, soft
X-rays respond first on timescales associated with their typical
emitting radii, while harder X-rays come later but evolve on faster
timescales. Hence, at any given time, we observe the superposition of
X-ray emission from different annuli on the disc, with hotter region
having already responded, and colder ones still rising, which does not
produce an obvious, global $L_{\rm bol}\propto T^4$. The
accretion-propagation scenario is in fact consistent with the energy
dependence of QPEs, with higher energy X-rays peaking earlier and
evolving faster than lower energy ones during QPEs as shown in Fig.~2
of \citet{2019Natur.573..381M}. On the other hand, the accretion
timescale must be extremely fast, which is only possible if QPEs are
induced by accretion from a narrow ring. This is consistent with a
compact orbiter with very high eccentricity (and hence small
pericenter distance) and, possibly, with a compact accretion flow
whose extension is limited by the repeated passages of the orbiting
star \citep{2020MNRAS.493L.120K,2022MNRAS.515.4344K}. A
compact donor on a highly eccentric orbit requires the black hole
mass to be of ${\rm few}\times 10^5~M_\sun$ at most for the object not
plunge directly into the black hole. As a side note, we point out that
X-ray flares with soft X-ray emission lasting longer than harder X-ray
one (as observed in QPEs) may originate from magnetic energy release
in the plunging region, that is at radii smaller than $R_{\rm isco}$, as
discussed by \citep{2003ApJ...585..429M} making use of
three-dimensional magnetohydrodynamical simulations, see in particular
their Fig.~17.

Within the context of mass transfer scenarios, the alternating
strong-weak intensities and long-short recurrence times have been
addressed by \citet{2022MNRAS.515.4344K} who propose that the very
fast mass transfer timescale at pericenter induce an oscillatory
behaviour around the evolutionary mean (basically set by gravitational
radiation). We point out that such an idea is consistent with the
properties shown in Fig.~\ref{fig:AversusT} as intensities and
recurrence times appear to indeed oscillate around the
equal-intensity, equal-recurrence time point that is naturally
associated with the evolutionary mean. Another plausible mechanism,
where changes in intensities and recurrence times are associated to
the X-ray irradiation of the stellar surface at QPEs, has been instead
proposed by \citet{2022arXiv220902786K}.

Finally, we note that disc tearing resulting from instabilities in a
warped accretion flow may also naturally give rise to shocks between
misaligned neighbouring rings in the innermost regions
\citep{2021ApJ...909...82R}. Moreover, shocks are expected to drive
mass inflow on fast timescales associated with narrow rings which, as
discussed above, may be consistent with the X-ray spectral evolution
of QPEs. As mentioned, the variability timescales can match the
observed QPE period if the black hole mass is of the order of $\sim 10^6~M_\sun$, consistent with the black hole mass range in GSN~069. The dependency of the
instability on the accretion flow properties (as well as on black hole
spin) may produce enough diversity to account for the different
characteristics of the observed QPE sources and could explain why QPEs
are not always present during the evolution of GSN~069.

\subsection{The quiescent level QPO}
\label{sec:QPOcnstraints}

Proposed models should also account for the quiescent level QPO
discussed in Sect.~\ref{sec:QPO}. Disturbances in the accretion disc
are naturally produced in both the mass transfer and impacts scenarios
with a characteristic timescale of $\sim 9$~hr. In this context, the
QPE-QPO time delay we measure is not necessarily relevant, as the QPO
observed at a given epoch may be due to events having occurred long
time before. On the other hand, the stable QPE-QPO intensity ratio
(Fig.~\ref{fig:qpo_prop}), and the 1:1 correlation
(Fig.~\ref{fig:delaycorr}) suggest that each QPO is indeed associated
with the immediately preceding QPE.

Here we discuss one of the possible mechanisms that can produce the
observed QPO: we assume (quasi-periodic) shocks on the accretion flow
at radius $R_{\rm sh}$ producing QPE emission locally, and also
driving a perturbation that propagates in the flow giving rise to
enhanced quasi-periodic X-ray emission (the QPO). That outlined above
is a plausible but not unique QPO origin, so that the conclusions
reached below are model-dependent. Shocks at $R_{\rm sh}$ may be
produced either by impacts on the accretion flow (twice per orbit) or
by the interaction between incoming streams and the flow in the mass
transfer scenario (with mass transfer events occurring once per
orbit). In both cases, shocks are expected to roughly occur at radii
of the order of $R_{\rm sh} \simeq R_{\rm p} = (1-e)\,a$ on the flow,
where $a$ is the orbit semi-major axis, and $e$ the orbital
eccentricity.

We shall first discuss mass transfer scenarios in which the QPE and
QPO periods are equal to the orbital one, namely $P \simeq 32$~ks (see
Table~\ref{tab:qpefit}). We note that the timescale for the
propagation of the perturbation from $R_{\rm sh}$ inwards, $t_{\rm
  prop}$, must be shorter than the period $P$ of the perturbation in
order not to smear out the QPO. The observed QPE-QPO time delay of
$8-11$~ks could perhaps be taken as a plausible estimate of $t_{\rm
  prop}$, although the delay may also be affected by the time it takes
to produce the QPE from the shock. The propagation of perturbations in
the accretion flow is associated with the viscous timescale $t_{\rm
  visc}$. Faster timescales are reached by propagating cooling or heating
fronts in the disc \citep{2018MNRAS.480.4468R}, with $t_{\rm front}
\simeq (H/R)\,t_{\rm visc}$. Although such a fast timescale is
unlikely to be associated with the propagation of mass accretion rate
fluctuations, we adopt conservatively $t_{\rm front} \leq t_{\rm prop}
\leq t_{\rm visc}$, baring in mind that we are only considering
order-of-magnitude estimates here.

\begin{figure}
\centering
\includegraphics[width=8.8cm]{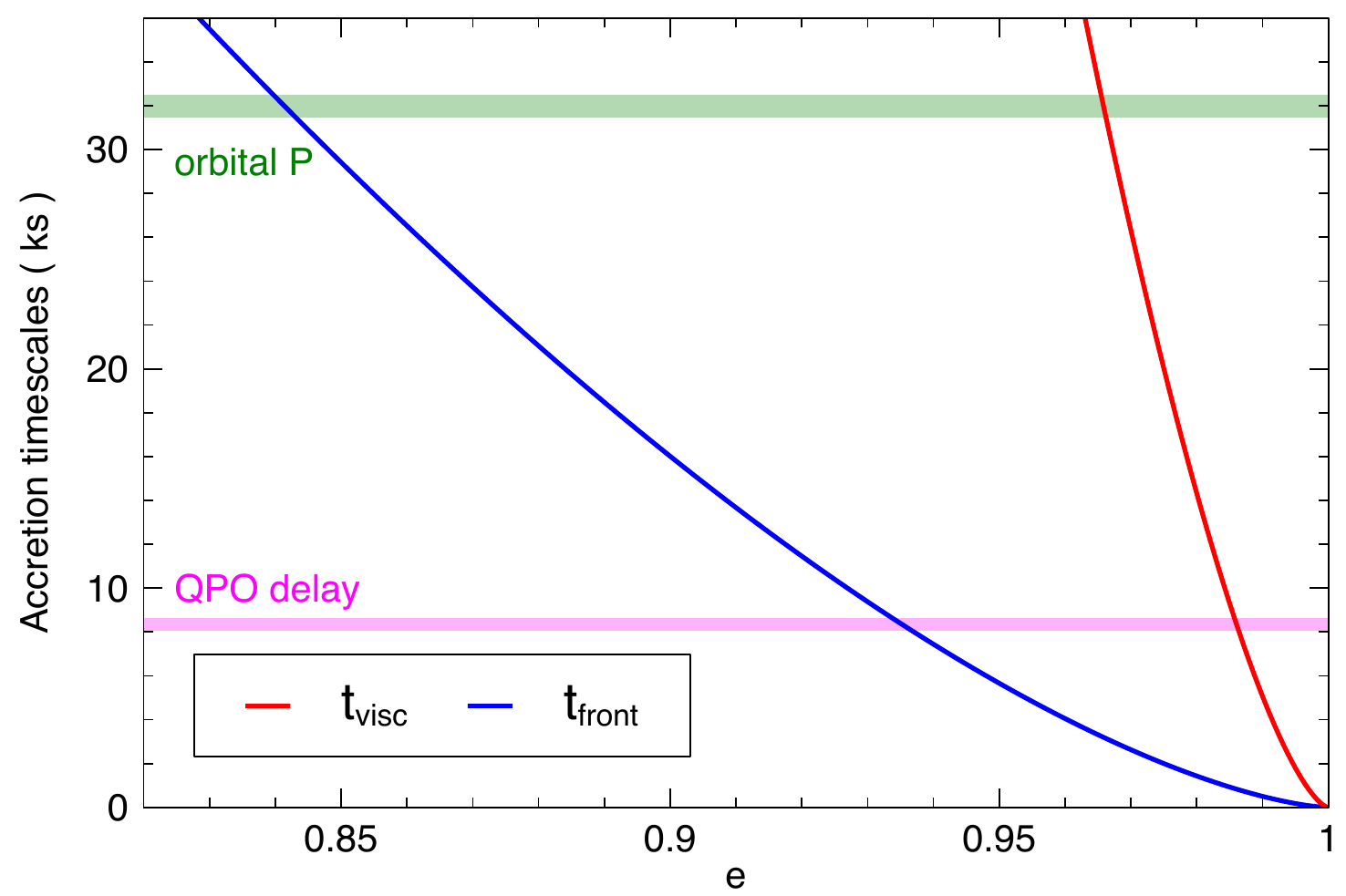}
\caption{Viscous and front propagation timescales $t_{\rm visc}$ and
  $t_{\rm front}$ evaluated at $R_{\rm sh}$ as a function of orbital
  eccentricity. The upper (green) horizontal shaded area represents
  the period of the shock-induced perturbation (equal to that of the QPE or QPO,
  i.e. to the orbital period in the assumed scenario), while the lower
  (magenta) one is the shortest observed time delay of the 
  QPO with respect to the immediately preceding QPE (XMM3).}
\label{fig:tvtf}
\end{figure}

All accretion timescales are proportional to the dynamical one $t_{\rm
  dyn} = (G\,M_{\rm BH})^{-1/2}R^{3/2}$ at radius $R$. When evaluated
at $R=R_{\rm sh} \simeq (1-e)\, a$, $t_{\rm dyn}$ does not depend
anymore on $M_{\rm BH}$, so that, for a given period $P$, $t_{\rm
  visc}$ and $t_{\rm front}$ are a function of the orbital
eccentricity $e$ only. In order to estimate $t_{\rm visc}$ and $t_{\rm
  front}$ at $R_{\rm sh}$, we set the viscosity parameter to $\alpha
=0.1$ and the flow scale-height to $H/R = 0.1$, as seems reasonable
for accretion flows produced by TDEs. The resulting $t_{\rm visc}$ and
$t_{\rm front}$ are shown in Fig.~\ref{fig:tvtf} as a function of
orbital eccentricity, where the upper (green) horizontal shaded area
is the perturbation period ($P\simeq 32$~ks) while the lower
(magenta) one is the shortest observed QPE-QPO time delay ($8.3\pm
0.3$~ks in XMM3) shown for completeness. The less stringent constraint
on orbital eccentricity is obtained by requiring that the propagation
timescale is shorter than the perturbation (in this case orbital)
period $P$ in order not to smear out the QPO (i.e. $t_{\rm prop}
\lesssim P$). Even using the faster of the two timescales ($t_{\rm
  front}$), the orbital eccentricity must be $e\geq 0.84$. If, perhaps
more reasonably, we adopt $t_{\rm prop} \sim t_{\rm visc}$, one has
$e\geq 0.965$.  A geometrically thinner accretion flow ($H/R
\leq 0.1$) requires even higher $e$. If the QPE-QPO time delay is
assumed to represents the actual $t_{\rm prop}$, higher eccentricities
are required, see Fig.~\ref{fig:tvtf}.

Hence, under the assumption that the QPO is due to a propagating
perturbation caused by each QPE and that QPEs are associated with an
orbiting body, the QPE-inducing body must describe a highly eccentric
orbit. In mass transfer scenarios, the orbiting object has to fill its
Roche lobe at pericenter. As discussed by
\citet{2020MNRAS.493L.120K,2022MNRAS.515.4344K},
\citet{2022ApJ...930..122C}, and \citet{2022ApJ...933..225W} such a
condition, when combined with a high eccentricity (i.e. small $R_{\rm
  p}$), implies a compact white-dwarf like object. In particular,
assuming a typical stellar core mass $M_* \simeq 0.3~M_\sun$, and the
less stringent constraint on eccentricity ($e\simeq 0.84$), one has
$R_*\simeq 0.1~R_\sun$, allowing for the presence of a significant
envelope bound to the core which may be then tidally stripped at
pericenter passage, giving rise to QPEs. Higher eccentricities result in
smaller $R_*$. Such a compact object in a highly eccentric orbit has
pericenter distance of only few$-$few tens of $R_g$
\citep{2020SSRv..216...39M,2022ApJ...929L..20C} which, as mentioned,
favour a low black hole mass in order for the object not to be
swallowed by the black hole. In any case, the orbit is highly
relativistic, and the flash of X-ray emission at each pericenter
passage (QPE) represents a probe sending back to us a signal (that can
be accurately timed) from a highly distorted region of spacetime. QPEs
may thus enable us to constrain the strong field regime of General
Relativity at distances from the SMBH event horizon that appear
difficult to probe by other means.

On the other hand, if QPEs are produced by impacts, the same arguments
apply with the exception that the orbital period is twice as large,
increasing all accretion timescales. Within this scenario, the
periodic perturbation (each impact) has still the same period of
$\simeq 32$~ks (as the model assumes two impacts per orbit), so that
the eccentricity needs to be even higher than in the mass transfer
scenario. However, one of the most attractive aspects of the impacts
model, is that it can explain the alternating recurrence times of QPEs
by invoking a small orbital eccentricity \citep{2021ApJ...921L..32X},
which is therefore inconsistent with the proposed QPO
interpretation. Moreover, and independently from the nature of the
QPO, we point out that, for small eccentricities, the pericenter distance
(which is also roughly the radial distance where impacts occur) is
consistent with the maximum allowed disc radius that is needed not to
over-predict the UV emission ($R_{\rm out} \simeq 140~R_g$) only for
$M_{\rm BH} \gtrsim 1\times 10^6~M_\sun$, while the orbital solution
of \citet{2021ApJ...921L..32X} implies a best-fitting black hole mass
of $3\times 10^5~M_\sun$. One further potential weakness of this model
is that, while it can likely explain the regular QPE patterns observed
in GSN~069 and eRO-QPE2, it appears difficult that it alone can
provide the diversity that is needed to account for the less regular
QPEs in RX~J301.9+2747 and eRO-QPE1
\citep{2020A&A...636L...2G,2022A&A...662A..49A}. Mass transfer
scenarios may be more naturally associated with different stellar and
orbital properties leading to a broader range of observable
properties, see for example \citet{2022ApJ...926..101M}.

As mentioned, the constraints derived above strongly depend on the adopted
interpretation of the QPO. As such, our conclusions should be treated
as model-dependent and therefore do not represent a robust, conclusive
result. Other possible QPO origins should be explored and contrasted
with the data. One possibility is that QPEs
and QPOs are produced by a combination of the QPE models proposed so
far. For instance, if the orbit is eccentric and the accretion flow
does not extend much beyond $R_{\rm p}$, star-disc collisions may
occur only once per orbit, depending on the relative inclination, and
could give rise to one of the two phenomena (QPE or QPO), with the
other being associated with mass transfer at pericenter. In this case,
the constraint on $e$ is relaxed, and mild eccentricities are
likely possible as well. Within this scenario, and depending on eccentricity
and relative orientation of the orbital and accretion planes, up to
three events per orbit might be expected (two impacts and one mass
transfer event) which may help explaining the complex timing behaviour
of some QPE sources, most notably RX~JJ301.9+2747
\citep{2020A&A...636L...2G}. In mass transfer scenarios, we also note
that low eccentricities would likely increase the fraction of the
orbit where the star over-fills its Roche lobe, perhaps accounting for
the much longer-lived QPEs observed in eRO-QPE~1
\citep{2022A&A...662A..49A} and, potentially, also for the QPO-like
variability in 2XMM~J123103.2+110648
\citep{2022MNRAS.tmp.3139W}. Tidal interactions close to pericenter
may also induce a time-dependent warp of the existing accretion flow
which then precesses and possibly accounts for the observed QPO. In
this context, disc tearing may be relevant
\citep{2021ApJ...909...82R}, and solutions in which QPEs are produced,
for example, by shocks between neighbouring rings should be explored
further. The QPO may be due to the modulation of the X-ray emission
from a dominant ring at its own precession timescale, or to the
propagation of mass accretion rate fluctuations in the inner regions
caused by ring-ring interactions. We encourage exploring these and
other possibilities theoretically and numerically to constrain the
different proposed models, deepening our understanding of the QPE
phenomenon in GSN~069 and the other QPE sources.

\subsection{The TDE-QPE connection}
\label{sec:evolutionall}

\begin{figure}
\centering
\includegraphics[width=8.8cm]{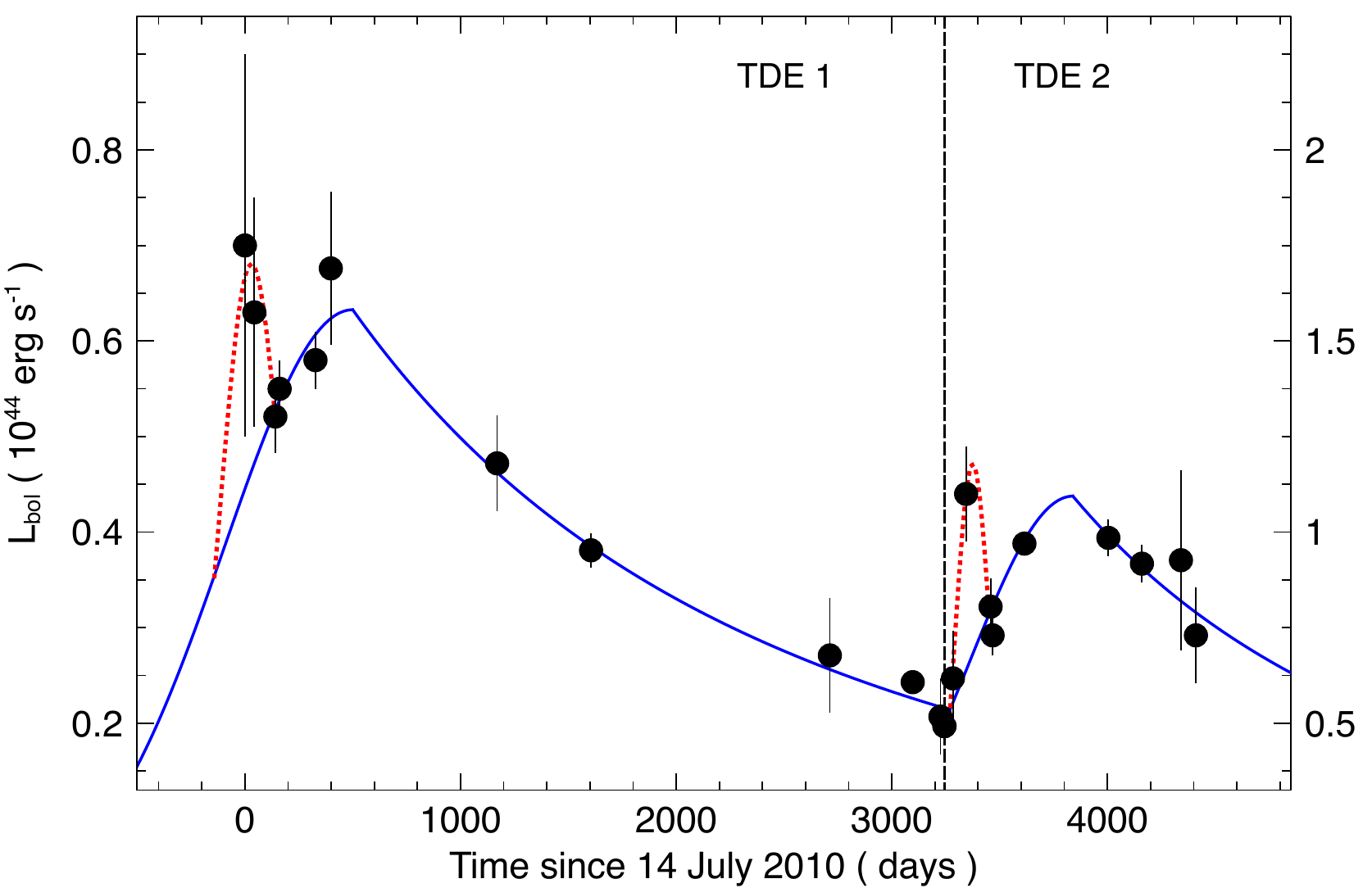}
\caption{Overall $L_{\rm bol}$ evolution over the $\sim 12$~yr probed
  by our \swift\ and \xmm\ campaign, together with the best-fitting
  models during Phase1 and Phase2, re-named here as TDE~1 and TDE~2
  respectively. The y-axis scales are the same as those
  defined in the caption of Fig.~\ref{fig:xmmLbolRout}.}
\label{fig:AllPhases}
\end{figure}

The overall evolution of GSN~069 over the $\simeq 12$~yr period probed
by our campaign is shown in Fig.~\ref{fig:AllPhases} together with the
best-fitting model discussed in the Sect~\ref{sec:DecayFits} and
\ref{sec:RebrighteningFits}. As shown there, Phase1 and Phase2 are
well described by the typical rise-and-decay light curve of TDEs, and we
therefore re-name Phase1 and Phase2 as TDE~1 and TDE~2
respectively. We also include in Fig.~\ref{fig:AllPhases} a possible
precursor X-ray flare before TDE~1: this is not a fit, but rather a
re-scaled version of the flare seen at the beginning of TDE~2 that
accounts well for the first two data points that, as noted in
Sect.~\ref{sec:DecayFits}, lie above the best-fitting model for TDE~1.

The evolution in Fig.~\ref{fig:AllPhases} strongly suggests repeating
TDEs in GSN~069 whose peaks (we ignore here the precursors) are
separated by $\simeq 3340$~d ($\simeq 9.1$~yr). Similar recurrent
  optical or X-ray flares that can be interpreted as repeating
  TDEs have been recently reported in ASASSN-14ko
  \citep{2021ApJ...910..125P}, eRASSt~J045650.3-203750
  \citep{2022arXiv220812452L}, and AT2018fyk
  \citep{2022arXiv220907538W}, with source-dependent timescales
  ranging from $\simeq 100-200$~d to $\simeq 3$~yr. One striking
property of both TDEs in GSN~069 is that their X-ray emission evolves
on much longer timescales than typical TDEs. The X-ray flux decays by
a factor of $\sim 6.8$ in $\sim 9$~yr, and the corresponding
bolometric luminosity drops by an even smaller factor of $\sim 3$ (see
Fig.~\ref{fig:xmmevolution} and \ref{fig:Lbolevolution}). This very
slow decay is rarely seen in TDEs \citep{2022ApJ...924L..35L}. When
the TDE~1 (TDE~2) decay is described with an exponential, its
e-folding timescale is $\tau \simeq 7.6$~yr ($\simeq 3.4$~yr), more
than one order of magnitude longer than typical for both X-ray and
optical TDEs \citep{2015JHEAp...7..148K,2021ApJ...908....4V}. The
width around the peak is also, in both cases, much longer than in typical
TDEs (about $\simeq 2.5-4$~yr and $\simeq 1.5-3$~yr for TDE~1 and
TDE~2 respectively). The years-long timescales characterising TDE~1
and TDE~2 can be compared with the so-called fallback time $t_{\rm
  fb}$ that sets the typical timescale of the fallback rate evolution
and that can be rewritten as \citep{2021ARA&A..59...21G}:
\begin{align}
t_{\rm fb} \simeq 41\,\left(\frac{R_*}{R_\sun}
\right)^{3/2}\,\left(\frac{M_*}{M_\sun}
\right)^{-1}\,\left(\frac{M_{\rm BH}}{10^6~M_\sun} \right)^{1/2}\,{\rm
  d}\,,
\label{eq:Tfb}
\end{align}
which, for the disruption of a Sun-like star, is $\simeq 20-80$~d in
the allowed range of black hole masses in GSN~069 ($0.3-4\times
10^6~M_\sun$). Although the timescales for partial disruptions are
likely a factor of a few longer than those characterising full TDEs
\citep{2013ApJ...767...25G,2021ApJ...922..168N}, the disruption of a
Sun-like star does not appear to be consistent with the observed
timescales. On the other hand, the disruption of an evolved star
comprising a compact core surrounded by a tenuous envelope may be
viable. Indeed, assuming a typical mass $M_*\sim 1~M_\sun$, the
disruption of a a star with $R_*\simeq 5-12~R_\sun$ (depending on the
assumed $M_{\rm BH}$), gives $t_{\rm fb} \simeq 1\,000$~d, roughly
consistent with the observed timescales. Hence, the years-long
timescales associated with TDE~1 indicate the likely disruption of an
evolved star extending out to at least a few times the Solar radius
\citep{2012ApJ...757..134M}.

The presence of TDE~2 suggests that TDE~1 was only partial and that a
remnant survived TDE~1. As shown by \citet{2012ApJ...757..134M}, the
mass loss for the disruption of the envelope of an evolved star ranges
from $\sim 0.1~M_\sun$ to $\sim 0.7~M_\sun$ depending on stellar
structure and penetration factor $\beta = R_{\rm t}/R_{\rm p}$, where
$R_{\rm p}$ is the pericenter distance of the orbit, and $R_{\rm t}$
is the tidal radius
\begin{align}
R_{\rm t} \simeq 47\,\left(\frac{M_*}{M_\sun} \right)^{-1/3}\,
\left(\frac{R_*}{R_\sun} \right)\, \left(\frac{M_{\rm
    BH}}{10^6~M_\sun} \right)^{-2/3}\,R_g\,.
\label{eq:Rt}
\end{align}
The disrupted mass we derive at TDE~1 ($0.46\pm 0.20~M_\sun$) is thus
fully consistent with the partial disruption of the envelope of an
evolved star.

As discussed in Sect.~\ref{sec:RebrighteningFits}, TDE~2 is also
characterised by years-long timescales, only slightly shorter than
those associated with TDE~1. At face value, this would suggest the
disruption of an extended object also at TDE~2, consistent with
partial/full disruption of the remnant of TDE~1, if the remnant still
comprises a significant envelope. TDE~1 and TDE~2 might then be due to
the repeated tidal stripping of the envelope of an evolved giant
star at each pericenter passage \citep{2013ApJ...777..133M}. If so,
the star is on a $\sim 9$~yr, likely eccentric orbit around the
central SMBH.

If the repeating TDEs in GSN~069 are due to tidal stripping of the
envelope of a star on a $\sim 9$~yr orbit, the remnant that survived
TDE~1 cannot be responsible for QPEs because of the different required
orbital periods ($\sim 9$~yr versus $9-18$~hr, depending on the
adopted QPE model). The system would then be a triple comprising the
SMBH, a giant star on a $\sim 9$~yr orbit whose envelope is tidally
stripped at each pericenter passage, and a further stellar companion
on a much shorter-period ($9-18$~hr) orbit producing QPEs. One further
possibility is that QPEs are not associated with an orbiting companion
and are due, for instance, to shocks between misaligned precessing
rings in the accretion flow, formed as a consequence of disc tearing
\citep{2021ApJ...909...82R}. In such a case, the system only comprises
the SMBH and an orbiting evolved star on a $\sim 9$~yr orbit. The
scenario in which the two repeating TDEs are associated with the
partial TDE of the envelope of an evolved star on a $\sim 9$~yr orbit
is testable, as it predicts a third TDE $\sim 9$~yr after
TDE~2. Monitoring GSN~069 in the soft X-rays during the next few years
is therefore crucial to confirm or falsify this possibility.

However, the partial nature of TDE~1, that is the presence of a
surviving remnant, suggests that a different scenario in which TDE~1,
QPEs, and TDE~2 are all associated with a common progenitor should be
explored. An evolved giant star could be partially disrupted at TDE~1,
leaving a remnant able to survive thousands of pericenter passages and
to give rise to QPEs before being driven to its own partial/full
disruption at TDE~2 by orbital evolution. If TDE~1 is the pTDE of an evolved star, as argued above, this is in
fact possible. The condition to have the envelope (or part of it)
disrupted at TDE~1 leaving a surviving remnant is that the penetration
factors of the progenitor star ($\beta_*$) and of the remnant
($\beta_{\rm rem}$) satisfy $\beta_*\geq 1$ and $\beta_{\rm rem} < 1$,
although this is only approximately true as the critical $\beta$ is
slightly different from unity in the case of pTDEs and, more
generally, $\beta$ also depends on the stellar structure
\citep{2013ApJ...767...25G,2017A&A...600A.124M}. Assuming that the
orbit of the remnant is unperturbed (i.e. the progenitor star and the
remnant have the same pericenter distance), one has
\begin{align}
\beta_{\rm rem}/\beta_* = \left(\frac{R_{\rm rem}}{R_*}
\right)\,\left(\frac{M_*}{M_{\rm rem}} \right)^{1/3}\,.
\label{eq:betacore}
\end{align}
In the limiting case of a bare compact core remnant (with say $M_{\rm
  rem}\simeq 0.3~M_\sun$ and $R_{\rm rem}\simeq 0.014~R_\sun$) and a
relatively common red giant progenitor star ($M_*\simeq 1~M_\sun$
and $R_* \simeq 12~R_\sun$), one has $\beta_{\rm rem}/\beta_* \simeq
1/574$, so that there is a significant range of parameters that can
satisfy the condition for which a remnant survives the next pericenter
passage even if part of the envelope is still bound to the core. This
also means that the disruption of the envelope alone (or part of it)
can occur even for $\beta_* \gg 1$, namely long before the star
reaches $R_{\rm p}$. The number of orbits that the remnant can
complete before being driven to disruption by further orbital (and
stellar size) evolution depends on the initial $\beta_{\rm rem}$ as
well as on the details of its subsequent evolution (gravitational wave
emission, mass transfer, tidal heating, etc). 

However, such an interpretation implies
that the initial evolved star is on a $\sim 9$~hr (or $\sim 18$~hr)
around the black hole. As noted by \citet{2022ApJ...929L..20C}, it is
difficult (basically impossible) to place a large star on an orbit with such a
short period even invoking the break-up of an initial binary system on
a very tight orbit \citep{1988Natur.331..687H}. One possibility is
that the evolved star was captured long ago in a significantly wider
(longer-period) orbit that hardened due to orbital evolution leading
the star to its tidal radius on an orbit with a $\sim 9-18$~hr period
long after being captured by the SMBH.

On the other hand, a significantly more compact object, such as a
white dwarf (WD) may be placed on a very short period, highly
eccentric orbit via the Hills mechanism \citep{2022ApJ...929L..20C},
although \citet{2022arXiv221008023L} argue that the rate of delivering
WDs to the required orbit is too small to explain the observed
QPEs. Within this context, \citet{2022ApJ...933..225W} have proposed
that TDE~1 may represent the disruption of a WD envelope significantly
inflated due tidal heating that can induce runaway fusion in the
envelope \citep{2012ApJ...756L..17F}. The inflated envelope may
produce TDE~1 at larger distances (and hence with longer timescales)
than expected from the partial disruption of an unperturbed WD. While
attractive because it potentially provides a link between TDE~1, QPEs,
and TDE~2, the model faces at least two challenges: firstly, the
disrupted mass that we estimate at TDE~1 ($0.46\pm 0.20~M_\sun$) is
much too large for the disruption of a typical WD envelope, and,
secondly, the surviving remnant would likely be just the WD itself,
whose further disruption at TDE~2 should in principle proceed on much
shorter timescales than observed \citep{2020SSRv..216...39M}. The
latter inconsistency may perhaps be cured by noting that TDE~2 does
not occur in a clean environment because of the presence of the
accretion flow from TDE~1. Hence, the TDE~2 timescales may not reflect
accurately the structure of the disrupted body, but could rather still
be dominated by the fallback rate from TDE~1 which is likely dominant
in terms of bound mass, as discussed in Sect~\ref{sec:DecayFits} and
Sect.~\ref{sec:RebrighteningFits}.

\begin{figure}
\centering
\includegraphics[width=8.8cm]{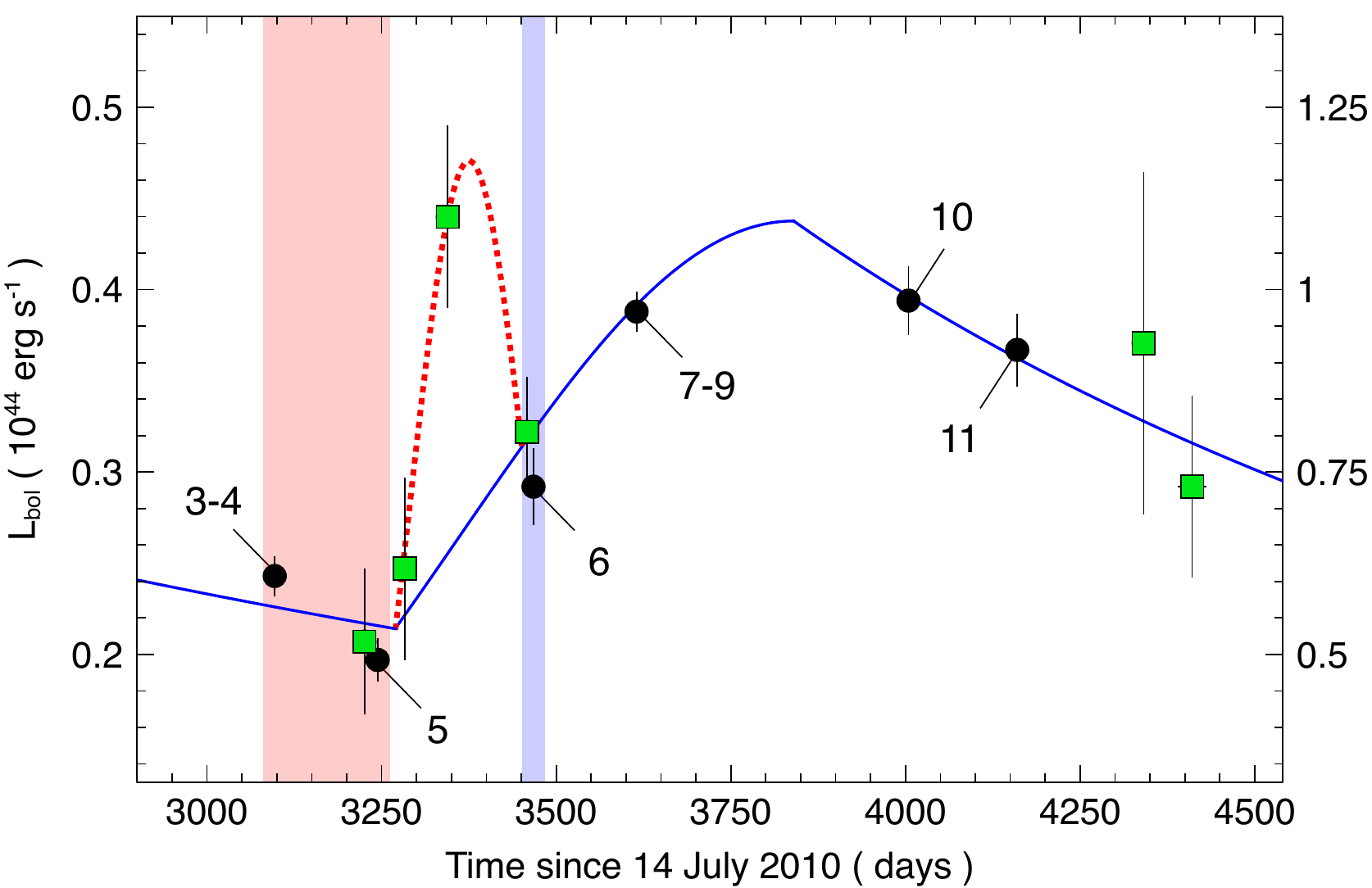}
\caption{Portion of the $L_{\rm bol}$ light curve comprising the final
  part of TDE~1 and the overall TDE~2 evolution. Black circles (green
  squares) denote \xmm\ (\swift) observations. Only \xmm\ observations
  (whose observation number is also indicated) are long enough to
  detect QPEs. The shaded area highlight time-intervals during which
  high-amplitude, regular QPE are detected (XMM3 to XMM5, red) and
  those in which weak QPEs with irregular recurrence times are
  observed (XMM6, blue). No QPEs are detected after XMM6, although the
  higher level of the quiescent emission might have decreased significantly their contrast in
  subsequent observations.}
\label{fig:TDE_QPE}
\end{figure}

As discussed above, weak and irregular QPEs are observed during
XMM6. As shown in Fig.~\ref{fig:TDE_QPE}, XMM6 is performed after
TDE~2, basically at the end of the precursor X-ray flare. The
irregular and weak nature of QPEs may be understood in terms of the
strong perturbation of the system (including the accretion flow)
caused by TDE~2. For example, if QPEs are produced by shocks on the
accretion flow (see e.g. Sect.~\ref{sec:QPEemission}), the disruption
(or, at least, disturbance) and subsequent re-formation of the
accretion flow due to TDE~2 is bound to perturb the regularity of
QPEs. Moreover, if QPEs are due to mass transfer at pericenter, they
are expected to be initially weak and then grow in amplitude with
time, as discussed by \citet{2010MNRAS.409L..25Z} in the case of a
white dwarf donor. The increase of the continuum level during the rise
of TDE~2 might have reduced the contrast of weak, XMM6-like QPEs with
respect to the quiescent continuum level, possibly preventing us from
detecting QPEs after XMM6\footnote{A few short-lived flares are
  present in some of the exposures after XMM6. They are weak, and
  possibly consistent with red-noise variability. A detailed analysis
  of these observations is on-going and will be reported in a
  forthcoming publication (Miniutti et al. in preparation).}. However,
as the continuum is now decaying again (and the QPE amplitude is
possibly growing), high-contrast QPEs may soon re-appear in GSN~069.

In mass transfer scenarios for QPEs, the detection of of a new QPE
phase in the future (of which QPEs in XMM6 may be considered the
beginning) would then imply that an orbiter is still present. The QPE
properties of the new (putative) QPE phase could be used to infer
whether the orbiter producing them is related to that inducing the
previous set of QPEs. If TDE~2 partially disrupted the object that is
responsible for previous QPEs, mass exchange at TDE~2 should result
into different QPE recurrence times (associated with the orbital
period), while if TDE~2 is due to tidal stripping of the envelope of a
giant star on a $\sim 9$~yr orbit and QPEs are associated with an
independent orbiting companion, the new QPE phase should have retained
the same recurrence times as before. Note, however, that the latter
scenario implies a triple system whose dynamical stability should be
further studied.

\subsection{The precursor X-ray flare}

The X-ray flare detected at the beginning of TDE~2 is a novel addition
to the rich phenomenology of X-ray TDEs evolution. Its peak luminosity
is similar to the subsequent TDE~2 peak (most likely associated with
the accretion phase), but its evolution timescales is much shorter
($50-90$~d as opposed to $\simeq 1-2$~yr). As mentioned, a precursor
X-ray flare preceding TDE~1 is also tentatively observed as shown in
Fig.~\ref{fig:AllPhases}.

Under the assumption that the repeating TDEs are due to the tidal
stripping of a giant star's envelope at pericenter, the two TDE light
curve should be similar, as observed. In this case, it is tempting to
associate the precursor flare (or flares) to the circularisation phase
in which the TDE tidal streams self-interact and circularise during
the formation of the accretion disc. As shown by
\citet{2016MNRAS.461.3760H}, if radiative cooling is efficient, the
debris circularise in a geometrically thin ring-like accretion flow
and the dissipated energy during circularisation can give rise to
observable precursor flares (with respect to the subsequent accretion
phase) that can be luminous enough to be detectable, especially for
parabolic TDEs (and likely highly eccentric ones), with an upper limit
comparable to the Eddington luminosity. A similar conclusion is also
reached by \citet{2014ApJ...788...99B} in the case of the pTDE of a
red giant envelope (which is likely relevant for GSN~069), showing
that the luminosity due to circularisation can exceed the fallback
luminosity at early times. Precursor X-ray flares are also found in
kinematic simulations by \citet{2021PhRvD.104j3019R} if the
circularisation timescale is significantly shorter than the
characteristic fallback timescale at which the accretion phase peaks.

On the other hand, assuming instead that TDEs and QPEs are all related
to the same progenitor, we note that the precursor flare relatively
short timescales are roughly consistent with the partial or full
disruption of a compact stellar core plus small remaining envelope
\citep{2012Natur.485..217G,2015MNRAS.454.2321S,2017ApJ...841..132L}. In
fact, timescales $\lesssim 100$~d, as observed for the precursor
flare, are also compatible with typical optical TDEs in general
\citep{2021ApJ...908....4V}. The subsequent (accretion-dominated)
evolution of TDE~2 might evolve on timescales still dictated by the
fallback rate from TDE~1 (dominant with respect to TDE~2 in terms of
disrupted mass), which could account for the years-long TDE~2 overall
timescales once the fact that TDE~2 does not occur in a clean
environment is considered. It is also worth to point out that the
disruption of a WD-like remnant may be associated with tidal
compression and subsequent ignition, leading to a thermonuclear
explosion accompanied by a relatively short-lived thermal-like soft
X-ray flare \citep{2009ApJ...695..404R}.

As mentioned, the two scenarios outlined above may be tested with
future observations. A third TDE $\sim 9$~yr after TDE~2 would confirm
the tidal stripping of the giant star's envelope at pericenter, and
the properties of a putative new QPE phase may be used to infer the
origin of the stellar-like companion that is responsible for QPEs, as
discussed in the previous Sect. Other theoretical origins of the
precursor flare are possible, and they should be explored in future
theoretical and numerical works adopting parameters that can account
for the overall evolution (and QPEs) in GSN~069. As discussed here,
understanding the origin of the precursor flare(s) may help
elucidate the relationship between QPEs and repeating TDEs in
GSN~069. We stress, once more, that the presence of a pre-existing
accretion flow due to the accretion of debris from TDE~1 should be
properly taken into account when describing the overall evolution and,
in particular, the properties of TDE~2.

\section{Conclusions}
\label{sec:conclusions}

We study the long-term evolution of GSN~069 and the properties of its
QPEs using 11 pointed \xmm, one \cha, and 34 \swift\ X-ray observations
obtained in the $12$~yr since its first X-ray detection in July 2010
during an \xmm\ slew. QPEs in GSN~069 are a transient phenomenon
observed between December 2018 and January 2020. However, considering
that QPEs might have been missed in observations with high
quiescent flux level, and accounting for observational gaps, the QPE
life-time could be considerably longer.

We report repeating TDEs in GSN~069 with two events separated by about
$9$~yr. The presence of TDE~2 suggests that the first event (TDE~1)
was a partial TDE. We show that TDE~1 is most likely produced by the
partial disruption of an evolved star, leaving a compact stellar core
possibly surrounded by a remaining envelope. We detect a precursor
X-ray flare $400-500$~d before the TDE~2 peak. The precursor might be
related to the circularisation phase during the formation of the
accretion flow. A similar precursor flare may be present $\simeq
500$~d before TDE~1, but its detection is only tentative.

The X-ray spectral evolution during QPEs suggests an origin in shocks,
consistent with a compact region that heats up rapidly, and then cools
down by emitting X-rays while expanding by a factor of $\sim 3$ in
radius in $\lesssim 1$~hr. Shocks may be produced by impacts between
the accretion flow and an orbiting stellar companion, the interaction
between the incoming stream(s) and the flow in mass transfer
scenarios, or the interaction between neighbouring misaligned rings in
a warped, unstable disc. A purely accretion-based origin of QPEs
cannot be discarded, and would imply a very compact, ring-like
emitting region for QPEs. In any case, we stress that the presence of
a pre-existing accretion flow from TDE~1 should be taken into account
in future realisations of QPE models.

During observations with QPEs, the quiescent level
emission exhibits a QPO with a period equal to the separation between
consecutive QPEs. Assuming that the QPO is produced by mass accretion
rate fluctuations caused by shocks and that QPEs are associated with
an orbiting body, the eccentricity of the QPE-inducing orbiter must be
$e\gtrsim 0.84$, likely ruling out star-disc collisions scenarios
associated with a low-eccentricity orbit. However, this conclusion
strongly depends on the assumed QPO origin and cannot be considered as
conclusive yet. We encourage further theoretical work that, informed
also by our observational results, should ideally be able to account
for the QPEs and QPO properties in a self-consistent manner.

As for the two TDEs that we observe $\sim 9$~yr apart, and their
connection with QPEs, we propose two possible scenarios. In the first,
all events (TDE~1, QPEs, and TDE~2) are associated with the same
progenitor (an evolved star), that is QPEs are due to the remnant of
TDE~1 that survives thousands of pericenter passages before being
partially or fully disrupted at TDE~2. In this case, as it is
difficult to place an evolved star on a very short-period orbit, the
progenitor could have been captured long before TDE~1, and been driven
to its tidal radius by subsequent orbital evolution. The long
timescales associated to TDE~2 (only marginally shorter than the TDE~1
timescales) are somewhat difficult to account for if TDE~2 disrupted a
significantly more compact object than TDE~1, as expected within this scenario. However, the disrupted
mass at TDE~2 likely interacts with the existing accretion flow (from
TDE~1). Since the disrupted mass at TDE~2 is likely comparable to that
of the region of the disc with which it interacts, the disc can
be significantly disturbed, or even disrupted. The disc
has then to re-stabilise, or re-form, by accreting not only the debris
from TDE~2 after circularisation, but also the remaining bound mass
from TDE~1. It is therefore possible that the TDE~2 timescales are
still dominated by the TDE~1 fallback rate, as the disrupted mass at
TDE~1 is likely dominant. 

A second scenario assumes that the two TDEs represent repeating
partial disruptions of the envelope of an evolved star at each
pericenter passage so that the star is on a $\sim 9$~yr orbit, thus
removing the difficulty to place a large star on a very short-period
orbit and also accounting naturally for the similar shape and
timescales of TDE~1 and TDE~2. If so, QPEs cannot be produced by the
remnant surviving TDE~1. A QPE-inducing orbiter could have been
captured, possibly through the Hills mechanism, at a different epoch
on its much shorter-period orbit ($P\simeq 9-18$~hr, depending on the
adopted QPE model). The stability of the putative triple system needs
to be studied. On the other hand, QPEs may be induced not by an
orbiter, but rather by shocks associated with disc tearing in a
warped, unstable disc, thus removing the need for a triple system in
GSN~069.

We note that, in all cases, as the tidal radius at TDE~2 is reached
smoothly from larger radii, TDE~2 is most likely another partial
rather than full TDE. If so, a third event is expected in the
future. Would a third TDE occur $\sim 9$~yr after TDE~2, the scenario
in which the two repeating TDEs in GSN~069 represent the partial
disruption of the envelope of an evolved star on a $\sim 9$~yr orbit
will be proved right. Future observations may also detect a new QPE
phase whose beginning could be identified with the last
observation in which QPEs are seen (XMM6, just after TDE~2). In mass
transfer scenarios, the detection of clear new QPE phase would signal
than an orbiter is still present after TDE~2. The properties of the
putative new QPE phase (most notably the QPE recurrence time,
i.e. orbital period) will significantly help to constrain the origin
of the orbiter, namely whether it is a new remnant that survived
TDE~2, or the same object that was producing previous QPEs. Changes in
the QPE recurrence time may also help constraining the alternative
disc tearing scenario, as the instability strength, location, and
timescales are expected to vary with disc properties.

Future X-ray observations of GSN~069 on both long and short timescales
will most likely allows us to pinpoint the origin of repeating TDEs,
of QPEs, and of the possible TDE-QPE connection in this galactic
nucleus, with consequences for QPE models in the other galaxies where
QPEs have been identified so far. Our work provides a comprehensive
summary of the rich phenomenology of GSN~069 that can be used to
inform further theoretical and numerical studies on the origin of
QPEs.  QPE models are evolving fast, and will ideally be able to not
only reproduce the most relevant properties of one specific source,
but also to account for the diversity of QPE phenomenology that is
starting to be revealed as new QPE sources are discovered and new data
are accumulated.

\begin{acknowledgements}

We thank the referee for a constructive review of our work that
contributed to improve our paper. GM warmly thanks Andrew King for
valuable discussions. GM, MG, and RA also thank A. Sesana,
E. Bortolas, A. Franchini, M. Bonetti, and A. Lupi for a pleasant and
useful QPE brain-storming session in Milan. Based on observations
obtained with \xmm\, an ESA science mission with instruments and
contributions directly funded by ESA Member States and NASA. This work
made use of data supplied by the UK \swift\ Science Data Centre at the
University of Leicester, and we thank the \swift\ science team for
organising and scheduling a series of unanticipated TOO observations
of GSN~069. The scientific results reported in this article are also
based on observations made by the \cha\ X-ray Observatory. MG is
supported by the ``Programa de Atracci\'on de Talento'' of the
Comunidad de Madrid, grant number 2018-T1/TIC-11733. SB acknowledges
financial support from the Italian Space Agency under grant ASI-INAF
2017-14-H.O.

\end{acknowledgements}

\bibliographystyle{aa}
\bibliography{biblio}

\begin{appendix}

\section{Light curves model}
\label{sec:tab}


\begin{table*}[htb!]
        \centering
        \caption{Best fitting parameters for fits to the 0.4--1~keV
          light curves from all observations with QPEs.}
        \label{tab:qpefit}
        \begin{tabular}{lcccccc} 
          \hline
          \T  &T$_{\mathrm{rec}}$ & C & N  & $\sigma$ & $\chi^2/{\mathrm{dof}}$ \\
          \T & [~s~] & [~cts~s$^{-1}$~] & [~cts~s$^{-1}$~] & [~s~] & \B \\
          \hline
          \T {\bf XMM3} & & & & & 414/246    \\
          \T QPE$^{\mathrm{(w)}}$ & $29757\pm 34$  & $0.046\pm 0.001$ & $0.96\pm 0.04$ & $810\pm 25$ \\
          \T QPE$^{\mathrm{(s)}}$ &              & $-$              & $1.37\pm 0.04$ & $809\pm 18$ \B \\
          \hline
          \T {\bf XMM4} & &  & & & 875/644  \\
          \T QPE$^{\text{(s)}}$ & $33089\pm 40$ & $0.0451\pm 0.0009$  & $1.24\pm 0.04$ & $818\pm 25$\\
          \T QPE$^{\text{(w)}}$ & $31266\pm 37$ & $-$                 & $0.87\pm 0.03$ & $817\pm 23$\\
          \T QPE$^{\text{(s)}}$ & $32592\pm 35$ & $-$                 & $1.10\pm 0.04$ & $805\pm 19$ \\
          \T QPE$^{\text{(w)}}$ & $31656\pm 35$ & $-$                 & $0.93\pm 0.04$ & $800\pm 22$ \\
          \T QPE$^{\text{(s)}}$ &             & $-$                 & $1.22\pm 0.04$ & $808\pm 18$ \B  \\
          \hline
          \T {\bf Chandra} & &&& & 14/61   \\
          \T QPE$^{\text{(s)}}$ & $33551\pm 191$ & $0.069 \pm 0.01$    & $1.59\pm 0.23$ & $828\pm 91$ \\
          \T QPE$^{\text{(w)}}$ & $31865\pm 211$ & $-$                 & $0.81\pm 0.17$ & $757\pm 119$ \\
          \T QPE$^{\text{(s)}}$ &              & $-$                 & $1.06\pm 0.19$ & $708\pm 99$ \B  \\
          \hline
          \T {\bf XMM5} &&&&& 867/636   \\
          \T QPE$^{\text{(w)}}$ & $30081\pm 42$ & $0.0242\pm 0.0006$  & $0.56\pm 0.03$ & $743\pm 29$ \\
          \T QPE$^{\text{(s)}}$ & $33458\pm 38$ & $-$                 & $0.93\pm 0.04$ & $810\pm 22$ \\
          \T QPE$^{\text{(w)}}$ & $31182\pm 38$ & $-$                 & $0.72\pm 0.03$ & $759\pm 22$ \\
          \T QPE$^{\text{(s)}}$ & $33373\pm 41$ & $-$                 & $0.89\pm 0.03$ & $831\pm 21$ \\
          \T QPE$^{\text{(w)}}$ &             & $-$                 & $0.76\pm 0.03$ & $854\pm 30$ \B  \\
          \hline
          \T {\bf XMM6} &&&&& 745/634   \\
          \T QPE$^{\text{(s)}}$ & $26428\pm 88$  & $0.070\pm 0.001$  & $0.28\pm 0.02$ & $611\pm 40$ \\
          \T QPE$^{\text{(w)}}$ & $28794\pm 88$  & $-$               & $0.22\pm 0.02$ & $749\pm 52$ \\
          \T QPE$^{\text{(s)}}$ & $35700\pm 76$  & $-$               & $0.41\pm 0.03$ & $851\pm 51$ \\
          \T QPE$^{\text{(w)}}$ &              & $-$               & $0.31\pm 0.02$ & $744\pm 50$ \B  \\
          \hline
        \end{tabular}
\tablefoot{We use time bins of 200~s and 500~s for \xmm\ and
  \cha\ observations respectively. Our model consists of an
  observation-dependent constant C representing the quiescent count
  rate and a series of Gaussian functions with normalisation N and
  width $\sigma$ describing the QPEs. Superscripts in the first column
  denote strong (s) and weak (w) QPEs respectively. The separation
  between a QPE and the next is denoted by $T_{\rm{rec}}$. The
  \cha\ light curve has been re-scaled to match the \xmm\ EPIC-pn
  detector effective area prior to fitting (see text for
  details). Errors represent the 1-$\sigma$ confidence intervals.}
\end{table*}
\FloatBarrier

\section{Quiescent level variability}
\label{app:QPE-QPO}

\begin{figure*}
\centering
\includegraphics[width=18cm]{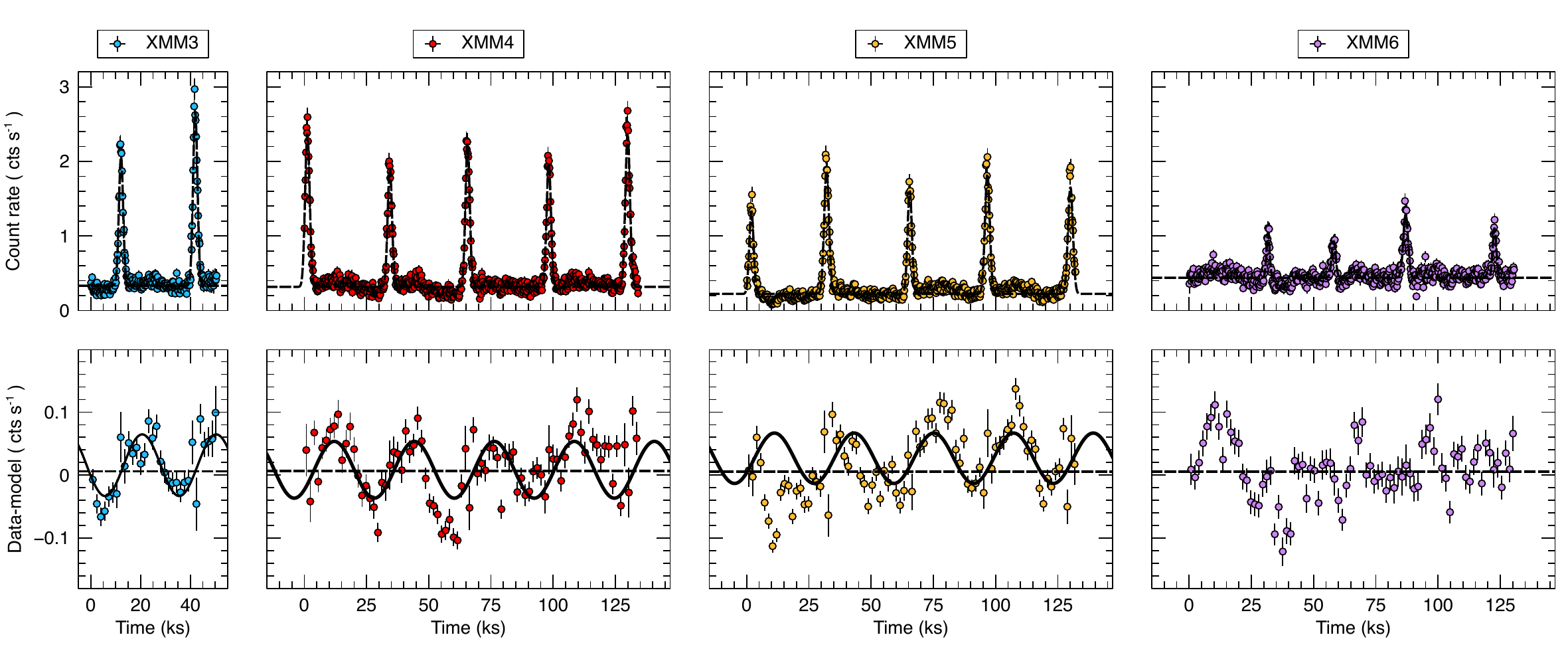}
{\vspace{-0.2cm}}
\caption{Quiescent level variability. In the upper panels, we show the
  0.2-1~keV light curves from all \xmm\ observations with QPEs,
  together with their best-fitting models comprising a constant and a
  series of Gaussian functions to describe the quiescent level and
  QPEs respectively. The lower panels show the corresponding residual
  light curves re-binned by a factor of 8, together with two models, a
  constant (dashed lines) and a sine function (solid lines) with
  a period equal to the average observation-dependent recurrence time.}
\label{fig:qpos}
\end{figure*}

In the upper panels of Fig.~\ref{fig:qpos}, we show light curves from
all \xmm\ observations in which QPEs are observed. We ignored the
\cha\ data as the quiescent level was barely detected, making it
difficult to study its variability. As all data are from the
\xmm\ EPIC-pn detector, there was no need to restrict the analysis to
energies above 0.4~keV, and we used the full 0.2-1~keV band. The light
curves were modelled with a constant and a series of Gaussian
functions describing the quiescence and QPEs respectively. The lower
panels of Fig.~\ref{fig:qpos} show the resulting residual light curves
that have been re-binned for visual clarity to highlight the
relatively long-term variations (tens of ks timescales) of the
quiescent level emission. In all cases but the XMM6 observation, the
quiescent level exhibits broad excess emission following most QPEs,
regardless of their strong or weak type. Excluding XMM6, an excess can
be clearly seen after nine out of the ten QPEs that are followed by a
significant exposure. The only QPE that is not followed by any excess
if the first during XMM5, which is also the weakest QPE detected in
the XMM3 to XMM5 observations (see Table~\ref{tab:qpefit}). The dashed and solid
lines represent models describing the residual light curves either
with a constant (dashed lines) or with an additional periodic function
(solid lines) with period fixed at the average observation-dependent
recurrence time between consecutive QPEs. The sinusoidal models
represent a very significant improvement with respect to the constant
one in all cases. In particular, re-fitting the original light curves
with the addition of a sine function improves the statistical result
by $\Delta \chi^2 =-114,\, -261,\, -63$ for the XMM3, XMM4, and XMM5
observations for $-2$ degrees of freedom, see Table~\ref{tab:qpefit}
for the statistical quality of the baseline model fits when the sine
function is not included.

\end{appendix}
 
\end{document}